\documentclass[a4paper,usenatbib,useAMS]{mnras}
\usepackage{txfonts}

\usepackage[T1]{fontenc}
\usepackage{ae,aecompl}

\usepackage{siunitx}
\usepackage{graphicx}	% Including figure files
\usepackage{multirow}
\usepackage{adjustbox}

\usepackage{bm}

\newcommand{\LCDM}{$\Lambda$CDM}
\newcommand{\nlenses}{13}

\usepackage{eso-pic}% http://ctan.org/pkg/eso-pic

\AddToShipoutPictureBG*{%
  \AtPageUpperLeft{%
    \hspace{0.75\paperwidth}%
    \raisebox{-3.5\baselineskip}{%
      \makebox[0pt][l]{\textnormal{DES 2018-0371}}
}}}%

\AddToShipoutPictureBG*{%
  \AtPageUpperLeft{%
    \hspace{0.75\paperwidth}%
    \raisebox{-4.5\baselineskip}{%
      \makebox[0pt][l]{\textnormal{FERMILAB PUB-18-331-AE}}
}}}%

\title[Uniform lens modelling]{Is every strong lens model unhappy in its own way? Uniform modelling of a sample of \nlenses\ quadruply$+$ imaged quasars}

\author[A. J. Shajib et al.]
{\parbox{\textwidth}
	{
	A. J. Shajib,$^{1}$\thanks{E-mail: ajshajib@astro.ucla.edu }
	S. Birrer,$^{1}$
	T. Treu,$^{1, 2}$
	M. W. Auger,$^{3}$ 
	A. Agnello,$^{4}$
	T. Anguita,$^{5, 6}$
	E. J. Buckley-Geer,$^{7}$
	J. H. H. Chan,$^{8}$
	T. E. Collett,$^{9}$
	F. Courbin,$^{8}$
	C. D. Fassnacht,$^{10}$
	J. Frieman$^{7, 11}$
	I. Kayo,$^{12}$
	C. Lemon,$^{3}$
	H. Lin,$^{7}$
	P. J. Marshall,$^{13}$
	R. McMahon,$^{3}$
	A. More,$^{14}$
	N. D. Morgan,$^{15}$
	V. Motta,$^{16}$
	M. Oguri,$^{17, 18, 19}$
	F. Ostrovski,$^{3}$
	C. E. Rusu,$^{20,21}$
	P. L. Schechter,$^{22}$
	T. Shanks,$^{23}$
	S. H. Suyu,$^{24, 25, 26}$
	G. Meylan,$^{8}$
	T.~M.~C.~Abbott,$^{27}$
	S.~Allam,$^{7}$
	J.~Annis,$^{7}$
	S.~Avila,$^{9}$
	E.~Bertin,$^{28, 29}$
	D.~Brooks,$^{30}$
	A.~Carnero~Rosell,$^{31, 32}$
	M.~Carrasco~Kind,$^{33, 34}$
	J.~Carretero,$^{35}$
	C.~E.~Cunha,$^{36}$
	L.~N.~da Costa,$^{31, 32}$
	J.~De~Vicente,$^{37}$
	S.~Desai,$^{38}$
	P.~Doel,$^{30}$
	B.~Flaugher,$^{7}$
	P.~Fosalba,$^{39, 40}$
	J.~Garc\'ia-Bellido,$^{41}$
	D.~W.~Gerdes,$^{42, 43}$
	D.~Gruen,$^{36, 44}$
	R.~A.~Gruendl,$^{33, 34}$
	G.~Gutierrez,$^{7}$
	W.~G.~Hartley,$^{30, 45}$
	D.~L.~Hollowood,$^{46}$
	B.~Hoyle,$^{47, 48}$
	D.~J.~James,$^{49}$
	K.~Kuehn,$^{50}$
	N.~Kuropatkin,$^{7}$
	O.~Lahav,$^{30}$
	M.~Lima,$^{51, 31}$
	M.~A.~G.~Maia,$^{31, 32}$
	M.~March,$^{52}$
	J.~L.~Marshall,$^{53}$
	P.~Melchior,$^{54}$
	F.~Menanteau,$^{33, 34}$
	R.~Miquel,$^{55}$
	A.~A.~Plazas,$^{56}$
	E.~Sanchez,$^{37}$
	V.~Scarpine,$^{7}$
	I.~Sevilla-Noarbe,$^{37}$
	M.~Smith,$^{57}$
	M.~Soares-Santos,$^{58}$
	F.~Sobreira,$^{59, 30}$
	E.~Suchyta,$^{60}$
	M.~E.~C.~Swanson,$^{34}$
	G.~Tarle,$^{43}$
	A.~R.~Walker$^{27}$
} \\
\vspace{0.0cm} \\
\parbox{\textwidth}
{Affiliations are listed at the end of the paper
}}

\date{Accepted 2018 December 9. Received 2018 November 2; in original form 2018 July 23}

\pubyear{2018}

\begin{document}

\label{firstpage}
\pagerange{\pageref{firstpage}--\pageref{lastpage}}
\maketitle

\begin{abstract}
Strong-gravitational lens systems with quadruply-imaged quasars (quads) are unique probes to address several fundamental problems in cosmology and astrophysics. Although they are intrinsically very rare, ongoing and planned wide-field deep-sky surveys are set to discover thousands of such systems in the next decade. It is thus paramount to devise a general framework to model strong-lens systems to cope with this large influx without being limited by expert investigator time. We propose such a general modelling framework (implemented with the publicly available software \textsc{Lenstronomy}) and apply it to uniformly model three-band \textit{Hubble Space Telescope} Wide Field Camera 3 images of \nlenses\ quads. This is the largest uniformly modelled sample of quads to date and paves the way for a variety of studies. To illustrate the scientific content of the sample, we investigate the alignment between the mass and light distribution in the deflectors. The position angles of these distributions are well-aligned, except when there is strong external shear. However, we find no correlation between the ellipticity of the light and mass distributions. We also show that the observed flux-ratios between the images depart significantly from the predictions of simple smooth models. The departures are strongest in the bluest band, consistent with microlensing being the dominant cause in addition to millilensing. Future papers will exploit this rich dataset in combination with ground based spectroscopy and time delays to determine quantities such as the Hubble constant, the free streaming length of dark matter, and the normalization of the initial stellar mass function.
\end{abstract}


%%%%%%%%%%%%%%%%%%%%%%%%%%%%%%%%%%%%%%%%%%%%%%%%%%

%%%%%%%%%%%%%%%%% BODY OF PAPER %%%%%%%%%%%%%%%%%%

\section{Introduction} \label{sec:intro}
 
Strong gravitational lensing is the effect where light from a background object is deflected by a foreground mass distribution (e.g. galaxy or galaxy cluster) and multiple images of the background object form.  Strong gravitational lenses are powerful probes to answer a variety of astrophysical and cosmological questions \citep[see, e.g.,][]{Treu10b}, as we discuss briefly below. 

According to the concordance model in cosmology, our Universe consists of 5 per cent baryonic matter, 26 per cent dark matter, and 69 per cent dark energy accounting for a cosmological constant $\Lambda$ \citep{PlanckCollaboration18}. This model is known as the $\Lambda$ cold dark matter (\LCDM) model. The predictions of the \LCDM\ model have been extensively tested with good agreement to observations spanning from the largest scale up to the horizon down to $\sim$1 Mpc \citep[e.g.][]{Dawson13, Shajib16, PlanckCollaboration18}. However, there also have been observations  that are in tension with the flat \LCDM\ paradigm. For instance, there is a tension at the $\gtrsim 3 \sigma$ level between the local measurement of $H_0$ from Type Ia supernovae \citep{Riess16, Riess18b, Riess18, Bernal16} and that extrapolated from the \textit{Planck} cosmic microwave background measurement for a flat \LCDM\ cosmology. This tension may arise from unknown systematic uncertainties in one or both of the measurements, or might point to new physics, e.g. additional species of relativistic particles, a non-flat cosmology, or dynamic dark energy. Therefore, it is crucial to have precise and independent measurements of $H_0$ to settle this discrepancy.

In a gravitational lens, if the background source is time-variable (typically a quasar, but also a supernova as originally proposed), the delay between the arrival time of photons for the different images can be used to measure the so-called `time-delay distance' \citep{Refsdal64, Suyu10}. This distance is inversely proportional to $H_0$, thus it can be used to constrain $H_0$ and other cosmological parameters \citep[for a detailed review, see][]{Treu16}. $H_0$ has been determined to 3.8 per cent precision using three lens systems in the flat \LCDM\ cosmology \citep{Suyu10, Suyu13, Suyu17, Sluse17, Rusu17, Wong17, Bonvin17, Tihhonova18}. With a large sample size of about 40 lenses, it is possible to measure $H_0$ with the per cent precision \citep{Jee16, Shajib18} necessary to resolve the $H_0$ tension and make the most of other dark energy probes \citep{Linder11,Suyu12,Weinberg13}. 

One of the baryonic components in dark matter is low-mass star. Surprisingly, recent studies have shown that the low-mass star contribution in massive elliptical galaxies is significantly underestimated if the stellar initial mass function (IMF) of the Milky Way is assumed \citep{Treu10, vanDokkum10, Auger10, Cappellari12, Schechter14}. Precise knowledge about the IMF is key in measuring almost any extragalactic quantity involving star and metal formation. Measuring the stellar mass-to-light ratio in the deflectors of quadruply imaged lensed quasars (henceforth quads) from microlensing statistics provides one of the most robust methods to constrain the IMF \citep[e.g.][]{Oguri14, Schechter14}.

% Furthermore, there have been discrepancies between the theoretical predictions of the \LCDM\ model and observations at the non-linear regime of the smallest scales. The major disagreements are observed in the mismatch between the abundance and density profiles of luminous satellite galaxies in the Milky Way \citep[e.g.][]{Kauffmann93, Klypin99, Moore99, Boylan-Kolchin11}. It is necessary to probe the small-scale mass distribution and structure formation to go beyond the \LCDM\ model to resolve these tensions. 

Quads also provide a unique test of small-scale structure formation \citep{Kauffmann93, Witt95, Klypin99, Moore99, Boylan-Kolchin11,Metcalf01, Dalal02, Yoo06, Keeton09, Moustakas09} by measuring the subhalo mass function \citep[][see also for studies involving extended source only, \citealt{Koopmans05, Vegetti09, Vegetti10, Vegetti12, Vegetti18, Hezaveh16}]{Metcalf02, Kochanek04, Amara06, Metcalf12, Nierenberg14, Nierenberg17, Xu15, Birrer17}, independent of their luminosity function. With a large sample of quads, \citet{Gilman18} demonstrate the possibility of constraining the free-streaming length of dark matter particles more precisely than current limits based on the Lyman-$\alpha$ forest \citep{Viel13}.

Until recently, all of these methods could only be applied to a small sample of known quads. However, such systems are currently being discovered at a rapidly increasing rate due to multiple strong-lens search efforts involving various large-area sky surveys \citep[e.g.][Treu et al. 2018, submitted]{Agnello15, Agnello18b, Agnello18c, Williams17, Williams18, Schechter17, Sonnenfeld18, Lemon18, Anguita18}. With more deep wide-field surveys, e.g. \textit{Wide-Field Infrared Survey Telescope}, Large Synoptic Survey Telescope, \textit{Euclid}, etc., coming online within the next decade, the sample size of quads is expected to increase by two orders of magnitude or more \citep{Oguri10, Collett15}. 

Modelling such lens systems has so far been carried out for individual systems while fine-tuning the modelling approach on a case-by-case basis. However, with the rapidly increasing rate of discovery, it is essential to develop a modelling technique that is applicable to a wide variety of quads to efficiently reduce the time and human labour necessary in this endeavour. Given the large diversity in the morphology and complexity of quads, this is an interesting problem to pose: is every quad different or `unhappy in its own way' that requires careful decision-making by a human in the modelling procedure, or are the quads similar or `happy' to some extent so that a uniform modelling technique can be applied to generate acceptable models without much human intervention?

Recently, some initial strides have been undertaken along the lines of solving this problem for strong lenses with extended sources. \citet{Nightingale18} devised an automated lens modelling procedure using Bayesian model comparison. \citet{Hezaveh17} and \citet{PerreaultLevasseur17} applied machine learning techniques to automatically model strong gravitational lenses and constrain the model parameters. In this paper, we devise a general framework or decision-tree that can be applied to model-fitting of quads both in a single band and simultaneously in multiple bands. We implement this uniform modelling approach using the publicly available lens-modelling software \textsc{Lenstronomy} \citep[][based on \citealt{Birrer15}]{Birrer18} to a sample of \nlenses\ quads from the \textit{Hubble Space Telescope} (\textit{HST}) data in three bands. \textsc{Lenstronomy} comes with sufficient modelling tools and the architecture allows a build-up in complexity as presented in this work. We report the model parameters and other derived quantities for these lens systems.

To demonstrate the scientific capabilities of such a sample of strong-lens systems, we study the properties of the deflector galaxy mass distribution, specifically the alignment of the mass and light distributions in them. The distribution of dark matter and baryons in galaxies can test predictions of \LCDM\ and galaxy formation theories \citep[e.g.][]{Dubinski94, Ibata01, Kazantzidis04, Maccio07, Debattista08, Lux12, Read14}. N-body simulations with only dark matter particles predict nearly triaxial, prolate haloes \citep{Dubinski91, Warren92, Navarro96, Jing02, Maccio07}. In the presence of baryons, the halos become rounder \citep{Dubinski91, Dubinski94, Warren92}. With a modestly-triaxial luminous galaxy embedded in the dark matter halo, large misalignments ($\sim 16 \pm 19 \degr$) between the projected light and mass major axes can be produced \citep{Romanowsky98}. For disk galaxies, the dark matter distribution is shown to be well-aligned with the light distribution \citep[][]{Katz91, Dubinski91, Debattista08}. %Dissipationless collapse produce highly elliptical dark matter halos \citep{Dubinski91, Warren92}. 

As the lensing effect is generated by mass, strong gravitational lenses give independent estimates of the mass distribution that can be compared with the observed light distribution. The deflectors in quads are typically massive ellipticals (with Einstein mass $M_{\rm E} \gtrsim 10^{11.5} M_{\odot}$). Most of the massive ellipticals are observed to be slow rotators with uniformly-distributed misalignments between the kinematic and photometric axes \citep{Ene18}. The uniform distribution of misalignments suggests these massive ellipticals to be intrinsically triaxial. Massive ellipticals can also have of stellar populations and dust distribution with different geometries producing isophotal twist which can create a misalignment between the mass and light distributions \citep{Goullaud18}. For lens systems, a tight alignment within $\pm 10 \degr$ between the major axes of the mass and light distribution has been observed for deflector galaxies with weak external shear, whereas galaxies with strong external shear can be highly misaligned \citep{Keeton98, Kochanek02, Treu09, Sluse12, Gavazzi12, Bruderer16}. However, there has been some conflict about the correlation between the ellipticity of the mass and light distributions with reports of both strong correlation \citep{Sluse12, Gavazzi12} and no correlation \citep{Keeton98, Ferreras08, Rusu16}. 

This paper is organized as follows. In Section \ref{sec:data}, we describe the data used in this study. We describe our methodology in Section \ref{sec:modeling} and the results in Section \ref{sec:result}. Finally, we summarize the paper followed by a discussion in Section \ref{sec:summary}. When necessary, we adopt a fiducial cosmology with $H_0=70$ km s$^{-1}$ Mpc$^{-1}$, $\Omega_{\rm m} = 0.3$, $\Omega_{\rm \Lambda}=0.7$, and $\Omega_{\rm r} = 0$. All magnitudes are given in the AB system.

\begin{table*}
	\centering
	\caption{Observation information and references for the lens systems.}
	\label{tab:lens_list}
	\begin{adjustbox}{width=\textwidth}
	\begin{tabular}{llccrl} % four columns, alignment for each
		\hline
		\multirow{3}{*}{System name} & \multirow{3}{*}{Observation date} & \multicolumn{3}{c}{Total exposure time} & \multirow{3}{*}{Reference} \\
		& & \multicolumn{3}{c}{(seconds)} \\
		     & & F160W & F814W & F475X \\
		\hline
	%\end{tabular}
	%\begin{tabular}{lccccl}
		PS J0147+4630     	& 2017 Sept 13	& 2196.9	& 1348.0	& 1332.0	& \cite{Berghea17} 	 \\
		SDSS J0248+1913   	& 2017 Sept 5	& 2196.9	& 1428.0	&  994.0	& Ostrovski et al. (in preparation), \citet{Delchambre18} \\
		ATLAS J0259-1635   	& 2017 Sept 7	& 2196.9	& 1428.0	&  994.0	& \citet{Schechter18} 	 \\
		DES J0405-3308    	& 2017 Sept 6	& 2196.9	& 1428.0	& 1042.0	& \citet{Anguita18} \\
		DES J0408-5354    	& 2018 Jan 17	& 2196.9	& 1428.0	& 1348.0	& \citet{Lin17, Diehl17, Agnello17} 	 \\
		DES J0420-4037    	& 2017 Nov 23	& 2196.9	& 1428.0	& 1158.0	& Ostrovski et al. (in preparation) 	 \\
		PS J0630-1201     	& 2017 Oct 5	& 2196.9	& 1428.0	&  980.0	& \citet{Ostrovski18, Lemon18} 	 \\
		SDSS J1251+2935   	& 2018 Apr 26	& 2196.9	& 1428.0	& 1010.0	& \citet{Kayo07}	 \\
		SDSS J1330+1810   	& 2018 Aug 15 	& 2196.9	& 1428.0	&  994.0	 & \citet{Oguri08} 	 \\
		SDSS J1433+6007   	& 2018 May 4	& 2196.9	& 1428.0	& 1504.0	& \citet{Agnello18}	 \\
		PS J1606-2333     	& 2017 Sept 1	& 2196.9	& 1428.0	&  994.0& \citet{Lemon18}	 \\
		DES J2038-4008    	& 2017 Aug 29	& 2196.9	& 1428.0	& 1158.0	& \citet{Agnello18c}	 \\
		WISE J2344-3056  	& 2017 Sept 9	& 2196.9	& 1428.0	& 1042.0	& \cite{Schechter17} 	 \\
		\hline
		\end{tabular}
		\end{adjustbox}
\end{table*}

\section{\textit{HST} sample} \label{sec:data}
Our sample consists of twelve quads and one five-image system. Some of these systems were discovered by the STRong-lensing Insights into the Dark Energy Survey (STRIDES)\footnote{STRIDES is a Dark Energy Survey Broad External Collabora- tion; PI: Treu. \url{http://strides.astro.ucla.edu}.} collaboration [STRIDES paper I \citet{Treu18}, paper II \citet{Anguita18}, and paper III Ostrovski et al. (in preparation)], some are recent discoveries by independent searches outside of the Dark Energy Survey (DES), and some are selected from the literature. In this section, we first describe the high-resolution imaging data obtained through \textit{HST}. We then briefly describe the lens systems in the sample.

\subsection{Data}
Images of the lenses were obtained using the \textit{HST} Wide Field Camera 3 (WFC3) in three filters: F160W in the infrared (IR) channel, and F814W and F475X in the ultraviolet-visual (UVIS) channel (ID 15320, PI Treu). In the IR channel filter, we used a 4-point dither pattern and \textsc{STEP100} readout sequence for the \textsc{MULTIACCUM} mode. This approach guarantees a sufficient dynamic range to expose both the bright lensed quasar images and the extended host galaxy. For the UVIS channel filters, we used a 2-point dither pattern. Two exposures at each position, one short and one long, were taken. Total exposure times for all the quads and the corresponding dates of observation are tabulated in Table \ref{tab:lens_list}. %Figure \ref{fig:lens_rgb} shows the RGB images of the lenses after combining the %three filters: F160W (red), F814W (green), and F475X (blue).

The data were reduced using \textsc{AstroDrizzle}. The pixel size after drizzling is 0\farcs08 in the F160W band, and 0\farcs04 in the F814W and F475X bands.

\subsection{Quads in the sample}
\begin{figure*}
	\includegraphics[width=2\columnwidth]{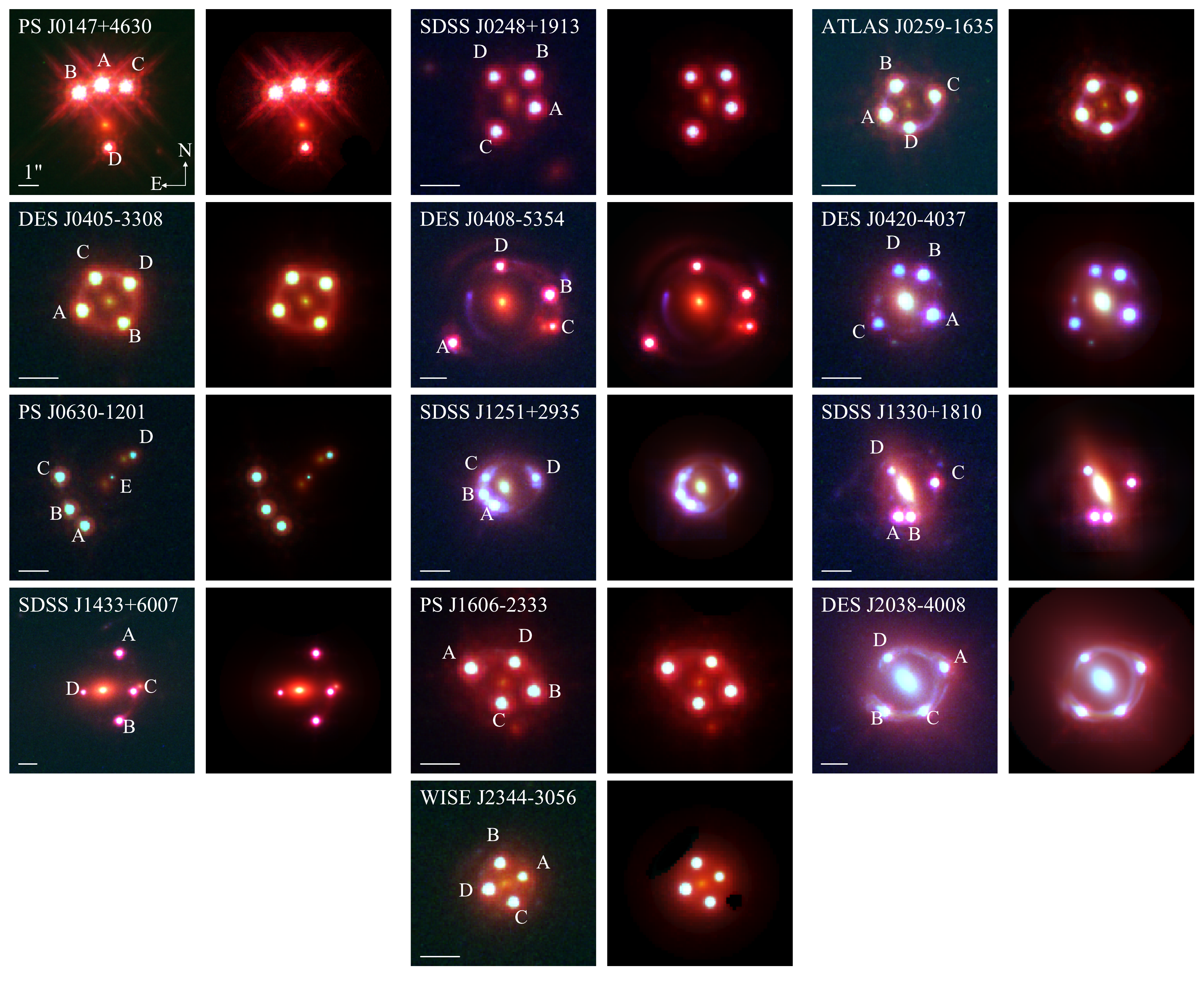}%{figures/montage.pdf} 
	\caption{
	Comparison between the observed (first, third and fifth columns) and reconstructed (second, fourth and sixth columns) strong-lens systems. The three \textit{HST} bands: F160W, F814W, and F475X are used in the red, green, and blue channels, respectively, to create the red-green-blue (RGB) images. Horizontal white lines for each system are rulers showing 1 arcsec. The relative intensities of the bands have been adjusted for each lens system for clear visualisation of the features in the system.  
	\label{fig:lens_rgb_w_model}
	}
	\end{figure*}

In this subsection, we give a brief description of each quad in our sample (Fig. \ref{fig:lens_rgb_w_model}).

\subsubsection{PS J0147+4630}
This quad was serendipitously discovered from the Panoramic Survey Telescope and Rapid Response System (Pan-STARRS) survey \citep{Berghea17}. The source redshift is $z_{\rm s} = 2.341 \pm 0.001$ \citep{Lee17} and the deflector redshift is $z_{\rm d} = 0.5716 \pm 0.0004$ \citep{Lee18}. Initial models from the Pan-STARRS data suggests a relatively large external shear $\gamma_{\rm ext} \sim 0.13$.

\subsubsection{SDSS J0248+1913}
This lens system was discovered in Sloan Digital Sky Survey (SDSS) imaging data using the  morphology-independent Gaussian-mixture-model supervised-machine-learning technique described in \citet{Ostrovski17} applied to SDSS u, g and i, and \textit{Wide-field Infrared Survey Explorer} (\textit{WISE}) W1 and W2 catalogue level  photometry (Ostrovski et al., in preparation).   The lensing nature was confirmed via optical spectroscopy with the Echellette Spectrograph and Imager (ESI) on the Keck telescope in 2016 December prior to the \textit{HST} observations presented here and will be described in Ostrovski et al. (in preparation). \citet{Delchambre18} report the independent discovery of this spectroscopically confirmed lensed system as a lensed quasar candidate using \textit{Gaia} observations. The lens system resides in a dense environment with several other galaxies within close proximity. Part of the lensed arc from the extended source is noticeable in the F160W band in IR.

\subsubsection{ATLAS J0259-1635}
This lens system was discovered in VLT Survey Telescope (VST)-ATLAS survey from candidates selected with quasar-like \textit{WISE} colours \citep{Schechter18}. The source for this system is at redshift $z_{\rm s} = 2.16$ \citep{Schechter18}.

\subsubsection{DES J0405-3308}
The discovery of this system is reported by \citet{Anguita18}. A complete or partial Einstein ring is noticeable in all the \textit{HST} bands. The source redshift is $z_{\rm s}=1.713 \pm 0.001$ \citep{Anguita18}.

\subsubsection{DES J0408-5354}
This system was discovered in the DES Year 1 data \citep{Lin17, Diehl17, Agnello17}. The deflector redshift is $z_{\rm d} = 0.597$ and the quasar redshift is $z_{\rm s} = 2.375$ \citep{Lin17}. This is a very complex lens system with multiple lensed arcs noticeable in addition to the quasar images. The sources of the lensed arcs can be components in the same source plane as the lensed quasar or they can be at different redshifts. This system has measured time-delays between the quasar images: $\Delta t_{\rm AB} = -112\pm2.1$ days, $\Delta t_{\rm AD} = -155.5\pm12.8$ days, and $\Delta t_{\rm BD} = -42.4 \pm 17.6$ days \citep{Courbin18}.

\subsubsection{DES J0420-4037}
This lens system was discovered in DES imaging data using the morphology-independent Gaussian-mixture-model supervised-machine-learning technique described in \citet{Ostrovski17} applied to DES g, r and i, Visible and Infrared Survey Telescope for Astronomy (VISTA) J and K, and \textit{WISE} W1 and W2 catalogue level photometry (Ostrovski et al., in preparation). Several small knots are noticeable near the quasar images that are possibly multiple images of extra components in the source plane.

\subsubsection{PS J0630-1201}
This system is a five-image lensed quasar system \citep{Ostrovski18}. The discovery was the result of a lens search from \textit{Gaia} data from a selection of lens candidates from Pan-STARRS and \textit{WISE}. The source redshift is $z_{\rm s} = 3.34$ \citep{Ostrovski18}.

\subsubsection{SDSS J1251+2935}
This quad was discovered from the SDSS Quasar Lens Search \citep[SQLS;][]{Oguri06, Inada12} \citep{Kayo07}. The source redshift is $z_{\rm s} = 0.802$ and the deflector redshift is $z_{\rm d} = 0.410$ measured from the SDSS spectra \citep{Kayo07}.

\subsubsection{SDSS J1330+1810}
This lens system was also discovered from the SQLS \citep{Oguri08}. The redshifts of the deflector and the source are $z_{\rm d} = 0.373$ and $z_{\rm s} = 1.393$, respectively \citep{Oguri08}.

\subsubsection{SDSS J1433+6007}
This lens system was discovered in the SDSS data release 12 photometric catalogue \citep{Agnello18}. The redshifts of the source and deflector are $z_{\rm s} = 2.737 \pm 0.003$ and $z_{\rm d} = 0.407 \pm 0.002$, respectively \citep{Agnello18}.

\subsubsection{PS J1606-2333}
This quad was discovered from \textit{Gaia} observations through a candidate search with quasar-like \textit{WISE} colours over the Pan-STARRS footprint \citep{Lemon18}. The main deflector has a noticeable companion near the South-most image.
	
\subsubsection{DES J2038-4008}
This lens system was discovered from a combined search in \textit{WISE} and \textit{Gaia} over the DES footprint \citep{Agnello18c}. The deflector and the source redshifts are $z_{\rm d}=0.230\pm0.002$ and $z_{\rm s}=0.777\pm0.001$, respectively \citep{Agnello18c}. This system has an intricate Einstein ring with complex features from the extended quasar host galaxy.

\subsubsection{WISE J2344-3056}
This lens system was discovered in the VST-ATLAS survey \citep{Schechter17}. This is a small-size quad with reported maximum image separation $\sim1\farcs1$. Several small and faint blobs are in close proximity, two of which are particularly noticeable near the North and East images.

\iffalse
\begin{figure*}
	\includegraphics[width=2\columnwidth]{figures/montage_10_w_names.png}
	\caption{
	RGB images of the lens systems.
	\label{fig:lens_rgb}
	}
\end{figure*}
\fi

\section{Lens modelling} \label{sec:modeling}

\begin{figure*} 
	\centering
	\includegraphics[width=2\columnwidth]{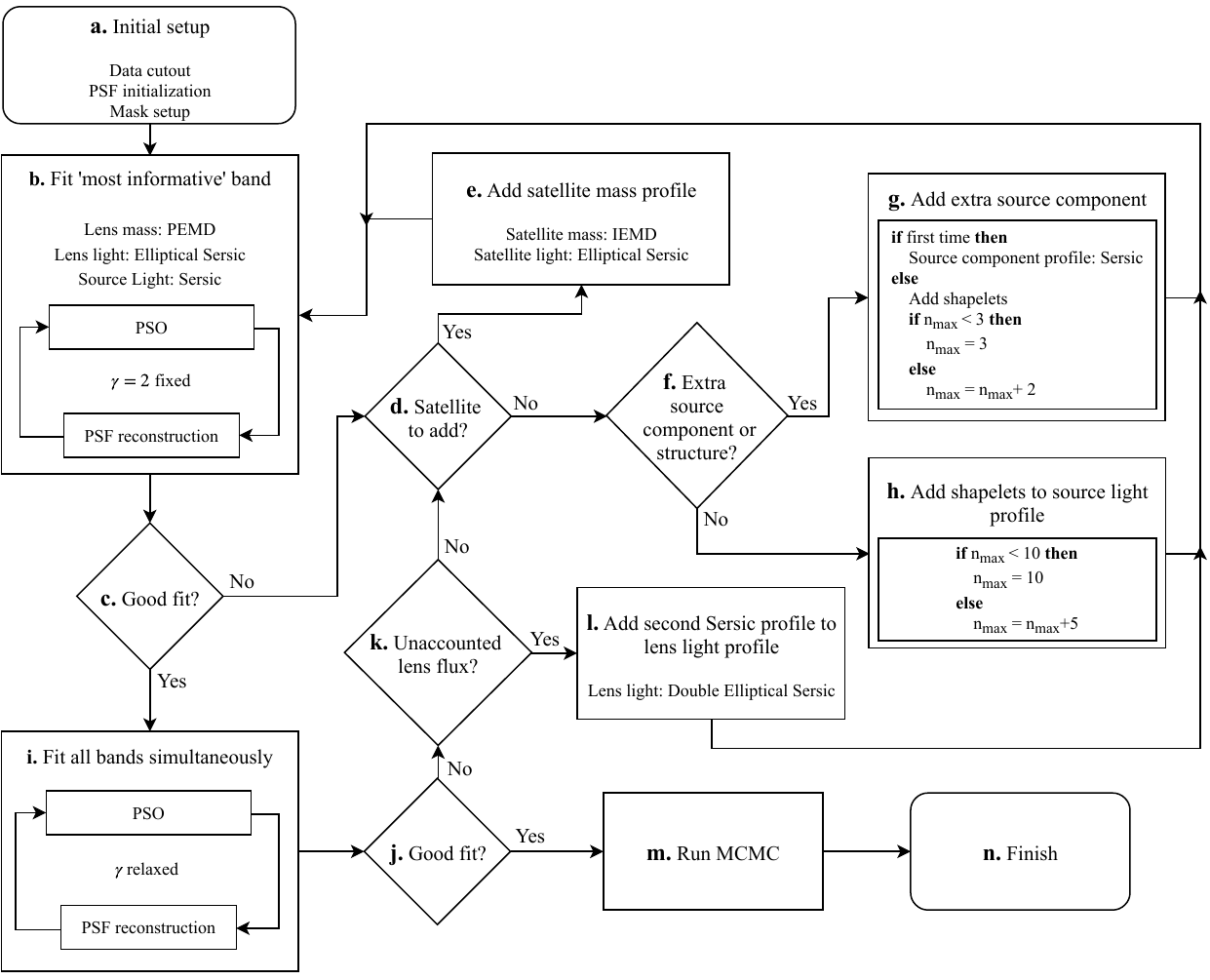}%{figures/flowchart.pdf}
	\caption{
		Flowchart showing the decision-tree for uniform modelling of quads to simultaneously fit multi-band data. After the initial setup (node a), the fitting is first done only with one band (node b) to iteratively choose the necessary level of complexity in the mass and light profiles (nodes d, e, f, g, h, k, l). A proposed model is accepted, if the power-law slope $\gamma$ does not diverge to a bound of the allowed range (nodes j) and the p-value $\ga 10^{-8}$ for the fit (nodes c, j). After deciding upon a set of profiles to simultaneously model the multi-band data (node i), the uncertainties on the model parameters are obtained by running a MCMC routine (node m).
		\label{fig:flowchart}
	}
\end{figure*}

To devise a uniform approach that will suit a wide range of quads that vary in size, configuration, light profiles, etc., we need to choose from the most general models for the lens mass profile and the light distributions. It is often required to fine-tune the choice of models by adding complexities to the lens model in a case-by-case basis to suit the purpose of the specific science driver of an investigator. However, such detailed lens-modelling is outside of the scope of this paper. We only require our models to satisfactorily ($\chi^2_{\rm red} \sim 1$) fit the data while being general enough to be applicable to a wide variety of lens systems.

We use the publicly available software package \textsc{Lenstronomy}\footnote{https://github.com/sibirrer/lenstronomy} \citep[][based on \citealt{Birrer15}]{Birrer18} to model the quads in our sample. Prior to this work, \textsc{Lenstronomy} was used to measure the Hubble constant \citep{Birrer16} and to quantify lensing substructure \citep{Birrer17}. We first adopt the simplest yet general set of profiles to model the deflector mass and light, and the source-light distributions (e.g. Section \ref{subsec:mass_parameterization}, \ref{subsec:light_parameterization}). Then, we run a particle swarm optimization (PSO) routine through \textsc{Lenstronomy} to find the maximum of the likelihood function. After the PSO routine, we check for the goodness-of-fit of the best fit model. If the adopted profiles can not produce an acceptable fit to the data, we gradually add more mass or light profiles to account for extra complexities in the lens system, e.g, presence of satellites, complex structure near the Einstein ring, or extra lensed source components. We run the PSO routine after each addition of complexity until a set of adopted mass and light profiles can produce an acceptable model. Next, we obtain the posterior probability distribution functions (PDFs) of the model parameters using a Markov chain Monte Carlo (MCMC) routine. The PSO and MCMC routines in \textsc{Lenstronomy} utilize the \textsc{cosmoHammer} package \citep{Akeret13}. \textsc{cosmoHammer} itself embeds \textsc{emcee} \citep{Foreman-Mackey13}, which is an affine-invariant ensemble sampler for MCMC \citep{Goodman10} written in \textsc{Python}.

In this section, we first describe the profiles used to parameterize the mass and light distributions. Then, we explain the decision-tree of the modelling procedure. 

\subsection{Mass profile parameterization} \label{subsec:mass_parameterization}
We adopt a power-law elliptical mass distribution (PEMD) for the lens mass profile. This profile is parameterized as
\begin{equation}
\kappa = \frac{3-\gamma}{2} \left( \frac{\theta_{\rm E}}{\sqrt{q\theta_1^2 + \theta_2^2/q}} \right)^{\gamma - 1},
\end{equation}
where $\gamma$ is the power-law slope, $\theta_{\rm E}$ is the Einstein radius, $q$ is the axis ratio. The coordinates $(\theta_1, \theta_2)$ depend on position angle $\phi$ through a rotational transformation of the on-sky coordinates that aligns the coordinate axes along the major and minor axes. 

We also add an external shear profile parameterized by two parameters, $\gamma_1$ and $\gamma_2$. The external-shear magnitude $\gamma_{\rm ext}$ and angle $\phi_{\rm ext}$ are related to these parameters by
\begin{equation}
	\gamma_{\rm ext} = \sqrt{\gamma_1^2 + \gamma_2^2}, \ \
	\tan 2 \phi_{\rm ext} = \frac{\gamma_2}{\gamma_1}.
\end{equation}

If there is a secondary deflector or a satellite of the main deflector, we choose  an isothermal elliptical mass distribution (IEMD), which is a PEMD with the power-law slope $\gamma$ fixed to 2. 

\subsection{Light profile parameterization} \label{subsec:light_parameterization}
We choose the elliptical S\'ersic function \citep{Sersic68} to model the deflector light profile. The S\'ersic function is parameterized as
\begin{equation}
	I(\theta_1, \theta_2) = I_{\rm e} \exp \left[ -k \left\{ \left( \frac{\sqrt{\theta_1^2 + \theta_2^2/q_{\rm L}^2}}{\theta_{\rm eff}} \right)^{1/n_{\rm Sersic}} - 1\right\} \right].
\end{equation}
 Here $I_{\rm e}$ is the amplitude, $k$ is a constant that normalizes $\theta_{\rm eff}$ so that it is the half-light radius, $q_{\rm L}$ is the axis ratio, and $n_{\rm Sersic}$ is the S\'ersic index. The coordinates $(\theta_1, \theta_2)$ also depend on the position angle $\phi_{\rm L}$ that rotationally transforms the on-sky coordinates to align the coordinate axes with the major and minor axes. We add a `uniform' light profile parameterized by only one parameter, the amplitude, that can capture unaccounted flux from the lens by a single S\'ersic profile.

The circular S\'ersic function (with $q_{\rm L} = 1$, $\phi_{\rm L} = 0$) is adopted to model the host-galaxy-light distribution. We limit $\theta_{\rm eff} > 0\farcs04$ (which is the pixel size in the UVIS bands) on the source plane to prevent the S\'ersic profile to be too pointy effectively mimicking a point source. For a typical source redshift $z_{\rm s}=2$, $0\farcs04$ corresponds to $\sim0.33$ kpc. This is a reasonable lower limit for the size of a lensed source hosting a supermassive black hole. If there are complex structures in the lensed arcs that can not be fully captured by a simple S\'ersic profile, we add a basis set of shapelets \citep{Refregier03, Birrer15} on top of the S\'ersic profile to reconstruct the source-light distribution. The basis set is parameterized by maximum order $n_{\rm max}$, and a characteristic scale $\beta$. The number of shapelets is given by $(n_{\rm max} + 1)(n_{\rm max} + 2)/2$.

The quasar images are modelled with point sources with a point spread function (PSF) on the image plane.

\subsection{Modelling procedure} \label{subsec:modeling}
We model the quads in a general framework to simultaneously fit the data from all three \textit{HST} bands. Fig. \ref{fig:flowchart} illustrates the flow of the modelling procedure. We describe the nodes of this flow-chart below. Each node is marked with a lowercase letter. Some of the decision nodes in Fig. \ref{fig:flowchart} are self-explanatory and need no further elaboration. \\ 

{\bf a. Initial setup:} We first pre-process the data in each band. A cutout with an appropriate field-of-view covering the lens and nearby environment from the whole image is chosen. The background flux estimated by \textsc{SExtractor} \citep{Bertin96} from the whole image is subtracted from the cutout. We also select four or more stars from the \textit{HST} images to estimate the initial PSF in each band. A circular mask with a suitable radius is chosen to only include the deflector-light distribution, and the lensed quasar-images and arcs. If there is a nearby galaxy or a star, we mask it out unless we specifically choose to model the light profile of a satellite or companion galaxy, e.g., for DES J0408-5354, PS J0630-1201, SDSS J1433+6007 and PS J1606-2333. As PS J0630-1201 is a five-image lens, we allow the model the flexibility to produce more than four images. \\

{\bf b. Fit the `most informative' band:} It is important to judiciously initiate any optimization routine, such as the PSO, to efficiently find the global extremum. Finding the global maximum of the joint likelihood from all the bands together from a random initial point is often very expensive in terms of time and computational resource. Therefore, we first only fit the `most informative' band to iteratively select the light and mass profiles necessary to account for the lens complexity. In this study, we choose F814W as the `most informative' band. It is easier to decompose the deflector and the source-light distributions in the F814W band than in the F160W band as the deflector does not have a large flux near or beyond the Einstein ring. The resolution in the F814W band is also twice as high as in the F160W band. Furthermore, the deflector flux in the F475X band is often too small to reliably model the deflector-light distribution without a good prior. At first, we fix the power-law slope for the lens mass model at $\gamma=2$ (i.e. the isothermal case). With each consecutive PSO routine, we narrow down the search region in the parameter space around the maximum of the likelihood. After each PSO routine, we iteratively reconstruct the PSF with the modelled-extended-light subtracted quasar images themselves. This is performed iteratively such that the extended light model updates its model with the new PSF to avoid biases and over-constraints on the PSF model. Similar procedures have been used in \citet{Chen16, Birrer17, Wong17}. The details are described in \citet{Birrer19} and the reconstruction routines are part of \textsc{Lenstronomy}. \\

{\bf c. Good fit?} We check for the goodness of fit by calculating the p-value for the total $\chi^2$ and degrees of freedom. We set p-value $\ga 10^{-8}$ as a criterion to accept a model. This low p-value is enough to point out substantial inadequacies in the model while applicable to the wide variety of the lens systems in our sample. Implementing a higher p-value would require noise-level modelling which is hard to achieve in a uniform framework. The total $\chi^2$ in this node is computed from the residuals in the F814W band only. \\

{\bf e. Add satellite mass profile:} We add an IEMD for the satellite or companion mass profile. The light distribution of the satellite is modelled with an elliptical S\'ersic profile. The initial centroid of the satellite is chosen approximately at the center of the brightest pixel in the satellite. \\

{\bf g. Add extra source component:} If there are extra lensed source components, e.g., blobs or arcs, that are not part of the primary source structure near the Einstein ring, we add extra light profiles in the same source plane of the lensed quasar. We only add one light profile for each set of conjugate components. It is easier to identify and constrain the positions of additional source components on the image plane. Among the identifiable conjugate components from visual inspection, if one component is a smaller blob, and the others form arcs, we choose the blob's position in the image plane as the initial guess. First, we only add one circular S\'ersic profile for each additional source component. For the second visit to this node, i.e. there is unaccounted structure or extra light near the additional lensed source components, we add shapelets with $n_{\rm max} = 3$ on top of the S\'ersic profile. For each subsequent visit, we increase $n_{\rm max}$ by 2. \\

{\bf h. Add shapelets to source-light profile:} If there are structures near the Einstein ring, we add a basis set of shapelets on top of the S\'ersic function to the primary source-light profile. We first add shapelets with $n_{\rm max} = 10$ and increase $n_{\rm max}$ by 5 for each future visit to this node. The characteristic scale $\beta$ of the shapelets is initiated with the best fit $\theta_{\rm eff}$ of the S\'ersic profile for the source. \\

{\bf i. Fit all bands simultaneously:} Before fitting all the bands simultaneously it is important to check astrometric alignment between the bands and correct accordingly if there is a misalignment. We align the data from the IR channel (F160W) with those from the UVIS band (F814W and F475X) by matching the positions of the four lensed quasar images. After that, we run PSO routines to fit all the bands simultaneously. Each PSO routine is followed by one iterative PSF reconstruction routine. During simultaneous fitting, only the intensities of the light profiles and shapelets are varied independently for different bands. All the other parameters, such as scalelength, ellipticity, position angle and S\'ersic index, are set to be common across wavelengths, which is a common practice for simultaneous fitting of multi-band data \citep[e.g.][]{Stoughton02, Lackner12}. As a result, for the case of a single S\'ersic profile the best fit parameters are effectively an average over the wavelengths. However, we find the resultant best-fit parameters from the simultaneous fitting to be within 1$\sigma$ systematic+statistical uncertainty of the ones from the individual fits of different bands for one representative system (DES J0405-3308) from our sample. Therefore, we assume that setting these parameters to be common across wavelengths is sufficient for the purpose of this study. For the case of shapelets or double S\'ersic profile, the relative intensities of the shapelets or S\'ersic components can freely vary across bands. This allows for more complex morphological variation across wavelengths and makes our assumption even more reasonable. \\

{\bf j. Good fit?} We check for the goodness of fit with the same criteria described in node c. In this node, the total $\chi^2$ is computed from the residuals in all the three bands. Moreover, we check that the power-law slope $\gamma$ has not diverged to the bound of the allowed values when $\gamma$ is relaxed in node i. This might happen if there is not enough complexity in the adopted model to reconstruct the observed fluxes. We also check if there is lens flux unaccounted by the single S\'ersic profile. If the total flux in the `uniform' light profile within the effective radius is more than one per cent of that for the elliptical S\'ersic profile, we decide that there is unaccounted lens flux. This can particularly happen in the F160W band as the lens light is more extended in the IR than in the UVIS channels and two concentric S\'ersic functions provide a better fit to the lens light \citep{Claeskens06, Suyu13}. If there is no unaccounted lens light, we discard the `uniform' profile from the set of lens-light profiles before moving to node m. \\

{\bf l. Add second S\'ersic function to lens-light profile:} If there is unaccounted lens flux, we discard the `uniform' light profile and add a second S\'ersic function on top of the first one with the same centroid. We fix the S\'ersic indices for the two S\'ersic profiles to $n_{\rm Sersic} = 4$ (de Vaucouleurs profile) and $n_{\rm Sersic} = 1$ (exponential). We fix these S\'ersic indices for numerical stability. These profile fits should not be interpreted as bulge-disk decompositions. For a proper bulge-disk decomposition, more robust methods should be adopted to detect the presence of multiple components, e.g., Bayesian model comparison \citep{DSouza14} and axis-ratio variation technique \citep{Oh17}.  \\

{\bf m. Run MCMC:} If the PSO fitting sequence finds an acceptable model for the quad, we run a MCMC routine. The initial positions of the walkers are centered around the best fit found by the PSO fitting sequence. \\

{\bf n. Finish:} After the MCMC routine, we check for the convergence of the chain. We accept the chain as converged, if the total number of steps is $\sim10$ times the autocorrelation length, and the median and variance of the walker positions at each step are stable for 1 autocorrelation length at the end of the chain. We then calculate the best-fit value for each model parameter from the median of the walker positions at the last step. Similarly, 1$\sigma$ confidence levels are computed from the 16- and 84-th percentiles in the last step.

\subsection{Systematics} \label{subsec:systematics}
We estimate the systematic uncertainties of the lens model parameters by marginalizing over several numerical settings. We performed the modelling technique described in Section \ref{subsec:modeling} with eleven different numerical settings: varying the lens-mask size, varying the mask size for extra quasar-images for PSF reconstruction, varying the sampling resolution of the reconstructed \textit{HST} image, without PSF reconstruction, and with different realisations of the reconstructed PSF. We checked for systematics for the lens system SDSS J0248+1913. This system was chosen for two reasons: (i) this system has relatively fainter arc compared to the point source and deflector brightness, thus providing a conservative estimate of the systematics, and (ii) the modelling procedure is one of the simplest ones that enables running the modelling procedure numerous times with different settings in relatively less time. We assume the systematics are the same order of magnitude for the other lens systems in the sample.

\section{Results} \label{sec:result}
In this section, we first describe the lens models and report the model parameters along with some derived parameters for all the quads. Then, we investigate the alignment between the mass and light profiles and report our findings. In Appendix \ref{app:lens_light_params}, \ref{app:convergence} and \ref{app:time_delay} we report additional inferred lens model parameters that are not directly relevant for the scientific investigation carried out here but may be of interest to some readers, especially in planning future follow-up and observations.

\subsection{Efficiency of the uniform framework}
All the \nlenses\ quads are reliably (p-value $\sim 1$, Table \ref{tab:lens_profiles}) modelled following the uniform approach described in Section \ref{sec:modeling}. The framework was designed and tuned from the experience gained from uniformly modelling the first ten observed quads in the sample. The three quads, SDSS J1251+2935, SDSS J1330+1810 and SDSS J1433+6007, were observed after the design phase. We effectively modelled these three lenses implementing the general framework, which validates its effectiveness. The total investigator time spent for these two lenses is $\sim3$ hours per lens including data reduction, initial setup and quality control of the model outputs. The number of CPU hours (on state-of-the-art machines\footnote{We utilized the Hoffman2 Shared Cluster provided by UCLA Institute for Digital Research and Education's Research Technology Group. \url{https://idre.ucla.edu/hoffman2}.}) per system ranges between 50 to 500 depending on the complexity of the model. 

\subsection{Lens models} \label{subsec:lens_models}
The set of profiles chosen through the decision-tree for modelling the quads along with the corresponding p-values are listed in Table \ref{tab:lens_profiles}. We show a breakdown of the best-fit models in each band for the quads, SDSS J0248+1913, DES J0408-5354, SDSS J1251+2935, SDSS J1433+6007, as examples, in Fig. \ref{fig:lens_model_breakdown}. Model breakdowns for the rest of lenses are provided in Appendix \ref{app:lens_models}. We show the red-green-blue (RGB) images produced from the \textit{HST} data alongside the reconstructed RGB images for all the quads in Fig. \ref{fig:lens_rgb_w_model}.

We checked the robustness of the estimated lens model parameters with and without PSF reconstructions. We find the Einstein radius $\theta_{\rm E}$, axis ratio $q$, mass position angle $\phi$, external shear $\gamma_{\rm ext}$ and shear angle $\phi_{\rm ext}$ to be robustly (within 1$\sigma$ systematic+statistical uncertainty) estimated. However, the power-law slope $\gamma$ is affected by $\geq$1$\sigma$ systematic+statistical uncertainty due to deviations of the reconstructed PSF. This is expected as $\gamma$ depends on the thickness of the Einstein ring and this thickness in the reconstructed model in turn depends on the PSF.

We investigated if setting the S\'ersic radius and index of the source light profile common across wavelength bands biases the measurement of the power-law slope. For one representative system (DES J0405+3308) from our sample, we find the power-law slope from the individual fits of different bands to agree within 1$\sigma$ systematic+statistical uncertainty of the one from the simultaneous fit. Therefore, we conclude that setting the scaling parameters of the source light profile except the intensity to be common across wavelengths does not significantly ($>1\sigma$) bias the power-law slope.

\begin{figure}
	\includegraphics[width=\columnwidth]{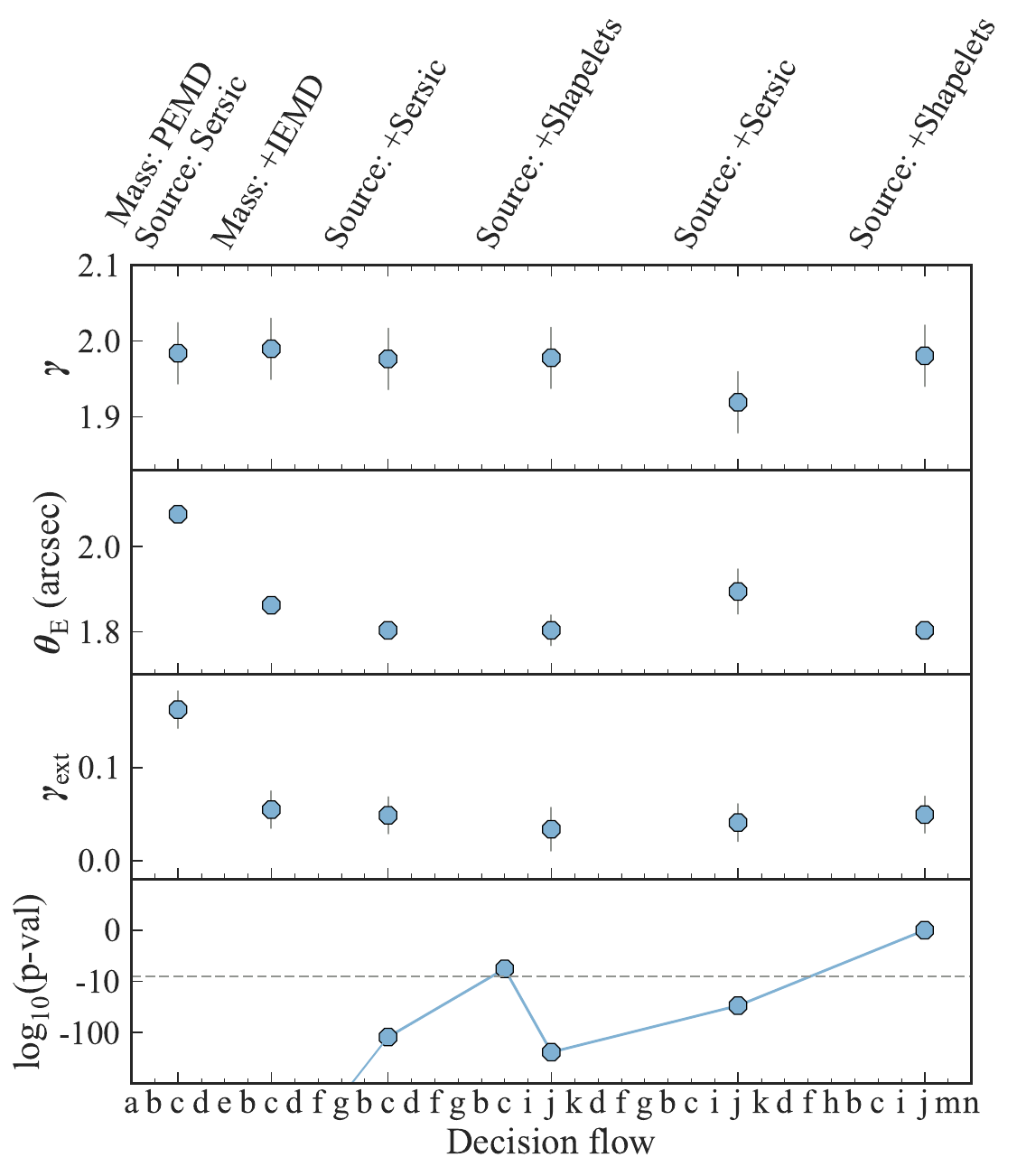}%{figures/complexity.pdf} 
	\caption{
	Stability of lens model parameters with increasing model complexity. The four panels show the power-law slope $\gamma$, Einstein radius $\theta_{\rm E}$, external shear $\gamma_{\rm ext}$, and logarithm of p-value of the reduced-$\chi^2$ of the model fit, top to bottom, along the decision-flow for the quad DES J0408-5354. The bottom-horizontal axis denotes the node identifiers along the decision flow as in Fig. \ref{fig:flowchart}. Short descriptions for added profiles at corresponding points along the decision flow are shown along the top-horizontal axis. Solid-grey lines attached to the blue circles show 1$\sigma$ systematic+statistical uncertainty. The dashed-grey line at the bottom panel marks the threshold p-value=$10^{-8}$ for accepting a model. The p-value decreases after crossing the threshold the first time due to addition of the other two bands for simultaneous fitting, which requires more complexity in the model.
	\label{fig:complexity}
	}
	\end{figure}

We checked if the lens model parameters are stable with increasing complexity in the model (Fig. \ref{fig:complexity}). The stability of the Einstein radius $\theta_{\rm E}$ and the external shear $\gamma_{\rm ext}$ improves if the mass profile of a satellite is explicitly modelled. For increasing complexity in modelling the source-light distribution, the power-law slope $\gamma$, the Einstein radius $\theta_{\rm E}$ and the external convergence $\gamma_{\rm ext}$ are stable.

We report the lens model parameters: Einstein radius $\theta_{\rm E}$, power-law slope $\gamma$, axis ratio $q$, position angle $\phi$, external shear $\gamma_{\rm ext}$, and shear angle $\phi_{\rm ext}$ and deflector light parameters: effective radius $\theta_{\rm eff}$, axis ratio $q_{\rm L}$, and position angle $\phi_{\rm L}$ in Table \ref{tab:lens_params}. For the deflectors fitted with double S\'ersic profiles, the ellipticity and position angles are computed by fitting isophotes to the double S\'ersic light distribution. We use the \textsc{Photutils}\footnote{\url{http://photutils.readthedocs.io}} package in \textsc{Python} for measuring the isophotes which implements an iterative method described by \citet{Jedrzejewski87}. We tabulate the astrometric positions of the deflector galaxy and the quasar images in Table \ref{tab:lens_astrometry}. The apparent magnitudes of the deflector galaxy and the quasar images in each of the three \textit{HST} bands are given in Table \ref{tab:lens_mags}.

\begin{table*}
	\centering
	\caption{Lens model profiles.}
	\label{tab:lens_profiles}
	\begin{tabular}{llllll}
		\hline
		System name & Mass profiles	& Lens-light profiles & Source-light profiles & p-value${}^{\ast}$ & Decision flow{${}^{\ast\ast}$} \\
		\hline
		PS J0147+4630     	& PEMD & Double elliptical S\'ersic & S\'ersic & 1.0 & abcijklbcijmn \\
							& & & Point source (image plane) \\
		\hline
		SDSS J0248+1913   	&  PEMD & Elliptical S\'ersic & S\'ersic & 1.0 & abcijmn \\
							& & & Point source (image plane) \\
		\hline
		ATLAS J0259-1635   	&  PEMD & Elliptical S\'ersic & S\'ersic & 1.0 & abcdfhbcijmn \\
							& & & Shapelets ($n_{\rm max} = 10$) \\
							& & & Point source (image plane) \\
		\hline
		DES J0405-3308    	&  PEMD & Elliptical S\'ersic & S\'ersic & 1.0 & abcijmn \\
							& & & Point source (image plane) \\
		\hline
		DES J0408-5354    	& PEMD & Elliptical S\'ersic & S\'ersic & 1.0 & abcdebcdfgbcdfgbcijkdf \\
							& IEMD${}^{\dagger}$ & Elliptical S\'ersic${}^{\dagger}$ & Shapelets ($n_{\rm max} = 10$) & & gbcijkdfhbcijmn \\
							& & & S\'ersic${}^{\dagger}$ \\
							& & & Shapelets${}^{\dagger}$ ($n_{\rm max} = 3$) \\
							& & & S\'ersic${}^{\dagger}$ \\
							& & & Point source (image plane) \\
		\hline
		DES J0420-4037    	& PEMD & Elliptical S\'ersic & S\'ersic & 1.0 & abcijkdfgbcijmn \\
							& &  & S\'ersic${}^{\dagger}$ \\
							& & & S\'ersic${}^{\dagger}$ \\
							& & & Point source (image plane) \\
		\hline
		PS J0630-1201     	& PEMD & Elliptical S\'ersic & S\'ersic & 1.0 & abcdebcijmn \\
							& IEMD${}^{\dagger}$ & Elliptical S\'ersic${}^{\dagger}$ & Point source (image plane) \\
							%& & & \\
		\hline
		SDSS J1251+2935 	    & PEMD & Double elliptical S\'ersic & S\'ersic & 1.0 & abcijklbcijkdfhbcijmn \\
							& & & Shapelets ($n_{\rm max}=10$) & \\
							& & & Point source (image plane) \\
		\hline
		SDSS J1330+1810 	    & PEMD & Double elliptical S\'ersic & S\'ersic & 0.005 & abcijklbcijkdfhbcijmn \\
							& & & Shapelets ($n_{\rm max}=10$) & \\
							& & & Point source (image plane) \\
		\hline
		SDSS J1433+6007     & PEMD & Double elliptical S\'ersic & S\'ersic & 1.0 & abcdebcijklbcijmn \\
							& IEMD${}^{\dagger}$ & Elliptical S\'ersic${}^{\dagger}$ & Point source (image plane) \\
							%& & & \\
		\hline
		PS J1606-2333     	& PEMD & Double elliptical S\'ersic & S\'ersic & 1.0 & abcdebcdfhbcijklbcijmn \\
							& IEMD${}^{\dagger}$ & Elliptical S\'ersic${}^{\dagger}$ & Shapelets ($n_{\rm max} = 10$) \\
							& & & Point source (image plane) \\
		\hline
		DES J2038-4008    	&  PEMD & Double elliptical S\'ersic &  S\'ersic & 1.0 & abcdfhbcijklbcijmn \\
							& & & Shapelets ($n_{\rm max} = 10$) \\
							& & & Point source (image plane) & & \\
		\hline
		WISE J2344-3056  	&  PEMD & Double elliptical S\'ersic & S\'ersic & 1.0 & abcijklbcijmn \\
							& & & Point source (image plane) \\
							%& & & \\
		\hline
		\multicolumn{5}{l}{${}^{\ast}$ The p-value is for the combined $\chi^2$ from all three bands.} \\
		\multicolumn{5}{l}{${}^{\ast\ast}$ Labels of nodes visited during the modelling procedure in the flow-chart shown in Fig. \ref{fig:flowchart}.} \\
		\multicolumn{5}{l}{${}^{\dagger}$ Satellite or extra source component separate from the central source.}
	\end{tabular}
\end{table*}

\begin{table*}
	\centering
	\caption{Lens model parameters. The reported uncertainties are systematic and statistical uncertainties added in quadrature.}
	\label{tab:lens_params}
	\begin{adjustbox}{width=\textwidth}
	\begin{tabular}{%{lrrrrrrrrrrr} % four columns, alignment for each
		@{\extracolsep{\fill}}
  		l
	  	S[table-format=1.3(1)]
	  	S[table-format=-1.2(1)]
	  	S[table-format=-1.2(1)]
	  	S[table-format=-2.0(2)]
	  	S[table-format=-1.2(1)]
	  	S[table-format=-2.0(1)]
	  	S[table-format=-1.2(1)]
	  	S[table-format=-1.2(1)]
	  	S[table-format=-2.0(1)]
	  	@{}
	}
		\hline
		System name & \multicolumn{1}{c}{$\theta_{\rm E}$} & \multicolumn{1}{c}{$\gamma$} & \multicolumn{1}{c}{$q$} & \multicolumn{1}{c}{$\phi$ (E of N)} & \multicolumn{1}{c}{$\gamma_{\rm ext}$} & \multicolumn{1}{c}{$\phi_{\rm ext}$ (E of N)} & \multicolumn{1}{c}{$\theta_{\rm eff}{}^{\dagger}$} & \multicolumn{1}{c}{$q_{\rm L}{}^{\dagger}$} & \multicolumn{1}{c}{$\phi_{\rm L}$ (E of N)${}^{\dagger}$} \\
								&  \multicolumn{1}{c}{(arcsec)} & & & \multicolumn{1}{c}{(degree)} & & \multicolumn{1}{c}{(degree)} & \multicolumn{1}{c}{(arcsec)} & & \multicolumn{1}{c}{(degree)} \\ 
								
		\hline
		
PS J0147+4630   	& 1.90 \pm 0.01	& 2.00 \pm 0.05	& 0.81 \pm 0.04	& -55 \pm 6	& 0.16 \pm 0.02	& -72 \pm 3	& 3.45 \pm 0.10	& 0.93 \pm 0.06	& 49 \pm 16 \\
SDSS J0248+1913 	& 0.804 \pm 0.004	& 2.19 \pm 0.04	& 0.40 \pm 0.06	& 46 \pm 6	& 0.09 \pm 0.02	& 6 \pm 3	& 0.16 \pm 0.03	& 0.40 \pm 0.02	& 13 \pm 1 \\
ATLAS J0259-1635 	& 0.75 \pm 0.01	& 2.01 \pm 0.04	& 0.66 \pm 0.04	& 18 \pm 6	& 0.00 \pm 0.02	& -30 \pm 3	& 1.00 \pm 0.09	& 0.38 \pm 0.04	& 20 \pm 4 \\
DES J0405-3308  	& 0.70 \pm 0.01	& 1.99 \pm 0.04	& 0.95 \pm 0.05	& 41 \pm 12	& 0.01 \pm 0.02	& -79 \pm 5	& 0.44 \pm 0.09	& 0.55 \pm 0.05	& 37 \pm 4 \\
DES J0408-5354  	& 1.80 \pm 0.01	& 1.98 \pm 0.04	& 0.62 \pm 0.04	& 18 \pm 6	& 0.05 \pm 0.02	& -15 \pm 3	& 2.15 \pm 0.09	& 0.82 \pm 0.04	& 28 \pm 4 \\
DES J0420-4037  	& 0.83 \pm 0.01	& 1.97 \pm 0.04	& 0.87 \pm 0.04	& 24 \pm 6	& 0.03 \pm 0.02	& -20 \pm 4	& 0.44 \pm 0.09	& 0.61 \pm 0.04	& 27 \pm 4 \\
PS J0630-1201   	& 1.02 \pm 0.01	& 2.00 \pm 0.04	& 0.53 \pm 0.04	& -27 \pm 6	& 0.14 \pm 0.02	& -2 \pm 3	& 1.64 \pm 0.09	& 0.79 \pm 0.04	& 12 \pm 4 \\
SDSS J1251+2935 	& 0.84 \pm 0.01	& 1.97 \pm 0.04	& 0.71 \pm 0.04	& 28 \pm 6	& 0.07 \pm 0.02	& -88 \pm 3	& 1.02 \pm 0.09	& 0.67 \pm 0.04	& 23 \pm 4 \\
SDSS J1330+1810 	& 0.954 \pm 0.005	& 2.00 \pm 0.04	& 0.59 \pm 0.06	& 24 \pm 6	& 0.07 \pm 0.02	& 8 \pm 3	& 0.40 \pm 0.03	& 0.28 \pm 0.02	& 24 \pm 1 \\
SDSS J1433+6007 	& 1.71 \pm 0.01	& 1.96 \pm 0.04	& 0.51 \pm 0.04	& -81 \pm 6	& 0.09 \pm 0.02	& -30 \pm 3	& 1.10 \pm 0.09	& 0.56 \pm 0.04	& -88 \pm 4 \\
PS J1606-2333   	& 0.63 \pm 0.01	& 1.97 \pm 0.04	& 0.88 \pm 0.05	& 41 \pm 10	& 0.16 \pm 0.02	& 53 \pm 3	& 1.36 \pm 0.09	& 0.60 \pm 0.07	& -24 \pm 5 \\
DES J2038-4008  	& 1.38 \pm 0.01	& 2.35 \pm 0.04	& 0.61 \pm 0.04	& 38 \pm 6	& 0.09 \pm 0.02	& -58 \pm 3	& 2.85 \pm 0.09	& 0.67 \pm 0.04	& 38 \pm 4 \\
WISE J2344-3056 	& 0.52 \pm 0.01	& 1.95 \pm 0.05	& 0.51 \pm 0.06	& -70 \pm 6	& 0.06 \pm 0.02	& -68 \pm 8	& 2.61 \pm 0.19	& 0.76 \pm 0.03	& -69 \pm 4 \\
		\hline
		\multicolumn{8}{l}{${}^{\dagger}$ Calculated from the F160W band for the lenses with double S\'ersic fit for the lens light.} \\
	\end{tabular}
	\end{adjustbox}
\end{table*}

\begin{table*}
	\centering
	\caption{Astrometric positions of the deflector light centroid and quasar images. The reported uncertainties are on relative astrometry and they are systematic and statistical uncertainties added in quadrature.}
	\label{tab:lens_astrometry}
	\begin{adjustbox}{width=\textwidth}
	\begin{tabular}{%{lrrrrrrrrrr} % eleven columns, alignment for each
		@{\extracolsep{\fill}}
  		l
	  	r
	  	r
	  	S[table-format=-1.4(1)]
	  	S[table-format=-1.4(1)]
	  	S[table-format=-1.4(1)]
	  	S[table-format=-1.4(1)]
	  	S[table-format=-1.4(1)]
	  	S[table-format=-1.4(1)]
	  	S[table-format=-1.4(1)]
	  	S[table-format=-1.4(1)]
	  	@{}
	}
		\hline
		\multirow{3}{*}{System name} &  \multicolumn{2}{c}{Deflector} & \multicolumn{2}{c}{Image A} & \multicolumn{2}{c}{Image B} & \multicolumn{2}{c}{Image C} & \multicolumn{2}{c}{Image D}   \\
								&  \multicolumn{1}{c}{$\alpha$} & \multicolumn{1}{c}{$\delta$} & \multicolumn{1}{c}{$\Delta \alpha$} & \multicolumn{1}{c}{$\Delta\delta$} & \multicolumn{1}{c}{$\Delta \alpha$} & \multicolumn{1}{c}{$\Delta\delta$} & \multicolumn{1}{c}{$\Delta \alpha$} & \multicolumn{1}{c}{$\Delta\delta$} & \multicolumn{1}{c}{$\Delta \alpha$} & \multicolumn{1}{c}{$\Delta\delta$} \\
								& \multicolumn{1}{c}{(degree)} & \multicolumn{1}{c}{(degree)} & \multicolumn{1}{c}{(arcsec)} & \multicolumn{1}{c}{(arcsec)} & \multicolumn{1}{c}{(arcsec)} & \multicolumn{1}{c}{(arcsec)} & \multicolumn{1}{c}{(arcsec)} & \multicolumn{1}{c}{(arcsec)} & \multicolumn{1}{c}{(arcsec)} & \multicolumn{1}{c}{(arcsec)} \\ 
								
		\hline
		PS J0147+4630   	& 26.792331	& 46.511559	& 0.1553 \pm 0.0002 & 2.0513 \pm 0.0001 & 1.3270 \pm 0.0002 & 1.6419 \pm 0.0001 & -1.0840 \pm 0.0002 & 1.9580 \pm 0.0002 & -0.1862 \pm 0.0005 & -1.1696 \pm 0.0003 \\
SDSS J0248+1913 	& 42.203099	& 19.225246	& -0.647 \pm 0.001 & -0.204 \pm 0.001 & -0.505 \pm 0.001 & 0.629 \pm 0.001 & 0.351 \pm 0.001 & -0.821 \pm 0.001 & 0.401 \pm 0.001 & 0.590 \pm 0.001 \\
ATLAS J0259-1635 	& 44.928561	& -16.595376 & 0.683 \pm 0.003 & -0.303 \pm 0.001 & 0.357 \pm 0.001 & 0.571 \pm 0.001 & -0.801 \pm 0.001 & 0.253 \pm 0.001 & -0.043 \pm 0.001 & -0.700 \pm 0.001 \\
DES J0405-3308  	& 61.498964	& -33.147417	& 0.694 \pm 0.001 & -0.238 \pm 0.001 & -0.375 \pm 0.001 & -0.561 \pm 0.002 & 0.344 \pm 0.002 & 0.603 \pm 0.002 & -0.525 \pm 0.001 & 0.454 \pm 0.004 \\
DES J0408-5354  	& 62.090451	& -53.899816	& 1.931 \pm 0.002 & -1.594 \pm 0.001 & -1.825 \pm 0.001 & 0.270 \pm 0.001 & -1.944 \pm 0.002 & -0.954 \pm 0.002 & 0.091 \pm 0.001 & 1.367 \pm 0.002 \\
DES J0420-4037  	& 65.194858	& -40.624081	& -0.697 \pm 0.001 & -0.350 \pm 0.001 & -0.457 \pm 0.001 & 0.683 \pm 0.001 & 0.711 \pm 0.001 & -0.568 \pm 0.001 & 0.172 \pm 0.002 & 0.788 \pm 0.002 \\
PS J0630-1201$^\dagger$   	& 97.537601	& -12.022037	& 0.686 \pm 0.001 & -1.426 \pm 0.001 & 1.204 \pm 0.001 & -0.859 \pm 0.001 & 1.543 \pm 0.001 & 0.260 \pm 0.001 & -0.977 \pm 0.002 & 1.006 \pm 0.001 \\
SDSS J1251+2935 	& 192.781427	& 29.594652	& 0.3460 \pm 0.0005 & -0.6163 \pm 0.0005 & 0.707 \pm 0.001 & -0.257 \pm 0.001 & 0.637 \pm 0.001 & 0.335 \pm 0.001 & -1.080 \pm 0.001 & 0.319 \pm 0.002 \\
SDSS J1330+1810 	& 202.577755 & 18.175788 & 0.274 \pm 0.001 & -0.978 \pm 0.001 & -0.152 \pm 0.001 & -1.002 \pm 0.001 & -0.985 \pm 0.001 & 0.180 \pm 0.002 & 0.5212 \pm 0.0004 & 0.597 \pm 0.002 \\
SDSS J1433+6007 	& 218.345420 & 60.120777 & -0.941 \pm 0.002 & 2.058 \pm 0.003 & -0.943 \pm 0.003 & -1.691 \pm 0.003 & -1.721 \pm 0.002 & -0.083 \pm 0.002 & 1.075 \pm 0.003 & -0.138 \pm 0.003 \\
PS J1606-2333   	& 241.500982	& -23.556114 & 0.833 \pm 0.001 & 0.373 \pm 0.001 & -0.793 \pm 0.001 & -0.223 \pm 0.001 & 0.040 \pm 0.001 & -0.541 \pm 0.001 & -0.296 \pm 0.001 & 0.524 \pm 0.001 \\
DES J2038-4008  	& 309.511379	& -40.137024 	& -1.482 \pm 0.001 & 0.499 \pm 0.001 & 0.8340 \pm 0.0005 & -1.212 \pm 0.001 & -0.688 \pm 0.001 & -1.182 \pm 0.001 & 0.704 \pm 0.001 & 0.864 \pm 0.001 \\
WISE J2344-3056 	& 356.070739 & -30.940633	& -0.452 \pm 0.001 & 0.179 \pm 0.001 & 0.133 \pm 0.001 & 0.530 \pm 0.001 & -0.212 \pm 0.001 & -0.478 \pm 0.001 & 0.421 \pm 0.001 & -0.140 \pm 0.001 \\
		\hline
		\multicolumn{11}{l}{$^{\dagger}$ The relative positions of the image E are $\Delta \alpha = -0\farcs257 \pm 0\farcs003$ and $\Delta \delta = 0\farcs249 \pm 0\farcs002$.}

	\end{tabular}
	\end{adjustbox}
\end{table*}

\begin{table*}
	\centering
	\caption{Photometry of the deflector and quasar images. The deflector magnitudes are calculated from the total flux within a $5\arcsec\times5\arcsec$ square aperture. Magnitudes are given in the AB system. The reported uncertainties are systematic and statistical uncertainties added in quadrature.}
	\label{tab:lens_mags}
	\begin{tabular}{llccccc} % four columns, alignment for each
		\hline
		System name &  Filter & Deflector & A & B & C & D \\
		\hline
\multirow{3}{*}{ PS J0147+4630 } 	& F160W	& $18.3 \pm 0.1$	& $15.46 \pm 0.03$ & $15.78 \pm 0.03$ & $16.18 \pm 0.03$ & $18.05 \pm 0.03$  \\
                         	& F814W	& $19.4 \pm 0.1$	& $15.79 \pm 0.03$ & $16.09 \pm 0.03$ & $16.45 \pm 0.03$ & $18.21 \pm 0.03$  \\
                         	& F475X	& $21.8 \pm 0.3$	& $16.39 \pm 0.03$ & $16.67 \pm 0.03$ & $17.13 \pm 0.03$ & $18.74 \pm 0.03$  \\
\hline
\multirow{3}{*}{ SDSS J0248+1913 } 	& F160W	& $20.8 \pm 0.1$	& $19.88 \pm 0.04$ & $20.41 \pm 0.04$ & $19.91 \pm 0.03$ & $20.13 \pm 0.04$  \\
                         	& F814W	& $22.7 \pm 0.1$	& $20.20 \pm 0.03$ & $20.23 \pm 0.03$ & $20.43 \pm 0.03$ & $20.66 \pm 0.03$  \\
                         	& F475X	& $26.4 \pm 0.3$	& $21.14 \pm 0.03$ & $21.18 \pm 0.03$ & $21.35 \pm 0.03$ & $21.80 \pm 0.03$  \\
\hline
\multirow{3}{*}{ ATLAS J0259-1635 } 	& F160W	& $20.7 \pm 0.1$	& $18.48 \pm 0.03$ & $18.57 \pm 0.04$ & $19.06 \pm 0.03$ & $19.30 \pm 0.04$  \\
                         	& F814W	& $22.7 \pm 0.1$	& $19.00 \pm 0.03$ & $19.16 \pm 0.03$ & $19.62 \pm 0.03$ & $19.70 \pm 0.03$  \\
                         	& F475X	& {--}	& $21.08 \pm 0.03$ & $20.81 \pm 0.03$ & $21.50 \pm 0.04$ & $21.33 \pm 0.03$  \\
\hline
\multirow{3}{*}{ DES J0405-3308 } 	& F160W	& $20.2 \pm 0.1$	& $19.43 \pm 0.07$ & $19.58 \pm 0.04$ & $19.60 \pm 0.04$ & $19.33 \pm 0.03$  \\
                         	& F814W	& $22.0 \pm 0.1$	& $20.22 \pm 0.04$ & $20.60 \pm 0.04$ & $20.33 \pm 0.03$ & $20.09 \pm 0.03$  \\
                         	& F475X	& $25.0 \pm 0.3$	& $22.16 \pm 0.04$ & $22.81 \pm 0.04$ & $22.04 \pm 0.03$ & $21.91 \pm 0.03$  \\
\hline
\multirow{3}{*}{ DES J0408-5354 } 	& F160W	& $18.6 \pm 0.1$	& $20.18 \pm 0.03$ & $19.79 \pm 0.04$ & $20.33 \pm 0.04$ & $20.82 \pm 0.04$  \\
                         	& F814W	& $19.9 \pm 0.1$	& $20.38 \pm 0.03$ & $20.00 \pm 0.03$ & $21.66 \pm 0.03$ & $20.87 \pm 0.03$  \\
                         	& F475X	& $22.6 \pm 0.3$	& $21.20 \pm 0.03$ & $21.34 \pm 0.03$ & $23.16 \pm 0.03$ & $21.86 \pm 0.04$  \\
\hline
\multirow{3}{*}{ DES J0420-4037 } 	& F160W	& $18.6 \pm 0.1$	& $20.18 \pm 0.03$ & $21.03 \pm 0.04$ & $21.85 \pm 0.04$ & $21.96 \pm 0.05$  \\
                         	& F814W	& $19.5 \pm 0.1$	& $20.44 \pm 0.03$ & $20.96 \pm 0.03$ & $21.71 \pm 0.03$ & $21.98 \pm 0.04$  \\
                         	& F475X	& $21.5 \pm 0.3$	& $20.66 \pm 0.03$ & $21.25 \pm 0.03$ & $22.09 \pm 0.03$ & $22.09 \pm 0.03$  \\
\hline
\multirow{3}{*}{ PS J0630-1201$^{\dagger}$} 	& F160W	& $20.4 \pm 0.1$	& $18.71 \pm 0.03$ & $18.82 \pm 0.03$ & $18.74 \pm 0.03$ & $21.01 \pm 0.04$  \\
                         	& F814W	& $22.5 \pm 0.1$	& $19.70 \pm 0.03$ & $19.67 \pm 0.03$ & $19.71 \pm 0.03$ & $21.67 \pm 0.03$  \\
                         	& F475X	& $26.7 \pm 0.3$	& $21.06 \pm 0.03$ & $20.92 \pm 0.03$ & $21.10 \pm 0.03$ & $23.03 \pm 0.03$  \\
\hline
\multirow{3}{*}{ SDSS J1251+2935 } 	& F160W	& $18.3 \pm 0.1$	& $19.35 \pm 0.03$ & $20.25 \pm 0.05$ & $21.30 \pm 0.06$ & $21.02 \pm 0.05$  \\
                         	& F814W	& $19.4 \pm 0.1$	& $20.01 \pm 0.03$ & $20.80 \pm 0.04$ & $22.80 \pm 0.06$ & $21.66 \pm 0.04$  \\
                         	& F475X	& $21.4 \pm 0.3$	& $20.01 \pm 0.03$ & $20.73 \pm 0.04$ & $22.73 \pm 0.04$ & $21.95 \pm 0.04$  \\
\hline
\multirow{3}{*}{ SDSS J1330+1810 } 	& F160W	& $17.9 \pm 0.1$	 & $19.17 \pm 0.03$ & $19.36 \pm 0.03$ & $20.00 \pm 0.03$ & $21.24 \pm 0.05$ \\
                         	& F814W	& $19.1 \pm 0.1$	& $20.11 \pm 0.03$ & $20.03 \pm 0.03$ & $20.48 \pm 0.03$ & $20.56 \pm 0.03$  \\
                         	& F475X	& $21.4 \pm 0.3$	 & $20.31 \pm 0.03$ & $20.82 \pm 0.04$ & $21.24 \pm 0.03$ & $21.58 \pm 0.04$  \\
\hline
\multirow{3}{*}{ SDSS J1433+6007 } 	& F160W	& $18.1 \pm 0.1$	& $20.43 \pm 0.03$ & $20.47 \pm 0.04$ & $20.55 \pm 0.04$ & $21.56 \pm 0.04$  \\
                         	& F814W	& $19.2 \pm 0.1$	& $20.25 \pm 0.03$ & $20.17 \pm 0.03$ & $20.45 \pm 0.03$ & $21.74 \pm 0.03$  \\
                         	& F475X	& $21.2 \pm 0.3$	& $20.31 \pm 0.03$ & $20.16 \pm 0.03$ & $20.49 \pm 0.03$ & $21.93 \pm 0.04$  \\
\hline
\multirow{3}{*}{ PS J1606-2333 } 	& F160W	& $19.5 \pm 0.1$	& $19.59 \pm 0.03$ & $19.65 \pm 0.04$ & $19.99 \pm 0.03$ & $19.47 \pm 0.03$  \\
                         	& F814W	& $20.6 \pm 0.1$	& $19.06 \pm 0.03$ & $19.22 \pm 0.03$ & $19.38 \pm 0.03$ & $19.52 \pm 0.03$  \\
                         	& F475X	& $21.8 \pm 0.3$	& $19.52 \pm 0.04$ & $19.76 \pm 0.04$ & $19.97 \pm 0.03$ & $20.48 \pm 0.04$  \\
\hline
\multirow{3}{*}{ DES J2038-4008 } 	& F160W	& $16.4 \pm 0.1$	& $18.48 \pm 0.03$ & $18.27 \pm 0.03$ & $18.60 \pm 0.03$ & $19.49 \pm 0.04$  \\
                         	& F814W	& $17.4 \pm 0.1$	& $20.25 \pm 0.03$ & $19.99 \pm 0.03$ & $20.05 \pm 0.03$ & $20.88 \pm 0.03$  \\
                         	& F475X	& $19.1 \pm 0.3$	& $21.02 \pm 0.03$ & $20.89 \pm 0.03$ & $20.71 \pm 0.03$ & $21.43 \pm 0.03$  \\
\hline
\multirow{3}{*}{ WISE J2344-3056 } 	& F160W	& $19.0 \pm 0.1$	& $21.36 \pm 0.05$ & $20.94 \pm 0.04$ & $21.16 \pm 0.06$ & $20.78 \pm 0.04$  \\
                         	& F814W	& $20.0 \pm 0.1$	& $21.76 \pm 0.03$ & $21.20 \pm 0.03$ & $21.27 \pm 0.03$ & $20.76 \pm 0.03$  \\
                         	& F475X	& $21.6 \pm 0.3$	& $22.79 \pm 0.03$ & $21.68 \pm 0.03$ & $21.66 \pm 0.03$ & $21.13 \pm 0.03$  \\
\hline
		\multicolumn{7}{l}{${}^{\dagger}$ The magnitudes of image E are $22.51\pm0.10$, $23.40\pm.04$, and $24.77\pm0.04$ in the F160W, F814W,}\\
		\multicolumn{7}{l}{and F475X bands, respectively.}
	\end{tabular}
\end{table*}

\begin{figure*}
	\includegraphics[width=1.05\columnwidth]{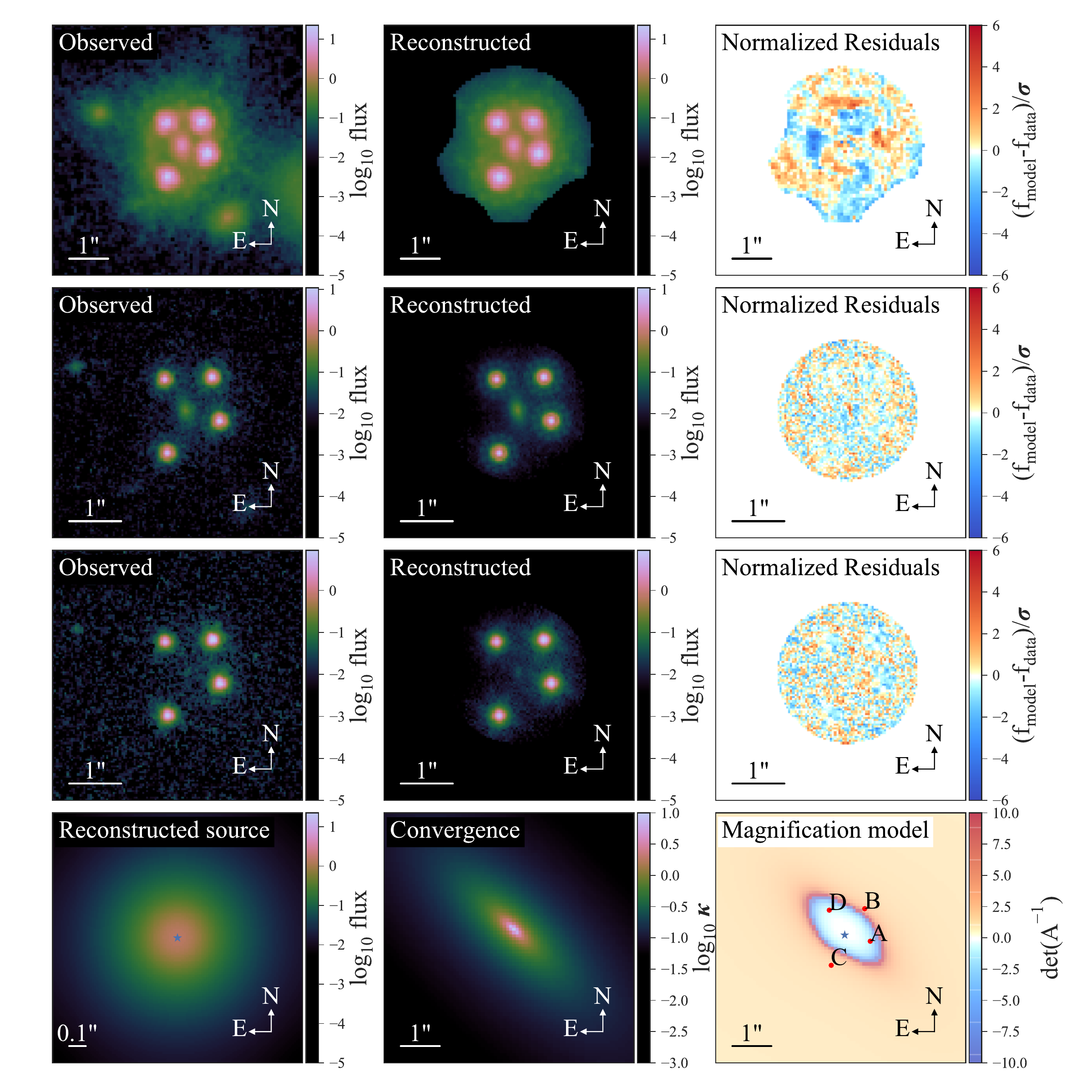}%{figures/SDSSJ0248+1913_model.pdf}
	\includegraphics[width=1.05\columnwidth]{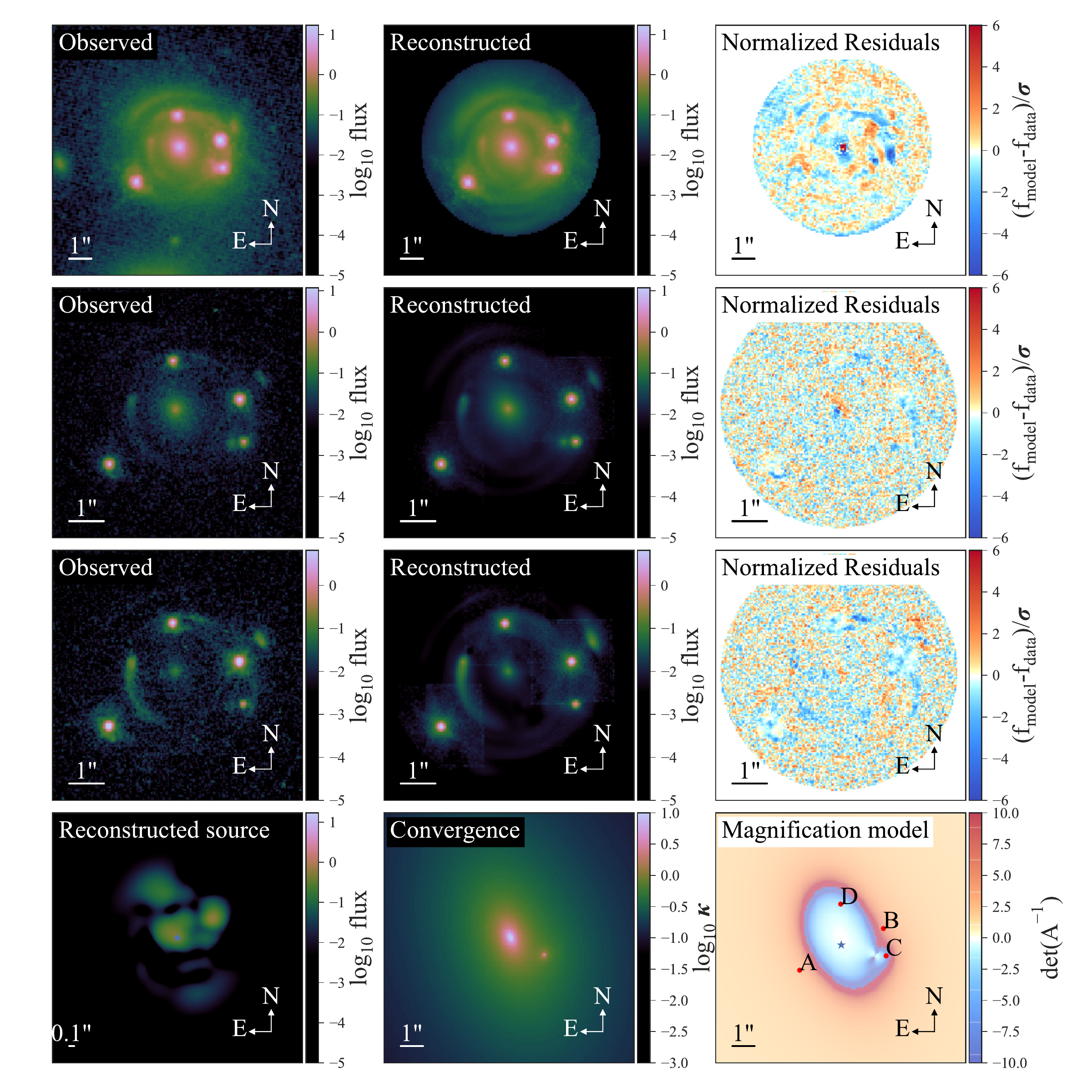} \\ %{figures/DESJ0408-5354_model.pdf} \\
	\includegraphics[width=1.05\columnwidth]{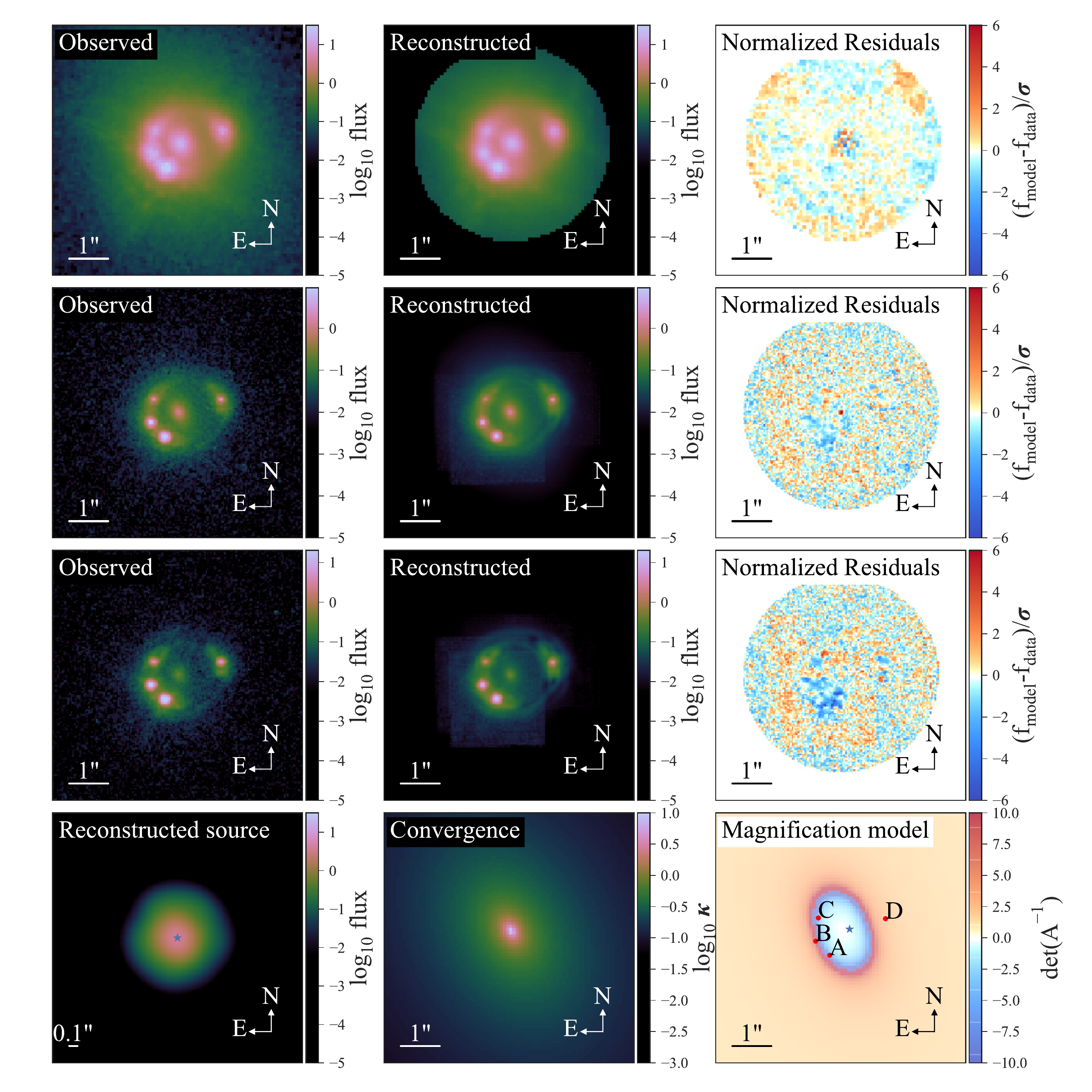}%{figures/SDSSJ1251+2935_model.pdf}
	\includegraphics[width=1.05\columnwidth]{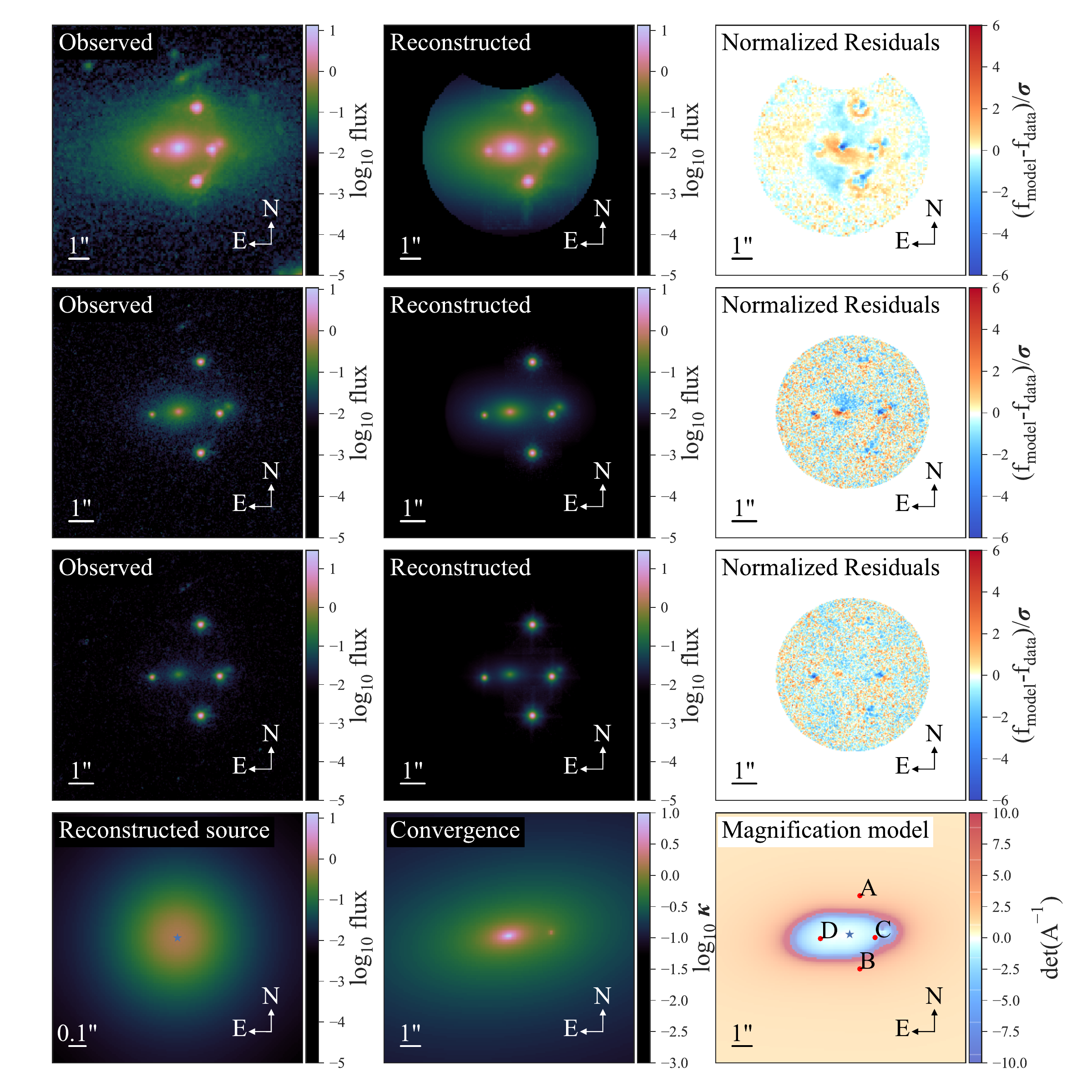}%{figures/SDSSJ1433+6007_model.pdf} 
	\caption{
	Best fit models for SDSS J0248+1913 (top left), DES J0408-5354 (top right), SDSS J1251+2935 (bottom left), and SDSS J1433+6007 (bottom right). The first three rows for each lens system show the observed image, reconstructed lens image, and the normalized residuals in three \textit{HST} bands: F160W, F814W, and F475X, respectively. The fourth row shows the reconstructed source in the F160W band, the convergence, and the magnification model. The models for the rest of the sample are shown in Appendix \ref{app:lens_models} (Figure \ref{fig:lens_model_breakdown_2}, \ref{fig:lens_model_breakdown_3}).
	\label{fig:lens_model_breakdown}
	}
	\end{figure*}

\subsection{Alignment between mass and light distributions} \label{subsec:alignment}

In this subsection, we report our results on the alignment between the mass and light distributions in our sample of quads (Fig. \ref{fig:mass_light_alignment}).

\subsubsection{Centroid}
The centers of the mass and light distributions match very well for most of the quads with a root-mean-square (RMS) of $0\farcs04$ excluding three outliers (Fig. \ref{fig:mass_light_alignment}a). The three outliers are PS J0147+4630, DES J0408-5354 and PS J0630-1201. In PS J0630-1201, there are two deflectors with comparable mass creating a total of five images. If the two deflectors are embedded in the same dark matter halo, the center of the luminous part of the deflector can have an offset from the center of the halo mass. The other two outliers also have nearby companions possibly biasing the centroid estimation. 

\subsubsection{Ellipticity} 
We find a weak correlation between the ellipticity parameters of the mass and light distribution for the whole sample (Fig. \ref{fig:mass_light_alignment}b). We calculate the Pearson correlation coefficient between the axis ratios $q$ and $q_{\rm L}$ of the mass and light distributions, respectively, in the following way. We sample 1000 points from a two-dimensional Gaussian distribution that is centered on the axis ratio pair ($q$, $q_{\rm L}$) for each quad. We take the standard deviation for this Gaussian distribution along each axis equal to the 1$\sigma$ systematic+statistical uncertainty. We take the covariance between the sampled points for each lens as zero as we observe no degeneracy in the posterior PDF of the axis ratios for individual lenses. The Pearson correlation coefficient for the distribution of the sampled points from all the quads is $r=0.2$ (weak correlation).

\subsubsection{Position angle}
\begin{figure*}
	\includegraphics[width=\textwidth]{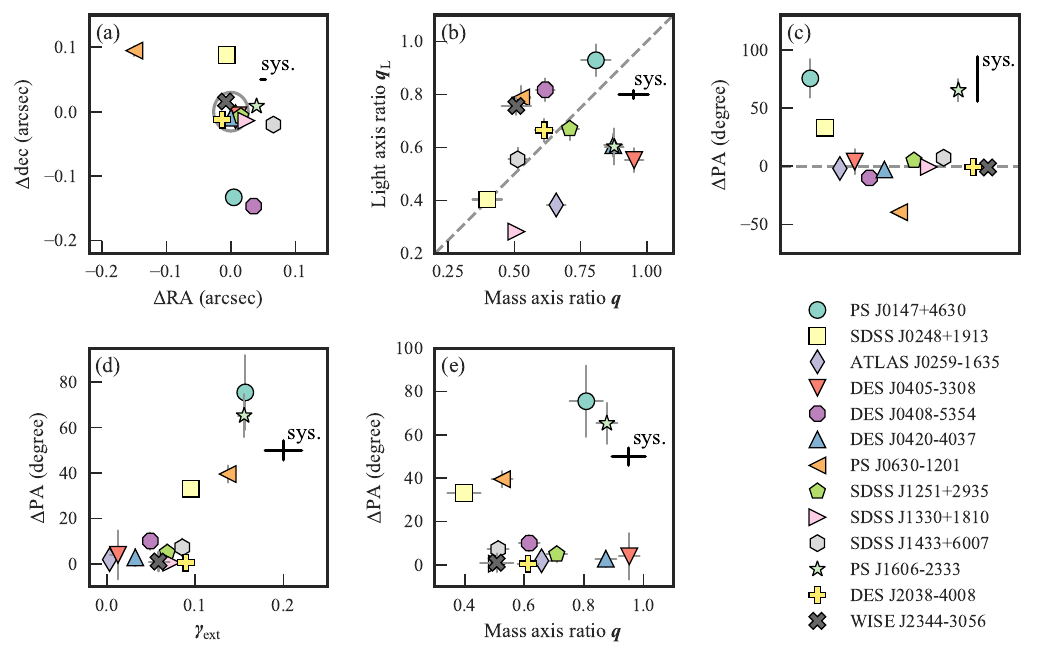}%{figures/alignments.pdf} 
	\caption{
		Mass and light alignments in the deflector galaxies: comparison between (a) the mass and light centroids, (b) the axis ratios of the light and mass profiles, (c) the misalignment angle (between the mass and light profiles' position angles), (d) the misalignment angle and the external shear, and (e) the misalignment angle and the mass profile axis ratio. The thin-solid-grey lines attached to the data points show 1$\sigma$ statistical uncertainty and the thick-solid-black bars annotated with `sys.' in each figure show the $1\sigma$  systematic uncertainty. The systematic uncertainty is estimated by marginalizing over various numerical settings for the system SDSS J0248+1913 as described in Section \ref{subsec:systematics}. In (a) the solid-grey ellipse centered at (0, 0) shows the root-mean-square (RMS) spread of $\Delta$RA and $\Delta$dec for nine lens systems excluding the systems with large deviations: PS J0147+4630, DES J0408-5354 and PS J0630-1201. This RMS spread can be taken as the upper limit of the systematics. The dashed grey line traces the perfect 1-to-1 correlation in (b) and the zero misalignment in (c) to aid visualisation. The centers of the mass and light distributions match very well (a). The systems with large offsets between the mass and light centroids have satellites or comparable-mass companions possibly biasing the centroid estimate. The axis ratios of the light and mass distributions are only weakly correlated (b). The position angles align very well within $\pm 12\degr$ for eight out of the 12 systems (c). Systems with large misalignment have larger values of external shear (d). However, there is very weak to no correlation between the position angle misalignment and mass ellipticity (e).		  
		\label{fig:mass_light_alignment}
	}
\end{figure*}

The position angles of the elliptical mass and light distributions are well aligned for nine out of \nlenses\ quads. The standard deviation of the misalignment in position angles for these eight lenses is $11\degr$.  (Fig. \ref{fig:mass_light_alignment}c). The systems with large misalignment also have large external shear. We find a strong correlation between the misalignment angle and the external shear magnitude ($r=0.74$, Fig. \ref{fig:mass_light_alignment}d). We find weak correlation between the misalignment angle and the mass axis ratio $q$ ($r=0.21$, Fig. \ref{fig:mass_light_alignment}e).

%We sample 1000 points from a Gaussian distribution with mean value equal to the misalignment angle $\Delta \phi$ between the mass and light distribution major axes. The standard deviation for each Gaussian distribution is taken as the $1\sigma$ statistical+systematic uncertainty. The mean and standard deviation of the cumulative distribution of $\Delta \phi$ from all the quads are \textbf{$\mu_{\Delta \phi} = 9.8\degr$, $\sigma_{\Delta \phi}=20\degr$}. 

\subsection{Deviation of flux ratios from macro-model} \label{subsec:flux_ratio}
Stars or dark subhalos in the deflector can produce additional magnification or de-magnification of the quasar images through microlensing and millilensing, respectively \citep[for detailed description, see][]{Schneider06}. In that case, the flux ratios of the quasar images will be different than those predicted by the smooth macro-model. Deviation of the flux ratios can also be produced by baryonic structures \citep{Gilman17} or disks \citep{Hsueh16, Hsueh17}, quasar variability with a time delay, and dust extinction \citep{Yonehara08, Anguita08}. We quantify this deviation of the flux ratios in the quasar images as a $\chi^2$-value by
\begin{equation} \label{eq:chi_square}
	\chi^2_{f} = \sum_{{\rm I, J} \in\{ {\rm A,\ B,\ C,\ D} \}}^{{\rm I} \neq {\rm J}} \frac{\left(f_{\rm IJ,\ observed} - f_{\rm IJ,\ model}\right)^2}{\sigma_{f_{\rm IJ}}^2},
\end{equation}
where $f_{\rm IJ} = F_{\rm I}/F_{\rm J}$ is the flux ratio between the images I and J. We assume 20 per cent flux error giving $\sigma_{f_{\rm IJ}} = 0.28 f_{\rm IJ}$. We set this error level considering the typical order of magnitude for intrinsic variability of quasars \citep[e.g.][]{Bonvin17, Courbin18}. Although, many of the quads in our sample have short predicted time-delays (Table \ref{tab:time_delays}), where intrinsic variability is not a major source of deviation in flux-ratios, we take 20 per cent as a conservative error estimate for these lenses.

\begin{figure}
	\includegraphics[width=\columnwidth]{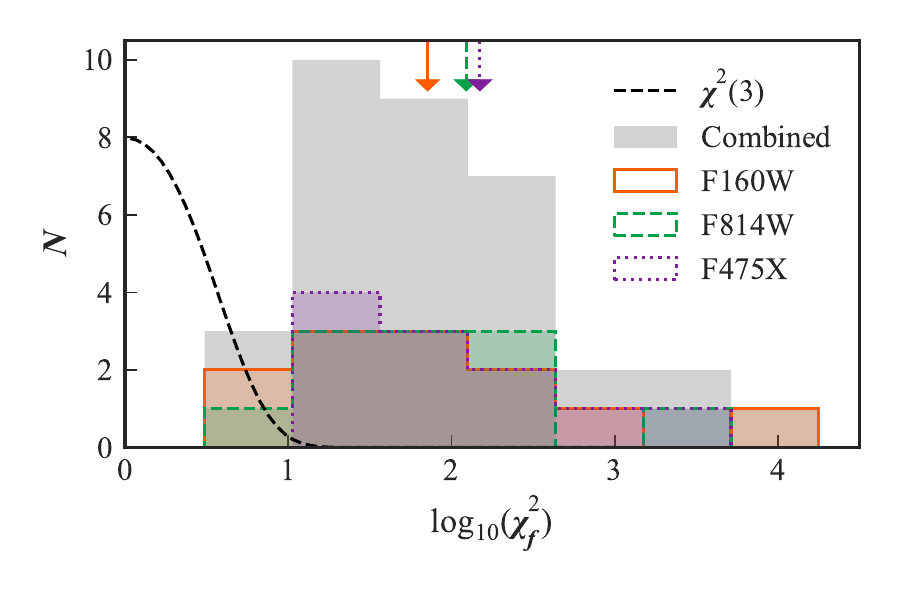}%{figures/flux_ratio.pdf} 
	\caption{
		Distribution of $\chi^2_f$ for flux-ratio anomalies from Equation (\ref{eq:chi_square}) assuming 20 per cent error in flux. The distribution is for twelve quads in our sample excluding PS J0630-1201. The $\chi^2_f$-distributions in bands F160W, F814W, and F475X are shown in orange, green, and purple shaded regions, respectively. The $\chi^2_f$-distribution from all the bands combined is shown as the grey shaded region. The black dashed line shows the expected $\chi^2_f \sim \chi^2(3)$ distribution in the absence of microlensing. The arrows on the top show the mean of the $\chi_f^2$-values in bands F160W (solid orange), F814W (dashed green), and F475X (dotted purple). The combined $\chi^2_f$-distribution is shifted to higher values. The shift is higher for shorter wavelengths, as the quasar size gets smaller with decreasing wavelength making it more susceptible to microlensing.
		\label{fig:flux_ratio_anomaly}
	}
\end{figure}

If the flux ratios are consistent with the macro-model, $\chi^2_f$ is expected to follow the  $\chi^2 (3)$ distribution, i.e. $\chi^2_f \sim \chi^2(3)$, as only three out of the six flux ratios are independent producing three degrees of freedom. However, the $\chi^2_f$-distribution is shifted toward a higher value than $\chi^2(3)$ (Fig. \ref{fig:flux_ratio_anomaly}). The mean of the combined distribution of $\log_{10}{\chi^2_f}$ from all the three \textit{HST} bands is 2.04. A Kolmogorov-Smirnov test of whether the observed $\chi^2_f$-distribution matches with the $\chi^2(3)$-distribution yields a p-value of $\sim0$. The shift is higher in shorter wavelengths. The mean of the $\log_{10}\chi_f^2$'s in the F160W, F814W, and F475X bands are 1.85, 2.09, and 2.17, respectively. This is expected, as the quasar size is smaller in shorter wavelengths making it more affected by microlensing, and as shorter wavelengths are also more affected by dust extinction.

\section{Summary and Discussion}\label{sec:summary}

We presented a general framework to uniformly model large samples of quads while attempting to minimize investigator time. We apply this framework to model a sample of \nlenses\ quads and simultaneously fit imaging data from three \textit{HST} WFC3 bands. All the quads are satisfactorily (p-value $\ga 10^{-8}$) modelled in our uniform framework.  We choose the p-value threshold to be suitably low to be applicable to our quad sample with large morphological variation while being able to point out deficiencies in the modelling choice of profiles along the decision tree. In the end, most of the lens systems in our sample are modelled with p-value $\sim$ 1 (Table \ref{tab:lens_profiles}). Thus, we showed that a large variety of quads can be modelled with a basic set of mass and light profiles under our framework, i.e. all the quads in our sample are `happy' (or, at least `content').

Only one of the quads in our sample, DES J0408-5354, has measured time delays: $\Delta t_{\rm AB}^{\rm observed} = -112 \pm 2.1$ days, $\Delta t_{\rm AD}^{\rm observed} = -155.5 \pm 12.8$ days \citep{Courbin18}. The predicted time delays: $\Delta t_{\rm AB}^{\rm predicted} = -100 \pm 9$ and $\Delta t_{\rm AD}^{\rm predicted} = -140 \pm 13$ days (Appendix \ref{app:time_delay}) are in good agreement with the measured values, although the measured values were not used as constraints in the modelling procedure.

In order to make the problem computationally tractable for much larger samples we made some simplifying assumptions. Thus, whereas some of the lensing quantities, such as Einstein radius, deflector center of mass, position angle and ellipticity, and image flux ratios, are robustly determined, our models are not appropriate for all applications. In particular, science cases requiring high precision might require more sophisticated modelling for each individual lens system.  

The main simplifying assumptions in our work are: 1) we restricted our
models to simple yet general profiles to describe the mass and light
distributions. 2) we assume no colour gradient in the deflector and
source fluxes. Thus, we use the same scalelengths and ellipticity in
the deflector- and source-light profiles in different bands while
fitting simultaneously. Some straightforward ways to further improve the
lens modelling are to allow for
colour dependency of the light distribution of the source and
deflector, explicitly including mass distribution of more nearby
companions or satellites, increasing the number of shapelets
($n_{\max}$), and consider composite mass models consisting of both stellar and dark matter components.

We illustrate the information content of this large sample of quads by investigating the alignment between the light and mass distributions in the deflector galaxies, and the distribution of so-called flux ratio anomalies. Our key results are as follows:

\begin{enumerate}
	\item The centers of the mass and light distributions match very well (the RMS of the offsets is $0\farcs04$).
	\item We find the correlation between the ellipticity of the mass and light distributions to be weak (Pearson correlation coefficient, $r=0.2$).
	\item The position angles of the major axes of the mass and light distributions are well-aligned within $\pm11\degr$ for nine out of 13 lenses.
	\item Systems with high ($>30\degr$) misalignment angle between the light and mass also have large external shear ($\gamma_{\rm ext} \ga 0.1$). The Pearson correlation coefficient between the misalignment angle and the external shear is $r=0.74$.	
    \item The measured flux ratios between the images depart significantly from those predicted by our simple mass models. These flux ratio anomalies are strongest in the bluest band, consistent with microlensing being the main physical driver, in addition to millilensing associated with unseen satellites.
\end{enumerate}

%\pagebreak 
Our finding of weak correlation between the light and mass ellipticity slightly agrees with \citet{Keeton01}, \citet{Ferreras08} and \citet{Rusu16} who find no correlation. However, we do not find a strong correlation as \citet{Sluse12} and \citet{Gavazzi12} report.  The weak correlation between the mass and light ellipticity in our study is consistent with the hierarchical formation scenario of elliptical galaxies where the remnants in the simulation of multiple mergers are shown to have no correlation between the halo and light ellipticity \citep{Weil96}. Moreover, some of the deflectors in our sample are disky galaxies. The projected ellipticity of disky galaxies will not be correlated with the halo ellipticity if viewed from arbitrary orientations.

Moreover, dark matter halos are expected to be rounder than the stellar distribution from simulation \citep{Dubinski91, Warren92, Dubinski94} with reported agreements to observations \citep{Bruderer16, Rusu16}. In our sample, the majority of the systems follow this prediction. Only three systems have significantly flatter mass distribution than the light distribution (DES J0408-5354, PS J0630-1201 and WISE J2344-3056). All these systems have satellites or comparable-mass companions and thus are not the typically relaxed systems where we expect this to hold. In contrast, four systems in our sample are significantly rounder in mass than in light:  ATLAS J0259-1635, DES J0405-3308, DES J0420-4037 and PS J1606-2333. These are likely to be disky galaxies from visual inspection of their shapes. This explains the large difference in ellipticity between the mass and light.

To reliably compare the ellipticity of the light and mass distribution, the ellipticity needs to be estimated within the same aperture, or within an aperture large enough beyond which the ellipticity does not significantly evolve. From a strong-lens system, only the total (projected) mass within the Einstein radius can be estimated. If the Einstein radius is much smaller than the effective radius of the deflector galaxy, the comparison of ellipticity between light and mass may not be representative of the entire galaxy. 

%\citet{Ferreras08} reconstructed non-parametric pixelated maps of stellar-mass and lensing-mass density to compare the ellipticity between light and mass distributions, which does not suffer the aforementioned limitation in our study. However, the agreement between \citet{Ferreras08} and our study in there being weak to no correlation in the ellipticity is reassuring.

We find a strong alignment between the mass and light position angles, which agree very well with previous reports \citep{Kochanek02, Ferreras08, Treu09, Gavazzi12, Sluse12, Bruderer16}. Our result is also in agreement with \citet{Bruderer16} that the systems with high misalignment ($>30\degr$) also have strong external shear ($\gamma_{\rm ext} \ga 0.1$). \iffalse However, a large external shear does not guarantee a high misalignment between light and mass [e.g. PS J1606-2333 in our sample and B1422+231 in the sample of \citet{Keeton98}].\fi The absence of systems with high misalignment but low external shear is in agreement with the prediction of galaxy formation models. Orbits that are highly misaligned in isolated galaxies (thus with low external shear) are shown to be rare and unstable \citep{Heiligman79, Martinet88, Adams07, Debattista15}. The misalignment in isolated galaxies can only be sustained by a constant gas-inflow in blue starburst galaxies \citep{Debattista15}. 

Furthermore, for systems with $\theta_{\rm E}/\theta_{\rm eff} < 1$, the lensing mass is likely to be dominated by the stellar mass. In that case, relatively stronger correlation between the mass and the light distributions is naturally expected. A comparison between the dark matter and luminous matter distribution would be more interesting in regard to directly testing \LCDM\ and galaxy formation theories. However, broadly speaking, large deviations in ellipticity and alignment in our sample have to be explained by the presence of dark matter. However, direct comparison between the dark and luminous mass distributions requires composite mass models with dark and luminous components as adopted by \citet{Bruderer16}. \citet{Gomer18} find that two elliptical mass distributions corresponding to the dark matter and baryon with an offset can better reproduce the image positions in quads than just one smooth elliptical mass distribution with external shear. Those kinds of mass models are beyond the scope of this paper and left for future studies.

The departures of flux-ratios from the smooth model in the disky galaxies in our sample are not at the extreme of the $\chi_f^2$-distribution. This further supports microlensing by foreground stars being the dominant source of the flux-ratio anomaly.

Detailed follow-up of this sample is under way, to measure redshift and velocity dispersion of the deflectors as well as the time delays between the quasars and the properties of the environment. Once follow-up is completed, we will use this sample to address fundamental questions such as the determination of the Hubble Constant \citep[e.g.][]{Bonvin17}, the nature of dark matter \citep[e.g.][]{Gilman18}, and the normalization of the stellar initial mass function in massive galaxies \citep[e.g.][]{Schechter14}.

\section*{Acknowledgements}
AJS, SB, TT, and CDF acknowledge support by NASA through STSCI grant HST-GO-15320, and by the Packard Foundation through a  Packard Fellowship to TT. Support for Program HST-GO-15320  was provided by NASA through a grant from the Space Telescope Science Institute, which is operated by the Association of Universities for Research in Astronomy, Incorporated, under NASA contract NAS5-26555. TA acknowledges support by the Ministry for the Economy, Development, and Tourism's Programa Inicativa Cient\'{i}fica Milenio through grant IC 12009, awarded to The Millennium Institute of Astrophysics (MAS). FC and GM acknowledge support from the Swiss National Science Foundation \#200020\_172712. IK is supported by JSPS KAKENHI Grant Number JP15H05896. PJM is supported by the U.S. Department of Energy under contract number DE-AC02-76SF00515. MO is supported in part by World Premier International Research Center Initiative (WPI Initiative), MEXT, Japan, and JSPS KAKENHI Grant Number JP18K03693 and JP15H05892. FO acknowledges the joint support of CAPES (the Science without Borders program) and the Cambridge Commonwealth Trust during part of this research. SHS thanks the Max Planck Society for support through the Max Planck Research Group.

This work is partly based on data from the ESO VLT Survey Telecope  at Paranal Observatory under program ID 177.A-3011. This work used computational and storage services associated with the Hoffman2 Shared Cluster provided by UCLA Institute for Digital Research and Education's Research Technology Group. 

Funding for the DES Projects has been provided by the U.S. Department of Energy, the U.S. National Science Foundation, the Ministry of Science and Education of Spain, 
the Science and Technology Facilities Council of the United Kingdom, the Higher Education Funding Council for England, the National Center for Supercomputing 
Applications at the University of Illinois at Urbana-Champaign, the Kavli Institute of Cosmological Physics at the University of Chicago, 
the Center for Cosmology and Astro-Particle Physics at the Ohio State University,
the Mitchell Institute for Fundamental Physics and Astronomy at Texas A\&M University, Financiadora de Estudos e Projetos, 
Funda{\c c}{\~a}o Carlos Chagas Filho de Amparo {\`a} Pesquisa do Estado do Rio de Janeiro, Conselho Nacional de Desenvolvimento Cient{\'i}fico e Tecnol{\'o}gico and 
the Minist{\'e}rio da Ci{\^e}ncia, Tecnologia e Inova{\c c}{\~a}o, the Deutsche Forschungsgemeinschaft and the Collaborating Institutions in the Dark Energy Survey. 

The Collaborating Institutions are Argonne National Laboratory, the University of California at Santa Cruz, the University of Cambridge, Centro de Investigaciones Energ{\'e}ticas, 
Medioambientales y Tecnol{\'o}gicas-Madrid, the University of Chicago, University College London, the DES-Brazil Consortium, the University of Edinburgh, 
the Eidgen{\"o}ssische Technische Hochschule (ETH) Z{\"u}rich, 
Fermi National Accelerator Laboratory, the University of Illinois at Urbana-Champaign, the Institut de Ci{\`e}ncies de l'Espai (IEEC/CSIC), 
the Institut de F{\'i}sica d'Altes Energies, Lawrence Berkeley National Laboratory, the Ludwig-Maximilians Universit{\"a}t M{\"u}nchen and the associated Excellence Cluster Universe, 
the University of Michigan, the National Optical Astronomy Observatory, the University of Nottingham, The Ohio State University, the University of Pennsylvania, the University of Portsmouth, 
SLAC National Accelerator Laboratory, Stanford University, the University of Sussex, Texas A\&M University, and the OzDES Membership Consortium.

Based in part on observations at Cerro Tololo Inter-American Observatory, National Optical Astronomy Observatory, which is operated by the Association of 
Universities for Research in Astronomy (AURA) under a cooperative agreement with the National Science Foundation.

The DES data management system is supported by the National Science Foundation under Grant Numbers AST-1138766 and AST-1536171.
The DES participants from Spanish institutions are partially supported by MINECO under grants AYA2015-71825, ESP2015-66861, FPA2015-68048, SEV-2016-0588, SEV-2016-0597, and MDM-2015-0509, 
some of which include ERDF funds from the European Union. IFAE is partially funded by the CERCA program of the Generalitat de Catalunya.
Research leading to these results has received funding from the European Research
Council under the European Union's Seventh Framework Program (FP7/2007-2013) including ERC grant agreements 240672, 291329, and 306478.
We  acknowledge support from the Australian Research Council Centre of Excellence for All-sky Astrophysics (CAASTRO), through project number CE110001020, and the Brazilian Instituto Nacional de Ci\^encia
e Tecnologia (INCT) e-Universe (CNPq grant 465376/2014-2).

This manuscript has been authored by Fermi Research Alliance, LLC under Contract No. DE-AC02-07CH11359 with the U.S. Department of Energy, Office of Science, Office of High Energy Physics. The United States Government retains and the publisher, by accepting the article for publication, acknowledges that the United States Government retains a non-exclusive, paid-up, irrevocable, world-wide license to publish or reproduce the published form of this manuscript, or allow others to do so, for United States Government purposes.

This research made use of \textsc{NumPy} \citep{Oliphant15}, \textsc{SciPy} \citep{Jones01}, \textsc{Astropy}, a community-developed core \textsc{Python} package for Astronomy \citep{AstropyCollaboration13}, \textsc{Jupyter} \citep{Kluyver16}, \textsc{Matplotlib} \citep{Hunter07}, and \textsc{draw.io} at \url{https://www.draw.io}.

%%%%%%%%%%%%%%%%%%%%%%%%%%%%%%%%%%%%%%%%%%%%%%%%%%



\bibliographystyle{mnras}
\bibliography{../ajshajib}

\begin{thebibliography}{}
\makeatletter
\relax
\def\mn@urlcharsother{\let\do\@makeother \do\$\do\&\do\#\do\^\do\_\do\%\do\~}
\def\mn@doi{\begingroup\mn@urlcharsother \@ifnextchar [ {\mn@doi@}
  {\mn@doi@[]}}
\def\mn@doi@[#1]#2{\def\@tempa{#1}\ifx\@tempa\@empty \href
  {http://dx.doi.org/#2} {doi:#2}\else \href {http://dx.doi.org/#2} {#1}\fi
  \endgroup}
\def\mn@eprint#1#2{\mn@eprint@#1:#2::\@nil}
\def\mn@eprint@arXiv#1{\href {http://arxiv.org/abs/#1} {{\tt arXiv:#1}}}
\def\mn@eprint@dblp#1{\href {http://dblp.uni-trier.de/rec/bibtex/#1.xml}
  {dblp:#1}}
\def\mn@eprint@#1:#2:#3:#4\@nil{\def\@tempa {#1}\def\@tempb {#2}\def\@tempc
  {#3}\ifx \@tempc \@empty \let \@tempc \@tempb \let \@tempb \@tempa \fi \ifx
  \@tempb \@empty \def\@tempb {arXiv}\fi \@ifundefined
  {mn@eprint@\@tempb}{\@tempb:\@tempc}{\expandafter \expandafter \csname
  mn@eprint@\@tempb\endcsname \expandafter{\@tempc}}}

\bibitem[\protect\citeauthoryear{{Adams}, {Bloch}, {Butler}, {Druce}  \&
  {Ketchum}}{{Adams} et~al.}{2007}]{Adams07}
{Adams} F.~C.,  {Bloch} A.~M.,  {Butler} S.~C.,  {Druce} J.~M.,   {Ketchum}
  J.~A.,  2007, \mn@doi [\apj] {10.1086/522581}, \href
  {http://adsabs.harvard.edu/abs/2007ApJ...670.1027A} {670, 1027}

\bibitem[\protect\citeauthoryear{{Agnello}, {Kelly}, {Treu}  \&
  {Marshall}}{{Agnello} et~al.}{2015}]{Agnello15}
{Agnello} A.,  {Kelly} B.~C.,  {Treu} T.,   {Marshall} P.~J.,  2015, \mn@doi
  [\mnras] {10.1093/mnras/stv037}, \href
  {http://adsabs.harvard.edu/abs/2015MNRAS.448.1446A} {448, 1446}

\bibitem[\protect\citeauthoryear{{Agnello} et~al.,}{{Agnello}
  et~al.}{2017}]{Agnello17}
{Agnello} A.,  et~al., 2017, \mn@doi [\mnras] {10.1093/mnras/stx2242}, \href
  {http://adsabs.harvard.edu/abs/2017MNRAS.472.4038A} {472, 4038}

\bibitem[\protect\citeauthoryear{{Agnello}, {Grillo}, {Jones}, {Treu},
  {Bonamigo}  \& {Suyu}}{{Agnello} et~al.}{2018a}]{Agnello18}
{Agnello} A.,  {Grillo} C.,  {Jones} T.,  {Treu} T.,  {Bonamigo} M.,   {Suyu}
  S.~H.,  2018a, \mn@doi [\mnras] {10.1093/mnras/stx2950}, \href
  {http://adsabs.harvard.edu/abs/2018MNRAS.474.3391A} {474, 3391}

\bibitem[\protect\citeauthoryear{{Agnello} et~al.,}{{Agnello}
  et~al.}{2018b}]{Agnello18b}
{Agnello} A.,  et~al., 2018b, \mn@doi [\mnras] {10.1093/mnras/stx3226}, \href
  {http://adsabs.harvard.edu/abs/2018MNRAS.475.2086A} {475, 2086}

\bibitem[\protect\citeauthoryear{{Agnello} et~al.,}{{Agnello}
  et~al.}{2018c}]{Agnello18c}
{Agnello} A.,  et~al., 2018c, \mn@doi [\mnras] {10.1093/mnras/sty1419}, \href
  {http://adsabs.harvard.edu/abs/2017arXiv171103971A} {479, 4345}

\bibitem[\protect\citeauthoryear{Akeret, Seehars, Amara, Refregier  \&
  Csillaghy}{Akeret et~al.}{2013}]{Akeret13}
Akeret J.,  Seehars S.,  Amara A.,  Refregier A.,   Csillaghy A.,  2013,
  \mn@doi [Astronomy and Computing] {10.1016/j.ascom.2013.06.003}, 2, 27

\bibitem[\protect\citeauthoryear{{Amara}, {Metcalf}, {Cox}  \&
  {Ostriker}}{{Amara} et~al.}{2006}]{Amara06}
{Amara} A.,  {Metcalf} R.~B.,  {Cox} T.~J.,   {Ostriker} J.~P.,  2006, \mn@doi
  [\mnras] {10.1111/j.1365-2966.2006.10053.x}, \href
  {http://adsabs.harvard.edu/abs/2006MNRAS.367.1367A} {367, 1367}

\bibitem[\protect\citeauthoryear{{Anguita}, {Faure}, {Yonehara}, {Wambsganss},
  {Kneib}, {Covone}  \& {Alloin}}{{Anguita} et~al.}{2008}]{Anguita08}
{Anguita} T.,  {Faure} C.,  {Yonehara} A.,  {Wambsganss} J.,  {Kneib} J.-P.,
  {Covone} G.,   {Alloin} D.,  2008, \mn@doi [\aap]
  {10.1051/0004-6361:20077306}, \href
  {http://adsabs.harvard.edu/abs/2008A%26A...481..615A} {481, 615}

\bibitem[\protect\citeauthoryear{{Anguita} et~al.,}{{Anguita}
  et~al.}{2018}]{Anguita18}
{Anguita} T.,  et~al., 2018, \mn@doi [\mnras] {10.1093/mnras/sty2172}, \href
  {http://adsabs.harvard.edu/abs/2018arXiv180512151A} {480, 5017}

\bibitem[\protect\citeauthoryear{{Astropy Collaboration} et~al.,}{{Astropy
  Collaboration} et~al.}{2013}]{AstropyCollaboration13}
{Astropy Collaboration} et~al., 2013, \mn@doi [\aap]
  {10.1051/0004-6361/201322068}, \href
  {http://adsabs.harvard.edu/abs/2013A%26A...558A..33A} {558, A33}

\bibitem[\protect\citeauthoryear{{Auger}, {Treu}, {Gavazzi}, {Bolton},
  {Koopmans}  \& {Marshall}}{{Auger} et~al.}{2010a}]{Auger10}
{Auger} M.~W.,  {Treu} T.,  {Gavazzi} R.,  {Bolton} A.~S.,  {Koopmans}
  L.~V.~E.,   {Marshall} P.~J.,  2010a, \mn@doi [\apjl]
  {10.1088/2041-8205/721/2/L163}, \href
  {http://adsabs.harvard.edu/abs/2010ApJ...721L.163A} {721, L163}

\bibitem[\protect\citeauthoryear{{Auger}, {Treu}, {Bolton}, {Gavazzi},
  {Koopmans}, {Marshall}, {Moustakas}  \& {Burles}}{{Auger}
  et~al.}{2010b}]{Auger10b}
{Auger} M.~W.,  {Treu} T.,  {Bolton} A.~S.,  {Gavazzi} R.,  {Koopmans}
  L.~V.~E.,  {Marshall} P.~J.,  {Moustakas} L.~A.,   {Burles} S.,  2010b,
  \mn@doi [\apj] {10.1088/0004-637X/724/1/511}, \href
  {http://adsabs.harvard.edu/abs/2010ApJ...724..511A} {724, 511}

\bibitem[\protect\citeauthoryear{{Berghea}, {Nelson}, {Rusu}, {Keeton}  \&
  {Dudik}}{{Berghea} et~al.}{2017}]{Berghea17}
{Berghea} C.~T.,  {Nelson} G.~J.,  {Rusu} C.~E.,  {Keeton} C.~R.,   {Dudik}
  R.~P.,  2017, \mn@doi [\apj] {10.3847/1538-4357/aa7aa6}, \href
  {http://adsabs.harvard.edu/abs/2017ApJ...844...90B} {844, 90}

\bibitem[\protect\citeauthoryear{{Bernal}, {Verde}  \& {Riess}}{{Bernal}
  et~al.}{2016}]{Bernal16}
{Bernal} J.~L.,  {Verde} L.,   {Riess} A.~G.,  2016, \mn@doi [\jcap]
  {10.1088/1475-7516/2016/10/019}, \href
  {http://adsabs.harvard.edu/abs/2016JCAP...10..019B} {10, 019}

\bibitem[\protect\citeauthoryear{{Bertin} \& {Arnouts}}{{Bertin} \&
  {Arnouts}}{1996}]{Bertin96}
{Bertin} E.,  {Arnouts} S.,  1996, \mn@doi [\aaps] {10.1051/aas:1996164}, \href
  {http://adsabs.harvard.edu/abs/1996A%26AS..117..393B} {117, 393}

\bibitem[\protect\citeauthoryear{Birrer \& Amara}{Birrer \&
  Amara}{2018}]{Birrer18}
Birrer S.,  Amara A.,  2018, preprint, \href
  {http://adsabs.harvard.edu/abs/2018arXiv180309746B} {} (\mn@eprint {ascl}
  {1804.012})

\bibitem[\protect\citeauthoryear{{Birrer}, {Amara}  \& {Refregier}}{{Birrer}
  et~al.}{2015}]{Birrer15}
{Birrer} S.,  {Amara} A.,   {Refregier} A.,  2015, \mn@doi [\apj]
  {10.1088/0004-637X/813/2/102}, \href
  {http://adsabs.harvard.edu/abs/2015ApJ...813..102B} {813, 102}

\bibitem[\protect\citeauthoryear{{Birrer}, {Amara}  \& {Refregier}}{{Birrer}
  et~al.}{2016}]{Birrer16}
{Birrer} S.,  {Amara} A.,   {Refregier} A.,  2016, \mn@doi [\jcap]
  {10.1088/1475-7516/2016/08/020}, \href
  {http://adsabs.harvard.edu/abs/2016JCAP...08..020B} {8, 020}

\bibitem[\protect\citeauthoryear{{Birrer}, {Amara}  \& {Refregier}}{{Birrer}
  et~al.}{2017}]{Birrer17}
{Birrer} S.,  {Amara} A.,   {Refregier} A.,  2017, \mn@doi [\jcap]
  {10.1088/1475-7516/2017/05/037}, \href
  {http://adsabs.harvard.edu/abs/2017JCAP...05..037B} {5, 037}

\bibitem[\protect\citeauthoryear{{Birrer} et~al.,}{{Birrer}
  et~al.}{2019}]{Birrer19}
{Birrer} S.,  et~al., 2019, arXiv e-prints, \href
  {http://adsabs.harvard.edu/abs/2018arXiv180901274B} {}

\bibitem[\protect\citeauthoryear{{Bonvin} et~al.,}{{Bonvin}
  et~al.}{2017}]{Bonvin17}
{Bonvin} V.,  et~al., 2017, \mn@doi [\mnras] {10.1093/mnras/stw3006}, \href
  {http://adsabs.harvard.edu/abs/2017MNRAS.465.4914B} {465, 4914}

\bibitem[\protect\citeauthoryear{{Boylan-Kolchin}, {Bullock}  \&
  {Kaplinghat}}{{Boylan-Kolchin} et~al.}{2011}]{Boylan-Kolchin11}
{Boylan-Kolchin} M.,  {Bullock} J.~S.,   {Kaplinghat} M.,  2011, \mn@doi
  [\mnras] {10.1111/j.1745-3933.2011.01074.x}, \href
  {http://adsabs.harvard.edu/abs/2011MNRAS.415L..40B} {415, L40}

\bibitem[\protect\citeauthoryear{{Bruderer}, {Read}, {Coles}, {Leier}, {Falco},
  {Ferreras}  \& {Saha}}{{Bruderer} et~al.}{2016}]{Bruderer16}
{Bruderer} C.,  {Read} J.~I.,  {Coles} J.~P.,  {Leier} D.,  {Falco} E.~E.,
  {Ferreras} I.,   {Saha} P.,  2016, \mn@doi [\mnras] {10.1093/mnras/stv2582},
  \href {http://adsabs.harvard.edu/abs/2016MNRAS.456..870B} {456, 870}

\bibitem[\protect\citeauthoryear{{Cappellari} et~al.,}{{Cappellari}
  et~al.}{2012}]{Cappellari12}
{Cappellari} M.,  et~al., 2012, \mn@doi [\nat] {10.1038/nature10972}, \href
  {http://adsabs.harvard.edu/abs/2012Natur.484..485C} {484, 485}

\bibitem[\protect\citeauthoryear{{Chen} et~al.,}{{Chen} et~al.}{2016}]{Chen16}
{Chen} G.~C.-F.,  et~al., 2016, \mn@doi [\mnras] {10.1093/mnras/stw991}, \href
  {http://adsabs.harvard.edu/abs/2016MNRAS.462.3457C} {462, 3457}

\bibitem[\protect\citeauthoryear{{Claeskens}, {Sluse}, {Riaud}  \&
  {Surdej}}{{Claeskens} et~al.}{2006}]{Claeskens06}
{Claeskens} J.-F.,  {Sluse} D.,  {Riaud} P.,   {Surdej} J.,  2006, \mn@doi
  [\aap] {10.1051/0004-6361:20054352}, \href
  {http://adsabs.harvard.edu/abs/2006A%26A...451..865C} {451, 865}

\bibitem[\protect\citeauthoryear{{Collett}}{{Collett}}{2015}]{Collett15}
{Collett} T.~E.,  2015, \mn@doi [\apj] {10.1088/0004-637X/811/1/20}, \href
  {http://adsabs.harvard.edu/abs/2015ApJ...811...20C} {811, 20}

\bibitem[\protect\citeauthoryear{{Courbin} et~al.,}{{Courbin}
  et~al.}{2018}]{Courbin18}
{Courbin} F.,  et~al., 2018, \mn@doi [\aap] {10.1051/0004-6361/201731461},
  \href {http://adsabs.harvard.edu/abs/2017arXiv170609424C} {609, A71}

\bibitem[\protect\citeauthoryear{{D'Souza}, {Kauffman}, {Wang}  \&
  {Vegetti}}{{D'Souza} et~al.}{2014}]{DSouza14}
{D'Souza} R.,  {Kauffman} G.,  {Wang} J.,   {Vegetti} S.,  2014, \mn@doi
  [\mnras] {10.1093/mnras/stu1194}, \href
  {http://adsabs.harvard.edu/abs/2014MNRAS.443.1433D} {443, 1433}

\bibitem[\protect\citeauthoryear{{Dalal} \& {Kochanek}}{{Dalal} \&
  {Kochanek}}{2002}]{Dalal02}
{Dalal} N.,  {Kochanek} C.~S.,  2002, \mn@doi [\apj] {10.1086/340303}, \href
  {http://adsabs.harvard.edu/abs/2002ApJ...572...25D} {572, 25}

\bibitem[\protect\citeauthoryear{{Dawson} et~al.,}{{Dawson}
  et~al.}{2013}]{Dawson13}
{Dawson} K.~S.,  et~al., 2013, \mn@doi [\aj] {10.1088/0004-6256/145/1/10},
  \href {http://adsabs.harvard.edu/abs/2013AJ....145...10D} {145, 10}

\bibitem[\protect\citeauthoryear{{Debattista}, {Moore}, {Quinn}, {Kazantzidis},
  {Maas}, {Mayer}, {Read}  \& {Stadel}}{{Debattista}
  et~al.}{2008}]{Debattista08}
{Debattista} V.~P.,  {Moore} B.,  {Quinn} T.,  {Kazantzidis} S.,  {Maas} R.,
  {Mayer} L.,  {Read} J.,   {Stadel} J.,  2008, \mn@doi [\apj]
  {10.1086/587977}, \href {http://adsabs.harvard.edu/abs/2008ApJ...681.1076D}
  {681, 1076}

\bibitem[\protect\citeauthoryear{{Debattista}, {van den Bosch}, {Ro{\v s}kar},
  {Quinn}, {Moore}  \& {Cole}}{{Debattista} et~al.}{2015}]{Debattista15}
{Debattista} V.~P.,  {van den Bosch} F.~C.,  {Ro{\v s}kar} R.,  {Quinn} T.,
  {Moore} B.,   {Cole} D.~R.,  2015, \mn@doi [\mnras] {10.1093/mnras/stv1563},
  \href {http://adsabs.harvard.edu/abs/2015MNRAS.452.4094D} {452, 4094}

\bibitem[\protect\citeauthoryear{{Delchambre} et~al.,}{{Delchambre}
  et~al.}{2018}]{Delchambre18}
{Delchambre} L.,  et~al., 2018, preprint, \href
  {http://adsabs.harvard.edu/abs/2018arXiv180702845D} {} (\mn@eprint {arXiv}
  {1807.02845})

\bibitem[\protect\citeauthoryear{{Diehl} et~al.,}{{Diehl}
  et~al.}{2017}]{Diehl17}
{Diehl} H.~T.,  et~al., 2017, \mn@doi [\apjs] {10.3847/1538-4365/aa8667}, \href
  {http://adsabs.harvard.edu/abs/2017ApJS..232...15D} {232, 15}

\bibitem[\protect\citeauthoryear{{Dubinski}}{{Dubinski}}{1994}]{Dubinski94}
{Dubinski} J.,  1994, \mn@doi [\apj] {10.1086/174512}, \href
  {http://adsabs.harvard.edu/abs/1994ApJ...431..617D} {431, 617}

\bibitem[\protect\citeauthoryear{{Dubinski} \& {Carlberg}}{{Dubinski} \&
  {Carlberg}}{1991}]{Dubinski91}
{Dubinski} J.,  {Carlberg} R.~G.,  1991, \mn@doi [\apj] {10.1086/170451}, \href
  {http://adsabs.harvard.edu/abs/1991ApJ...378..496D} {378, 496}

\bibitem[\protect\citeauthoryear{{Ene} et~al.,}{{Ene} et~al.}{2018}]{Ene18}
{Ene} I.,  et~al., 2018, \mn@doi [\mnras] {10.1093/mnras/sty1649}, \href
  {http://adsabs.harvard.edu/abs/2018MNRAS.479.2810E} {479, 2810}

\bibitem[\protect\citeauthoryear{{Ferreras}, {Saha}  \& {Burles}}{{Ferreras}
  et~al.}{2008}]{Ferreras08}
{Ferreras} I.,  {Saha} P.,   {Burles} S.,  2008, \mn@doi [\mnras]
  {10.1111/j.1365-2966.2007.12606.x}, \href
  {http://adsabs.harvard.edu/abs/2008MNRAS.383..857F} {383, 857}

\bibitem[\protect\citeauthoryear{{Foreman-Mackey}, {Hogg}, {Lang}  \&
  {Goodman}}{{Foreman-Mackey} et~al.}{2013}]{Foreman-Mackey13}
{Foreman-Mackey} D.,  {Hogg} D.~W.,  {Lang} D.,   {Goodman} J.,  2013, \mn@doi
  [\pasp] {10.1086/670067}, \href
  {http://adsabs.harvard.edu/abs/2013PASP..125..306F} {125, 306}

\bibitem[\protect\citeauthoryear{{Gavazzi}, {Treu}, {Marshall}, {Brault}  \&
  {Ruff}}{{Gavazzi} et~al.}{2012}]{Gavazzi12}
{Gavazzi} R.,  {Treu} T.,  {Marshall} P.~J.,  {Brault} F.,   {Ruff} A.,  2012,
  \mn@doi [\apj] {10.1088/0004-637X/761/2/170}, \href
  {http://adsabs.harvard.edu/abs/2012ApJ...761..170G} {761, 170}

\bibitem[\protect\citeauthoryear{{Gilman}, {Agnello}, {Treu}, {Keeton}  \&
  {Nierenberg}}{{Gilman} et~al.}{2017}]{Gilman17}
{Gilman} D.,  {Agnello} A.,  {Treu} T.,  {Keeton} C.~R.,   {Nierenberg} A.~M.,
  2017, \mn@doi [\mnras] {10.1093/mnras/stx158}, \href
  {http://adsabs.harvard.edu/abs/2017MNRAS.467.3970G} {467, 3970}

\bibitem[\protect\citeauthoryear{Gilman, Birrer, Treu, Keeton  \&
  Nierenberg}{Gilman et~al.}{2018}]{Gilman18}
Gilman D.,  Birrer S.,  Treu T.,  Keeton C.~R.,   Nierenberg A.,  2018, \mn@doi
  [\mnras] {10.1093/mnras/sty2261}, \href
  {http://adsabs.harvard.edu/abs/2017arXiv171204945G} {481, 819}

\bibitem[\protect\citeauthoryear{{Gomer} \& {Williams}}{{Gomer} \&
  {Williams}}{2018}]{Gomer18}
{Gomer} M.~R.,  {Williams} L.~L.~R.,  2018, \mn@doi [\mnras]
  {10.1093/mnras/stx3294}, \href
  {http://adsabs.harvard.edu/abs/2018MNRAS.475.1987G} {475, 1987}

\bibitem[\protect\citeauthoryear{Goodman \& Weare}{Goodman \&
  Weare}{2010}]{Goodman10}
Goodman J.,  Weare J.,  2010, \mn@doi [Communications in Applied Mathematics
  and Computational Science] {10.2140/camcos.2010.5.65}, 5, 65

\bibitem[\protect\citeauthoryear{{Goullaud}, {Jensen}, {Blakeslee}, {Ma},
  {Greene}  \& {Thomas}}{{Goullaud} et~al.}{2018}]{Goullaud18}
{Goullaud} C.~F.,  {Jensen} J.~B.,  {Blakeslee} J.~P.,  {Ma} C.-P.,  {Greene}
  J.~E.,   {Thomas} J.,  2018, \mn@doi [\apj] {10.3847/1538-4357/aab1f3}, \href
  {http://adsabs.harvard.edu/abs/2018ApJ...856...11G} {856, 11}

\bibitem[\protect\citeauthoryear{{Heiligman} \& {Schwarzschild}}{{Heiligman} \&
  {Schwarzschild}}{1979}]{Heiligman79}
{Heiligman} G.,  {Schwarzschild} M.,  1979, \mn@doi [\apj] {10.1086/157449},
  \href {http://adsabs.harvard.edu/abs/1979ApJ...233..872H} {233, 872}

\bibitem[\protect\citeauthoryear{{Hezaveh} et~al.,}{{Hezaveh}
  et~al.}{2016}]{Hezaveh16}
{Hezaveh} Y.~D.,  et~al., 2016, \mn@doi [\apj] {10.3847/0004-637X/823/1/37},
  \href {http://adsabs.harvard.edu/abs/2016ApJ...823...37H} {823, 37}

\bibitem[\protect\citeauthoryear{{Hezaveh}, {Levasseur}  \&
  {Marshall}}{{Hezaveh} et~al.}{2017}]{Hezaveh17}
{Hezaveh} Y.~D.,  {Levasseur} L.~P.,   {Marshall} P.~J.,  2017, \mn@doi [\nat]
  {10.1038/nature23463}, \href
  {http://adsabs.harvard.edu/abs/2017Natur.548..555H} {548, 555}

\bibitem[\protect\citeauthoryear{{Hsueh}, {Fassnacht}, {Vegetti}, {McKean},
  {Spingola}, {Auger}, {Koopmans}  \& {Lagattuta}}{{Hsueh}
  et~al.}{2016}]{Hsueh16}
{Hsueh} J.-W.,  {Fassnacht} C.~D.,  {Vegetti} S.,  {McKean} J.~P.,  {Spingola}
  C.,  {Auger} M.~W.,  {Koopmans} L.~V.~E.,   {Lagattuta} D.~J.,  2016, \mn@doi
  [\mnras] {10.1093/mnrasl/slw146}, \href
  {http://adsabs.harvard.edu/abs/2016MNRAS.463L..51H} {463, L51}

\bibitem[\protect\citeauthoryear{{Hsueh} et~al.,}{{Hsueh}
  et~al.}{2017}]{Hsueh17}
{Hsueh} J.-W.,  et~al., 2017, \mn@doi [\mnras] {10.1093/mnras/stx1082}, \href
  {http://adsabs.harvard.edu/abs/2017MNRAS.469.3713H} {469, 3713}

\bibitem[\protect\citeauthoryear{Hunter}{Hunter}{2007}]{Hunter07}
Hunter J.~D.,  2007, \mn@doi [Computing in Science and Engineering]
  {10.1109/MCSE.2007.55}, 9, 90

\bibitem[\protect\citeauthoryear{{Ibata}, {Lewis}, {Irwin}, {Totten}  \&
  {Quinn}}{{Ibata} et~al.}{2001}]{Ibata01}
{Ibata} R.,  {Lewis} G.~F.,  {Irwin} M.,  {Totten} E.,   {Quinn} T.,  2001,
  \mn@doi [\apj] {10.1086/320060}, \href
  {http://adsabs.harvard.edu/abs/2001ApJ...551..294I} {551, 294}

\bibitem[\protect\citeauthoryear{{Inada} et~al.,}{{Inada}
  et~al.}{2012}]{Inada12}
{Inada} N.,  et~al., 2012, \mn@doi [\aj] {10.1088/0004-6256/143/5/119}, \href
  {http://adsabs.harvard.edu/abs/2012AJ....143..119I} {143, 119}

\bibitem[\protect\citeauthoryear{{Jedrzejewski}}{{Jedrzejewski}}{1987}]{Jedrzejewski87}
{Jedrzejewski} R.~I.,  1987, \mn@doi [\mnras] {10.1093/mnras/226.4.747}, \href
  {http://adsabs.harvard.edu/abs/1987MNRAS.226..747J} {226, 747}

\bibitem[\protect\citeauthoryear{{Jee}, {Komatsu}, {Suyu}  \& {Huterer}}{{Jee}
  et~al.}{2016}]{Jee16}
{Jee} I.,  {Komatsu} E.,  {Suyu} S.~H.,   {Huterer} D.,  2016, \mn@doi [\jcap]
  {10.1088/1475-7516/2016/04/031}, \href
  {http://adsabs.harvard.edu/abs/2016JCAP...04..031J} {4, 031}

\bibitem[\protect\citeauthoryear{{Jing} \& {Suto}}{{Jing} \&
  {Suto}}{2002}]{Jing02}
{Jing} Y.~P.,  {Suto} Y.,  2002, \mn@doi [\apj] {10.1086/341065}, \href
  {http://adsabs.harvard.edu/abs/2002ApJ...574..538J} {574, 538}

\bibitem[\protect\citeauthoryear{Jones, Oliphant, Peterson  \& Others}{Jones
  et~al.}{2001}]{Jones01}
Jones E.,  Oliphant T.,  Peterson P.,   Others 2001, {SciPy}: Open source
  scientific tools for Python, \url {http://www.scipy.org/}

\bibitem[\protect\citeauthoryear{{Katz} \& {Gunn}}{{Katz} \&
  {Gunn}}{1991}]{Katz91}
{Katz} N.,  {Gunn} J.~E.,  1991, \mn@doi [\apj] {10.1086/170367}, \href
  {http://adsabs.harvard.edu/abs/1991ApJ...377..365K} {377, 365}

\bibitem[\protect\citeauthoryear{{Kauffmann}, {White}  \&
  {Guiderdoni}}{{Kauffmann} et~al.}{1993}]{Kauffmann93}
{Kauffmann} G.,  {White} S.~D.~M.,   {Guiderdoni} B.,  1993, \mn@doi [\mnras]
  {10.1093/mnras/264.1.201}, \href
  {http://adsabs.harvard.edu/abs/1993MNRAS.264..201K} {264, 201}

\bibitem[\protect\citeauthoryear{{Kayo} et~al.,}{{Kayo} et~al.}{2007}]{Kayo07}
{Kayo} I.,  et~al., 2007, \mn@doi [\aj] {10.1086/521652}, \href
  {http://adsabs.harvard.edu/abs/2007AJ....134.1515K} {134, 1515}

\bibitem[\protect\citeauthoryear{{Kazantzidis}, {Kravtsov}, {Zentner},
  {Allgood}, {Nagai}  \& {Moore}}{{Kazantzidis} et~al.}{2004}]{Kazantzidis04}
{Kazantzidis} S.,  {Kravtsov} A.~V.,  {Zentner} A.~R.,  {Allgood} B.,  {Nagai}
  D.,   {Moore} B.,  2004, \mn@doi [\apjl] {10.1086/423992}, \href
  {http://adsabs.harvard.edu/abs/2004ApJ...611L..73K} {611, L73}

\bibitem[\protect\citeauthoryear{{Keeton}}{{Keeton}}{2001}]{Keeton01}
{Keeton} C.~R.,  2001, ArXiv Astrophysics e-prints, \href
  {http://adsabs.harvard.edu/abs/2001astro.ph..2341K} {}

\bibitem[\protect\citeauthoryear{{Keeton} \& {Moustakas}}{{Keeton} \&
  {Moustakas}}{2009}]{Keeton09}
{Keeton} C.~R.,  {Moustakas} L.~A.,  2009, \mn@doi [\apj]
  {10.1088/0004-637X/699/2/1720}, \href
  {http://adsabs.harvard.edu/abs/2009ApJ...699.1720K} {699, 1720}

\bibitem[\protect\citeauthoryear{{Keeton}, {Kochanek}  \& {Falco}}{{Keeton}
  et~al.}{1998}]{Keeton98}
{Keeton} C.~R.,  {Kochanek} C.~S.,   {Falco} E.~E.,  1998, \mn@doi [\apj]
  {10.1086/306502}, \href {http://adsabs.harvard.edu/abs/1998ApJ...509..561K}
  {509, 561}

\bibitem[\protect\citeauthoryear{Kluyver et~al.,}{Kluyver
  et~al.}{2016}]{Kluyver16}
Kluyver T.,  et~al., 2016, in Loizides F.,  Schmidt B.,  eds, Positioning and
  Power in Academic Publishing: Players, Agents and Agendas. pp 87 -- 90

\bibitem[\protect\citeauthoryear{{Klypin}, {Kravtsov}, {Valenzuela}  \&
  {Prada}}{{Klypin} et~al.}{1999}]{Klypin99}
{Klypin} A.,  {Kravtsov} A.~V.,  {Valenzuela} O.,   {Prada} F.,  1999, \mn@doi
  [\apj] {10.1086/307643}, \href
  {http://adsabs.harvard.edu/abs/1999ApJ...522...82K} {522, 82}

\bibitem[\protect\citeauthoryear{{Kochanek}}{{Kochanek}}{2002}]{Kochanek02}
{Kochanek} C.~S.,  2002, in {Natarajan} P.,  ed., The Shapes of Galaxies and
  their Dark Halos. {WORLD} {SCIENTIFIC}, pp 62--71 (\mn@eprint {}
  {astro-ph/0106495}), \mn@doi{10.1142/9789812778017_0010}

\bibitem[\protect\citeauthoryear{{Kochanek} \& {Dalal}}{{Kochanek} \&
  {Dalal}}{2004}]{Kochanek04}
{Kochanek} C.~S.,  {Dalal} N.,  2004, \mn@doi [\apj] {10.1086/421436}, \href
  {http://adsabs.harvard.edu/abs/2004ApJ...610...69K} {610, 69}

\bibitem[\protect\citeauthoryear{{Koopmans}}{{Koopmans}}{2005}]{Koopmans05}
{Koopmans} L.~V.~E.,  2005, \mn@doi [\mnras]
  {10.1111/j.1365-2966.2005.09523.x}, \href
  {http://adsabs.harvard.edu/abs/2005MNRAS.363.1136K} {363, 1136}

\bibitem[\protect\citeauthoryear{{Lackner} \& {Gunn}}{{Lackner} \&
  {Gunn}}{2012}]{Lackner12}
{Lackner} C.~N.,  {Gunn} J.~E.,  2012, \mn@doi [\mnras]
  {10.1111/j.1365-2966.2012.20450.x}, \href
  {http://adsabs.harvard.edu/abs/2012MNRAS.421.2277L} {421, 2277}

\bibitem[\protect\citeauthoryear{{Lee}}{{Lee}}{2017}]{Lee17}
{Lee} C.-H.,  2017, \mn@doi [\aap] {10.1051/0004-6361/201731695}, \href
  {http://adsabs.harvard.edu/abs/2017A%26A...605L...8L} {605, L8}

\bibitem[\protect\citeauthoryear{{Lee}}{{Lee}}{2018}]{Lee18}
{Lee} C.-H.,  2018, \mn@doi [\mnras] {10.1093/mnras/sty078}, \href
  {http://adsabs.harvard.edu/abs/2018MNRAS.475.3086L} {475, 3086}

\bibitem[\protect\citeauthoryear{{Lemon}, {Auger}, {McMahon}  \&
  {Ostrovski}}{{Lemon} et~al.}{2018}]{Lemon18}
{Lemon} C.~A.,  {Auger} M.~W.,  {McMahon} R.~G.,   {Ostrovski} F.,  2018,
  \mn@doi [\mnras] {10.1093/mnras/sty911}, \href
  {http://adsabs.harvard.edu/abs/2018MNRAS.tmp..893L} {}

\bibitem[\protect\citeauthoryear{{Lin} et~al.,}{{Lin} et~al.}{2017}]{Lin17}
{Lin} H.,  et~al., 2017, \mn@doi [\apjl] {10.3847/2041-8213/aa624e}, \href
  {http://adsabs.harvard.edu/abs/2017ApJ...838L..15L} {838, L15}

\bibitem[\protect\citeauthoryear{{Linder}}{{Linder}}{2011}]{Linder11}
{Linder} E.~V.,  2011, \mn@doi [\prd] {10.1103/PhysRevD.84.123529}, \href
  {http://adsabs.harvard.edu/abs/2011PhRvD..84l3529L} {84, 123529}

\bibitem[\protect\citeauthoryear{{Lux}, {Read}, {Lake}  \& {Johnston}}{{Lux}
  et~al.}{2012}]{Lux12}
{Lux} H.,  {Read} J.~I.,  {Lake} G.,   {Johnston} K.~V.,  2012, \mn@doi
  [\mnras] {10.1111/j.1745-3933.2012.01276.x}, \href
  {http://adsabs.harvard.edu/abs/2012MNRAS.424L..16L} {424, L16}

\bibitem[\protect\citeauthoryear{{Macci{\`o}}, {Dutton}, {van den Bosch},
  {Moore}, {Potter}  \& {Stadel}}{{Macci{\`o}} et~al.}{2007}]{Maccio07}
{Macci{\`o}} A.~V.,  {Dutton} A.~A.,  {van den Bosch} F.~C.,  {Moore} B.,
  {Potter} D.,   {Stadel} J.,  2007, \mn@doi [\mnras]
  {10.1111/j.1365-2966.2007.11720.x}, \href
  {http://adsabs.harvard.edu/abs/2007MNRAS.378...55M} {378, 55}

\bibitem[\protect\citeauthoryear{{Martinet} \& {de Zeeuw}}{{Martinet} \& {de
  Zeeuw}}{1988}]{Martinet88}
{Martinet} L.,  {de Zeeuw} T.,  1988, \aap, \href
  {http://adsabs.harvard.edu/abs/1988A%26A...206..269M} {206, 269}

\bibitem[\protect\citeauthoryear{{Metcalf} \& {Amara}}{{Metcalf} \&
  {Amara}}{2012}]{Metcalf12}
{Metcalf} R.~B.,  {Amara} A.,  2012, \mn@doi [\mnras]
  {10.1111/j.1365-2966.2011.19982.x}, \href
  {http://adsabs.harvard.edu/abs/2012MNRAS.419.3414M} {419, 3414}

\bibitem[\protect\citeauthoryear{{Metcalf} \& {Madau}}{{Metcalf} \&
  {Madau}}{2001}]{Metcalf01}
{Metcalf} R.~B.,  {Madau} P.,  2001, \mn@doi [\apj] {10.1086/323695}, \href
  {http://adsabs.harvard.edu/abs/2001ApJ...563....9M} {563, 9}

\bibitem[\protect\citeauthoryear{{Metcalf} \& {Zhao}}{{Metcalf} \&
  {Zhao}}{2002}]{Metcalf02}
{Metcalf} R.~B.,  {Zhao} H.,  2002, \mn@doi [\apjl] {10.1086/339798}, \href
  {http://adsabs.harvard.edu/abs/2002ApJ...567L...5M} {567, L5}

\bibitem[\protect\citeauthoryear{{Moore}, {Quinn}, {Governato}, {Stadel}  \&
  {Lake}}{{Moore} et~al.}{1999}]{Moore99}
{Moore} B.,  {Quinn} T.,  {Governato} F.,  {Stadel} J.,   {Lake} G.,  1999,
  \mn@doi [\mnras] {10.1046/j.1365-8711.1999.03039.x}, \href
  {http://adsabs.harvard.edu/abs/1999MNRAS.310.1147M} {310, 1147}

\bibitem[\protect\citeauthoryear{{Moustakas} et~al.,}{{Moustakas}
  et~al.}{2009}]{Moustakas09}
{Moustakas} L.~A.,  et~al., 2009, in astro2010: The Astronomy and Astrophysics
  Decadal Survey.  (\mn@eprint {arXiv} {0902.3219}), \url
  {http://adsabs.harvard.edu/abs/2009astro2010S.214M}

\bibitem[\protect\citeauthoryear{{Navarro}, {Frenk}  \& {White}}{{Navarro}
  et~al.}{1996}]{Navarro96}
{Navarro} J.~F.,  {Frenk} C.~S.,   {White} S.~D.~M.,  1996, \mn@doi [\apj]
  {10.1086/177173}, \href {http://adsabs.harvard.edu/abs/1996ApJ...462..563N}
  {462, 563}

\bibitem[\protect\citeauthoryear{{Nierenberg}, {Treu}, {Wright}, {Fassnacht}
  \& {Auger}}{{Nierenberg} et~al.}{2014}]{Nierenberg14}
{Nierenberg} A.~M.,  {Treu} T.,  {Wright} S.~A.,  {Fassnacht} C.~D.,   {Auger}
  M.~W.,  2014, \mn@doi [\mnras] {10.1093/mnras/stu862}, \href
  {http://adsabs.harvard.edu/abs/2014MNRAS.442.2434N} {442, 2434}

\bibitem[\protect\citeauthoryear{{Nierenberg} et~al.,}{{Nierenberg}
  et~al.}{2017}]{Nierenberg17}
{Nierenberg} A.~M.,  et~al., 2017, \mn@doi [\mnras] {10.1093/mnras/stx1400},
  \href {http://adsabs.harvard.edu/abs/2017MNRAS.471.2224N} {471, 2224}

\bibitem[\protect\citeauthoryear{{Nightingale}, {Dye}  \&
  {Massey}}{{Nightingale} et~al.}{2018}]{Nightingale18}
{Nightingale} J.,  {Dye} S.,   {Massey} R.,  2018, \mn@doi [\mnras]
  {10.1093/mnras/sty1264}, \href
  {http://adsabs.harvard.edu/abs/2017arXiv170807377N} {478, 4738}

\bibitem[\protect\citeauthoryear{{Oguri} \& {Marshall}}{{Oguri} \&
  {Marshall}}{2010}]{Oguri10}
{Oguri} M.,  {Marshall} P.~J.,  2010, \mn@doi [\mnras]
  {10.1111/j.1365-2966.2010.16639.x}, \href
  {http://adsabs.harvard.edu/abs/2010MNRAS.405.2579O} {405, 2579}

\bibitem[\protect\citeauthoryear{{Oguri} et~al.,}{{Oguri}
  et~al.}{2006}]{Oguri06}
{Oguri} M.,  et~al., 2006, \mn@doi [\aj] {10.1086/506019}, \href
  {http://adsabs.harvard.edu/abs/2006AJ....132..999O} {132, 999}

\bibitem[\protect\citeauthoryear{{Oguri}, {Inada}, {Blackburne}, {Shin},
  {Kayo}, {Strauss}, {Schneider}  \& {York}}{{Oguri} et~al.}{2008}]{Oguri08}
{Oguri} M.,  {Inada} N.,  {Blackburne} J.~A.,  {Shin} M.-S.,  {Kayo} I.,
  {Strauss} M.~A.,  {Schneider} D.~P.,   {York} D.~G.,  2008, \mn@doi [\mnras]
  {10.1111/j.1365-2966.2008.14032.x}, \href
  {http://adsabs.harvard.edu/abs/2008MNRAS.391.1973O} {391, 1973}

\bibitem[\protect\citeauthoryear{{Oguri}, {Rusu}  \& {Falco}}{{Oguri}
  et~al.}{2014}]{Oguri14}
{Oguri} M.,  {Rusu} C.~E.,   {Falco} E.~E.,  2014, \mn@doi [\mnras]
  {10.1093/mnras/stu106}, \href
  {http://adsabs.harvard.edu/abs/2014MNRAS.439.2494O} {439, 2494}

\bibitem[\protect\citeauthoryear{{Oh}, {Greene}  \& {Lackner}}{{Oh}
  et~al.}{2017}]{Oh17}
{Oh} S.,  {Greene} J.~E.,   {Lackner} C.~N.,  2017, \mn@doi [\apj]
  {10.3847/1538-4357/836/1/115}, \href
  {http://adsabs.harvard.edu/abs/2017ApJ...836..115O} {836, 115}

\bibitem[\protect\citeauthoryear{Oliphant}{Oliphant}{2015}]{Oliphant15}
Oliphant T.~E.,  2015, Guide to NumPy, 2nd edn.
CreateSpace Independent Publishing Platform, USA

\bibitem[\protect\citeauthoryear{{Ostrovski} et~al.,}{{Ostrovski}
  et~al.}{2017}]{Ostrovski17}
{Ostrovski} F.,  et~al., 2017, \mn@doi [\mnras] {10.1093/mnras/stw2958}, \href
  {http://adsabs.harvard.edu/abs/2017MNRAS.465.4325O} {465, 4325}

\bibitem[\protect\citeauthoryear{{Ostrovski} et~al.,}{{Ostrovski}
  et~al.}{2018}]{Ostrovski18}
{Ostrovski} F.,  et~al., 2018, \mn@doi [\mnras] {10.1093/mnrasl/slx173}, \href
  {http://adsabs.harvard.edu/abs/2018MNRAS.473L.116O} {473, L116}

\bibitem[\protect\citeauthoryear{{Perreault Levasseur}, {Hezaveh}  \&
  {Wechsler}}{{Perreault Levasseur} et~al.}{2017}]{PerreaultLevasseur17}
{Perreault Levasseur} L.,  {Hezaveh} Y.~D.,   {Wechsler} R.~H.,  2017, \mn@doi
  [\apjl] {10.3847/2041-8213/aa9704}, \href
  {http://adsabs.harvard.edu/abs/2017ApJ...850L...7P} {850, L7}

\bibitem[\protect\citeauthoryear{{Planck Collaboration} et~al.,}{{Planck
  Collaboration} et~al.}{2018}]{PlanckCollaboration18}
{Planck Collaboration} et~al., 2018, preprint, \href
  {http://adsabs.harvard.edu/abs/2018arXiv180706209P} {} (\mn@eprint {arXiv}
  {1807.06209})

\bibitem[\protect\citeauthoryear{{Read}}{{Read}}{2014}]{Read14}
{Read} J.~I.,  2014, \mn@doi [Journal of Physics G Nuclear Physics]
  {10.1088/0954-3899/41/6/063101}, \href
  {http://adsabs.harvard.edu/abs/2014JPhG...41f3101R} {41, 063101}

\bibitem[\protect\citeauthoryear{{Refregier}}{{Refregier}}{2003}]{Refregier03}
{Refregier} A.,  2003, \mn@doi [\mnras] {10.1046/j.1365-8711.2003.05901.x},
  \href {http://adsabs.harvard.edu/abs/2003MNRAS.338...35R} {338, 35}

\bibitem[\protect\citeauthoryear{{Refsdal}}{{Refsdal}}{1964}]{Refsdal64}
{Refsdal} S.,  1964, \mn@doi [\mnras] {10.1093/mnras/128.4.307}, \href
  {http://adsabs.harvard.edu/abs/1964MNRAS.128..307R} {128, 307}

\bibitem[\protect\citeauthoryear{{Riess} et~al.,}{{Riess}
  et~al.}{2016}]{Riess16}
{Riess} A.~G.,  et~al., 2016, \mn@doi [\apj] {10.3847/0004-637X/826/1/56},
  \href {http://adsabs.harvard.edu/abs/2016ApJ...826...56R} {826, 56}

\bibitem[\protect\citeauthoryear{{Riess} et~al.,}{{Riess}
  et~al.}{2018a}]{Riess18}
{Riess} A.~G.,  et~al., 2018a, \mn@doi [\apj] {10.3847/1538-4357/aaadb7}, \href
  {http://adsabs.harvard.edu/abs/2018ApJ...855..136R} {855, 136}

\bibitem[\protect\citeauthoryear{Riess et~al.,}{Riess et~al.}{2018b}]{Riess18b}
Riess A.~G.,  et~al., 2018b, \mn@doi [\apj] {10.3847/1538-4357/aac82e}, \href
  {http://adsabs.harvard.edu/abs/2018arXiv180410655R} {861, 126}

\bibitem[\protect\citeauthoryear{{Romanowsky} \& {Kochanek}}{{Romanowsky} \&
  {Kochanek}}{1998}]{Romanowsky98}
{Romanowsky} A.~J.,  {Kochanek} C.~S.,  1998, \mn@doi [\apj] {10.1086/305151},
  \href {http://adsabs.harvard.edu/abs/1998ApJ...493..641R} {493, 641}

\bibitem[\protect\citeauthoryear{{Rusu} et~al.,}{{Rusu} et~al.}{2016}]{Rusu16}
{Rusu} C.~E.,  et~al., 2016, \mn@doi [\mnras] {10.1093/mnras/stw092}, \href
  {http://adsabs.harvard.edu/abs/2016MNRAS.458....2R} {458, 2}

\bibitem[\protect\citeauthoryear{{Rusu} et~al.,}{{Rusu} et~al.}{2017}]{Rusu17}
{Rusu} C.~E.,  et~al., 2017, \mn@doi [\mnras] {10.1093/mnras/stx285}, \href
  {http://adsabs.harvard.edu/abs/2017MNRAS.467.4220R} {467, 4220}

\bibitem[\protect\citeauthoryear{{Schechter}, {Pooley}, {Blackburne}  \&
  {Wambsganss}}{{Schechter} et~al.}{2014}]{Schechter14}
{Schechter} P.~L.,  {Pooley} D.,  {Blackburne} J.~A.,   {Wambsganss} J.,  2014,
  \mn@doi [\apj] {10.1088/0004-637X/793/2/96}, \href
  {http://adsabs.harvard.edu/abs/2014ApJ...793...96S} {793, 96}

\bibitem[\protect\citeauthoryear{{Schechter}, {Morgan}, {Chehade}, {Metcalfe},
  {Shanks}  \& {McDonald}}{{Schechter} et~al.}{2017}]{Schechter17}
{Schechter} P.~L.,  {Morgan} N.~D.,  {Chehade} B.,  {Metcalfe} N.,  {Shanks}
  T.,   {McDonald} M.,  2017, \mn@doi [\aj] {10.3847/1538-3881/aa6899}, \href
  {http://adsabs.harvard.edu/abs/2017AJ....153..219S} {153, 219}

\bibitem[\protect\citeauthoryear{{Schechter}, {Anguita}, {Morgan}, {Read}  \&
  {Shanks}}{{Schechter} et~al.}{2018}]{Schechter18}
{Schechter} P.~L.,  {Anguita} T.,  {Morgan} N.~D.,  {Read} M.,   {Shanks} T.,
  2018, Research Notes of the AAS, \href
  {http://adsabs.harvard.edu/abs/2018arXiv180501939S} {2, 21}

\bibitem[\protect\citeauthoryear{{Schneider}, {Kochanek}  \&
  {Wambsganss}}{{Schneider} et~al.}{2006}]{Schneider06}
{Schneider} P.,  {Kochanek} C.~S.,   {Wambsganss} J.,  2006, Gravitational
  Lensing: Strong, Weak and Micro.
 Saas-Fee Advanced Courses Vol. 33, Springer (\mn@eprint {}
  {astro-ph/0407232}), \url {http://adsabs.harvard.edu/abs/2006glsw.conf.....M}

\bibitem[\protect\citeauthoryear{{Sersic}}{{Sersic}}{1968}]{Sersic68}
{Sersic} J.~L.,  1968, {Atlas de Galaxias Australes}.
\url {http://adsabs.harvard.edu/abs/1968adga.book.....S}

\bibitem[\protect\citeauthoryear{{Shajib} \& {Wright}}{{Shajib} \&
  {Wright}}{2016}]{Shajib16}
{Shajib} A.~J.,  {Wright} E.~L.,  2016, \mn@doi [\apj]
  {10.3847/0004-637X/827/2/116}, \href
  {http://adsabs.harvard.edu/abs/2016ApJ...827..116S} {827, 116}

\bibitem[\protect\citeauthoryear{{Shajib}, {Treu}  \& {Agnello}}{{Shajib}
  et~al.}{2018}]{Shajib18}
{Shajib} A.~J.,  {Treu} T.,   {Agnello} A.,  2018, \mn@doi [\mnras]
  {10.1093/mnras/stx2302}, \href
  {http://adsabs.harvard.edu/abs/2018MNRAS.473..210S} {473, 210}

\bibitem[\protect\citeauthoryear{{Sluse}, {Chantry}, {Magain}, {Courbin}  \&
  {Meylan}}{{Sluse} et~al.}{2012}]{Sluse12}
{Sluse} D.,  {Chantry} V.,  {Magain} P.,  {Courbin} F.,   {Meylan} G.,  2012,
  \mn@doi [\aap] {10.1051/0004-6361/201015844}, \href
  {http://adsabs.harvard.edu/abs/2012A%26A...538A..99S} {538, A99}

\bibitem[\protect\citeauthoryear{{Sluse} et~al.,}{{Sluse}
  et~al.}{2017}]{Sluse17}
{Sluse} D.,  et~al., 2017, MNRAS, submitted, \href
  {http://adsabs.harvard.edu/abs/2016arXiv160700382S} {470, 4838}

\bibitem[\protect\citeauthoryear{{Sonnenfeld} et~al.,}{{Sonnenfeld}
  et~al.}{2018}]{Sonnenfeld18}
{Sonnenfeld} A.,  et~al., 2018, \mn@doi [\pasj] {10.1093/pasj/psx062}, \href
  {http://adsabs.harvard.edu/abs/2018PASJ...70S..29S} {70, S29}

\bibitem[\protect\citeauthoryear{{Stoughton} et~al.,}{{Stoughton}
  et~al.}{2002}]{Stoughton02}
{Stoughton} C.,  et~al., 2002, \mn@doi [\aj] {10.1086/324741}, \href
  {http://adsabs.harvard.edu/abs/2002AJ....123..485S} {123, 485}

\bibitem[\protect\citeauthoryear{{Suyu}, {Marshall}, {Auger}, {Hilbert},
  {Blandford}, {Koopmans}, {Fassnacht}  \& {Treu}}{{Suyu}
  et~al.}{2010}]{Suyu10}
{Suyu} S.~H.,  {Marshall} P.~J.,  {Auger} M.~W.,  {Hilbert} S.,  {Blandford}
  R.~D.,  {Koopmans} L.~V.~E.,  {Fassnacht} C.~D.,   {Treu} T.,  2010, \mn@doi
  [\apj] {10.1088/0004-637X/711/1/201}, \href
  {http://adsabs.harvard.edu/abs/2010ApJ...711..201S} {711, 201}

\bibitem[\protect\citeauthoryear{{Suyu} et~al.,}{{Suyu} et~al.}{2012}]{Suyu12}
{Suyu} S.~H.,  et~al., 2012, preprint, \href
  {http://adsabs.harvard.edu/abs/2012arXiv1202.4459S} {} (\mn@eprint {arXiv}
  {1202.4459})

\bibitem[\protect\citeauthoryear{{Suyu} et~al.,}{{Suyu} et~al.}{2013}]{Suyu13}
{Suyu} S.~H.,  et~al., 2013, \mn@doi [\apj] {10.1088/0004-637X/766/2/70}, \href
  {http://adsabs.harvard.edu/abs/2013ApJ...766...70S} {766, 70}

\bibitem[\protect\citeauthoryear{{Suyu} et~al.,}{{Suyu} et~al.}{2017}]{Suyu17}
{Suyu} S.~H.,  et~al., 2017, \mn@doi [\mnras] {10.1093/mnras/stx483}, \href
  {http://adsabs.harvard.edu/abs/2017MNRAS.468.2590S} {468, 2590}

\bibitem[\protect\citeauthoryear{{Tihhonova} et~al.,}{{Tihhonova}
  et~al.}{2018}]{Tihhonova18}
{Tihhonova} O.,  et~al., 2018, \mn@doi [\mnras] {10.1093/mnras/sty1040}, \href
  {http://adsabs.harvard.edu/abs/2018MNRAS.477.5657T} {477, 5657}

\bibitem[\protect\citeauthoryear{{Treu}}{{Treu}}{2010}]{Treu10b}
{Treu} T.,  2010, \mn@doi [\araa] {10.1146/annurev-astro-081309-130924}, \href
  {http://adsabs.harvard.edu/abs/2010ARA%26A..48...87T} {48, 87}

\bibitem[\protect\citeauthoryear{{Treu} \& {Marshall}}{{Treu} \&
  {Marshall}}{2016}]{Treu16}
{Treu} T.,  {Marshall} P.~J.,  2016, \mn@doi [The Astronomy and Astrophysics
  Review] {10.1007/s00159-016-0096-8}, \href
  {http://adsabs.harvard.edu/abs/2016arXiv160505333T} {24, 11}

\bibitem[\protect\citeauthoryear{{Treu}, {Gavazzi}, {Gorecki}, {Marshall},
  {Koopmans}, {Bolton}, {Moustakas}  \& {Burles}}{{Treu} et~al.}{2009}]{Treu09}
{Treu} T.,  {Gavazzi} R.,  {Gorecki} A.,  {Marshall} P.~J.,  {Koopmans}
  L.~V.~E.,  {Bolton} A.~S.,  {Moustakas} L.~A.,   {Burles} S.,  2009, \mn@doi
  [\apj] {10.1088/0004-637X/690/1/670}, \href
  {http://adsabs.harvard.edu/abs/2009ApJ...690..670T} {690, 670}

\bibitem[\protect\citeauthoryear{{Treu}, {Auger}, {Koopmans}, {Gavazzi},
  {Marshall}  \& {Bolton}}{{Treu} et~al.}{2010}]{Treu10}
{Treu} T.,  {Auger} M.~W.,  {Koopmans} L.~V.~E.,  {Gavazzi} R.,  {Marshall}
  P.~J.,   {Bolton} A.~S.,  2010, \mn@doi [\apj]
  {10.1088/0004-637X/709/2/1195}, \href
  {http://adsabs.harvard.edu/abs/2010ApJ...709.1195T} {709, 1195}

\bibitem[\protect\citeauthoryear{{Treu} et~al.,}{{Treu} et~al.}{2018}]{Treu18}
{Treu} T.,  et~al., 2018, \mn@doi [\mnras] {10.1093/mnras/sty2329}, \href
  {http://adsabs.harvard.edu/abs/2018MNRAS.481.1041T} {481, 1041}

\bibitem[\protect\citeauthoryear{{Vegetti} \& {Koopmans}}{{Vegetti} \&
  {Koopmans}}{2009}]{Vegetti09}
{Vegetti} S.,  {Koopmans} L.~V.~E.,  2009, \mn@doi [\mnras]
  {10.1111/j.1365-2966.2008.14005.x}, \href
  {http://adsabs.harvard.edu/abs/2009MNRAS.392..945V} {392, 945}

\bibitem[\protect\citeauthoryear{{Vegetti}, {Koopmans}, {Bolton}, {Treu}  \&
  {Gavazzi}}{{Vegetti} et~al.}{2010}]{Vegetti10}
{Vegetti} S.,  {Koopmans} L.~V.~E.,  {Bolton} A.,  {Treu} T.,   {Gavazzi} R.,
  2010, \mn@doi [\mnras] {10.1111/j.1365-2966.2010.16865.x}, \href
  {http://adsabs.harvard.edu/abs/2010MNRAS.408.1969V} {408, 1969}

\bibitem[\protect\citeauthoryear{{Vegetti}, {Lagattuta}, {McKean}, {Auger},
  {Fassnacht}  \& {Koopmans}}{{Vegetti} et~al.}{2012}]{Vegetti12}
{Vegetti} S.,  {Lagattuta} D.~J.,  {McKean} J.~P.,  {Auger} M.~W.,  {Fassnacht}
  C.~D.,   {Koopmans} L.~V.~E.,  2012, \mn@doi [\nat] {10.1038/nature10669},
  \href {http://adsabs.harvard.edu/abs/2012Natur.481..341V} {481, 341}

\bibitem[\protect\citeauthoryear{{Vegetti}, {Despali}, {Lovell}  \&
  {Enzi}}{{Vegetti} et~al.}{2018}]{Vegetti18}
{Vegetti} S.,  {Despali} G.,  {Lovell} M.~R.,   {Enzi} W.,  2018, \mn@doi
  [\mnras] {10.1093/mnras/sty2393}, \href
  {http://adsabs.harvard.edu/abs/2018arXiv180101505V} {481, 3661}

\bibitem[\protect\citeauthoryear{{Viel}, {Becker}, {Bolton}  \&
  {Haehnelt}}{{Viel} et~al.}{2013}]{Viel13}
{Viel} M.,  {Becker} G.~D.,  {Bolton} J.~S.,   {Haehnelt} M.~G.,  2013, \mn@doi
  [\prd] {10.1103/PhysRevD.88.043502}, \href
  {http://adsabs.harvard.edu/abs/2013PhRvD..88d3502V} {88, 043502}

\bibitem[\protect\citeauthoryear{{Warren}, {Quinn}, {Salmon}  \&
  {Zurek}}{{Warren} et~al.}{1992}]{Warren92}
{Warren} M.~S.,  {Quinn} P.~J.,  {Salmon} J.~K.,   {Zurek} W.~H.,  1992,
  \mn@doi [\apj] {10.1086/171937}, \href
  {http://adsabs.harvard.edu/abs/1992ApJ...399..405W} {399, 405}

\bibitem[\protect\citeauthoryear{{Weil} \& {Hernquist}}{{Weil} \&
  {Hernquist}}{1996}]{Weil96}
{Weil} M.~L.,  {Hernquist} L.,  1996, \mn@doi [\apj] {10.1086/176955}, \href
  {http://adsabs.harvard.edu/abs/1996ApJ...460..101W} {460, 101}

\bibitem[\protect\citeauthoryear{{Weinberg}, {Mortonson}, {Eisenstein},
  {Hirata}, {Riess}  \& {Rozo}}{{Weinberg} et~al.}{2013}]{Weinberg13}
{Weinberg} D.~H.,  {Mortonson} M.~J.,  {Eisenstein} D.~J.,  {Hirata} C.,
  {Riess} A.~G.,   {Rozo} E.,  2013, \mn@doi [\physrep]
  {10.1016/j.physrep.2013.05.001}, \href
  {http://adsabs.harvard.edu/abs/2013PhR...530...87W} {530, 87}

\bibitem[\protect\citeauthoryear{{Williams}, {Agnello}  \& {Treu}}{{Williams}
  et~al.}{2017}]{Williams17}
{Williams} P.,  {Agnello} A.,   {Treu} T.,  2017, \mn@doi [\mnras]
  {10.1093/mnras/stw3239}, \href
  {http://adsabs.harvard.edu/abs/2017MNRAS.466.3088W} {466, 3088}

\bibitem[\protect\citeauthoryear{{Williams} et~al.,}{{Williams}
  et~al.}{2018}]{Williams18}
{Williams} P.~R.,  et~al., 2018, \mn@doi [\mnras] {10.1093/mnrasl/sly043},
  \href {http://adsabs.harvard.edu/abs/2018MNRAS.tmpL..44W} {477, L70}

\bibitem[\protect\citeauthoryear{{Witt}, {Mao}  \& {Schechter}}{{Witt}
  et~al.}{1995}]{Witt95}
{Witt} H.~J.,  {Mao} S.,   {Schechter} P.~L.,  1995, \mn@doi [\apj]
  {10.1086/175499}, \href {http://adsabs.harvard.edu/abs/1995ApJ...443...18W}
  {443, 18}

\bibitem[\protect\citeauthoryear{{Wong} et~al.,}{{Wong} et~al.}{2017}]{Wong17}
{Wong} K.~C.,  et~al., 2017, \mn@doi [\mnras] {10.1093/mnras/stw3077}, \href
  {http://adsabs.harvard.edu/abs/2017MNRAS.465.4895W} {465, 4895}

\bibitem[\protect\citeauthoryear{{Xu}, {Sluse}, {Gao}, {Wang}, {Frenk}, {Mao},
  {Schneider}  \& {Springel}}{{Xu} et~al.}{2015}]{Xu15}
{Xu} D.,  {Sluse} D.,  {Gao} L.,  {Wang} J.,  {Frenk} C.,  {Mao} S.,
  {Schneider} P.,   {Springel} V.,  2015, \mn@doi [\mnras]
  {10.1093/mnras/stu2673}, \href
  {http://adsabs.harvard.edu/abs/2015MNRAS.447.3189X} {447, 3189}

\bibitem[\protect\citeauthoryear{{Yonehara}, {Hirashita}  \&
  {Richter}}{{Yonehara} et~al.}{2008}]{Yonehara08}
{Yonehara} A.,  {Hirashita} H.,   {Richter} P.,  2008, \mn@doi [\aap]
  {10.1051/0004-6361:20067014}, \href
  {http://adsabs.harvard.edu/abs/2008A%26A...478...95Y} {478, 95}

\bibitem[\protect\citeauthoryear{{Yoo}, {Kochanek}, {Falco}  \& {McLeod}}{{Yoo}
  et~al.}{2006}]{Yoo06}
{Yoo} J.,  {Kochanek} C.~S.,  {Falco} E.~E.,   {McLeod} B.~A.,  2006, \mn@doi
  [\apj] {10.1086/500968}, \href
  {http://adsabs.harvard.edu/abs/2006ApJ...642...22Y} {642, 22}

\bibitem[\protect\citeauthoryear{{van Dokkum} \& {Conroy}}{{van Dokkum} \&
  {Conroy}}{2010}]{vanDokkum10}
{van Dokkum} P.~G.,  {Conroy} C.,  2010, \mn@doi [\nat] {10.1038/nature09578},
  \href {http://adsabs.harvard.edu/abs/2010Natur.468..940V} {468, 940}

\makeatother
\end{thebibliography}
%%%%%%%%%%%%%%%%%%%%%%%%%%%%%%%%%%%%%%%%%%%%%%%%%%

%%%%%%%%%%%%%%%%% APPENDICES %%%%%%%%%%%%%%%%%%%%%

\appendix

\section{Lens light parameters} \label{app:lens_light_params}
We report the parameters of the best fit S\'ersic functions for the deflectors in Table \ref{tab:lens_light_params}.

\begin{table*}
	\centering
	\caption{Lens light parameters. The reported uncertainties are systematic and statistical uncertainties added in quadrature. The magnitudes are given in AB system.}
	\label{tab:lens_light_params}
	\begin{adjustbox}{width=\textwidth}
	\begin{tabular}{%{lccccccc} % four columns, alignment for each
		  @{\extracolsep{\fill}}
		  l
		  S[table-format=2.1(2)]
		  S[table-format=1.2(1)]
		  S[table-format=2.1(1)]
		  S[table-format=2.1(1)]
		  S[table-format=2.1(1)]
		  S[table-format=1.2(1)]
		  S[table-format=-2.0(1)]
		  @{}
	}
		\hline
		System name & \multicolumn{1}{c}{$n_{\rm Sersic}$} & \multicolumn{1}{c}{$\theta_{\rm eff}$} & \multicolumn{1}{c}{$I_{\rm e}$ (F160W)} & \multicolumn{1}{c}{$I_{\rm e}$ (F814W)} & \multicolumn{1}{c}{$I_{\rm e}$ (F475X)} & \multicolumn{1}{c}{$q_{\rm L}$} & \multicolumn{1}{c}{$\phi_{\rm L}$ (E of N)} \\
								&  & \multicolumn{1}{c}{(arcsec)} & \multicolumn{1}{c}{(mag arcsec$^{-2}$)} & \multicolumn{1}{c}{(mag arcsec$^{-2}$)} & \multicolumn{1}{c}{(mag arcsec$^{-2}$)} & & \multicolumn{1}{c}{(degree)} \\ 
								
		\hline	
PS J0147+4630   	& 4	& 4.97 \pm 0.03	& 29.7 \pm 1.0	& 32.2 \pm 1.0	& 34.4 \pm 1.5	& 0.81 \pm 0.01	& 18 \pm 1 \\
                	& 1	& 0.14 \pm 0.03	& 23.4 \pm 0.9	& 26.7 \pm 0.9	& 30.0 \pm 1.3	& 0.87 \pm 0.01	& 62 \pm 1 \\
\hline
SDSS J0248+1913 	& 2.4 \pm 1.4	& 0.16 \pm 0.03	& 24.5 \pm 0.4	& 27.8 \pm 0.4	& 31.6 \pm 0.6	& 0.40 \pm 0.02	& 13 \pm 1 \\
\hline
ATLAS J0259-1635 	& 11.8 \pm 1.4	& 1.00 \pm 0.03	& 28.7 \pm 1.0	& 32.2 \pm 1.0	& {--}	& 0.38 \pm 0.02	& 20 \pm 1 \\
\hline
DES J0405-3308  	& 7.6 \pm 1.4	& 0.44 \pm 0.03	& 26.8 \pm 1.0	& 30.1 \pm 1.0	& 33.1 \pm 1.5	& 0.55 \pm 0.02	& 37 \pm 1 \\
\hline
DES J0408-5354  	& 5.5 \pm 1.4	& 2.15 \pm 0.03	& 28.5 \pm 1.0	& 31.3 \pm 1.0	& 34.0 \pm 1.5	& 0.82 \pm 0.01	& 28 \pm 2 \\
\hline
DES J0420-4037  	& 4.0 \pm 1.4	& 0.44 \pm 0.03	& 25.1 \pm 1.0	& 27.5 \pm 1.0	& 29.5 \pm 1.5	& 0.61 \pm 0.01	& 27 \pm 1 \\
\hline
PS J0630-1201   	& 6.8 \pm 1.4	& 1.64 \pm 0.03	& 29.9 \pm 1.0	& 33.5 \pm 1.0	& 37.7 \pm 1.5	& 0.79 \pm 0.01	& 12 \pm 1 \\
\hline
SDSS J1251+2935 	& 4	& 0.53 \pm 0.03	& 25.5 \pm 1.0	& 28.1 \pm 1.0	& 30.4 \pm 1.5	& 0.67 \pm 0.01	& 23 \pm 1 \\
                	& 1	& 5.00 \pm 0.03	& 30.4 \pm 0.9	& 32.6 \pm 0.9	& 34.0 \pm 1.3	& 0.67 \pm 0.03	& 16 \pm 3 \\
\hline
SDSS J1330+1810 	& 4	& 0.75 \pm 0.03	& 24.8 \pm 0.4	& 27.7 \pm 0.4	& 31.9 \pm 0.6	& 0.26 \pm 0.01	& 24 \pm 1 \\
                	& 1	& 0.37 \pm 0.03	& 24.5 \pm 0.4	& 26.7 \pm 0.4	& 27.9 \pm 0.6	& 0.37 \pm 0.01	& 25 \pm 1 \\
\hline
SDSS J1433+6007 	& 4	& 0.56 \pm 0.03	& 25.4 \pm 1.0	& 28.2 \pm 1.0	& 30.4 \pm 1.5	& 0.56 \pm 0.02	& -88 \pm 2 \\
                	& 1	& 3.35 \pm 0.03	& 28.9 \pm 0.9	& 31.1 \pm 0.9	& 32.8 \pm 1.3	& 0.54 \pm 0.02	& -88 \pm 1 \\
\hline
PS J1606-2333   	& 4	& 0.11 \pm 0.03	& 25.2 \pm 1.0	& 27.7 \pm 1.0	& {--}	& 0.56 \pm 0.06	& -26 \pm 4 \\
                	& 1	& 1.66 \pm 0.04	& 28.5 \pm 0.9	& 31.0 \pm 0.9	& 32.1 \pm 1.3	& 0.77 \pm 0.02	& -11 \pm 2 \\
\hline
DES J2038-4008  	& 4	& 3.36 \pm 0.03	& 26.7 \pm 1.0	& 29.3 \pm 1.0	& 31.1 \pm 1.5	& 0.64 \pm 0.01	& 38 \pm 1 \\
                	& 1	& 4.99 \pm 0.03	& 29.2 \pm 0.9	& 31.3 \pm 0.9	& 32.6 \pm 1.3	& 0.47 \pm 0.01	& -62 \pm 1 \\
\hline
WISE J2344-3056 	& 4	& 0.61 \pm 0.04	& 26.9 \pm 1.0	& 30.9 \pm 1.0	& {--}	& 0.75 \pm 0.03	& -68 \pm 4 \\
                	& 1	& 4.67 \pm 0.05	& 30.1 \pm 0.9	& 31.7 \pm 0.9	& 33.1 \pm 1.3	& 0.80 \pm 0.07	& 65 \pm 10 \\
\hline
	\end{tabular}
	\end{adjustbox}
\end{table*}

\section{Convergence, shear and stellar convergence} \label{app:convergence}
The convergence $\kappa$, shear $\gamma$, and the stellar convergence $\kappa_{\star}$ at the image positions for each lens are given in Table \ref{tab:lens_kappas}. The convergence at the image position is given by the lens mass distribution. We assume a constant mass-to-light ratio to convert the surface brightness distribution into a stellar surface mass-density distribution. We choose the maximum normalization factor for the stellar convergence that meets these two criteria: (i) the stellar convergence is smaller than the convergence, and (ii) the integrated stellar convergence is smaller than two-thirds of the integrated convergence within half of the effective radius \citep{Auger10b}.

\begin{table}
	\centering
	\caption{Convergence, shear, and stellar convergence at the image positions. The reported uncertainties are systematic and statistical uncertainties added in quadrature. The stellar convergence, $\kappa_{\star}$, is estimated from the F160W band for the lenses with double S\'ersic fit for the lens light.}
	\label{tab:lens_kappas}
	\begin{adjustbox}{width=\columnwidth}
	\begin{tabular}{lcccc} % 13 columns, alignment for each
		\hline
		System name &  Image & $\kappa$ & $\gamma$ & $\kappa_{\star}/\kappa$ \\
		\hline
\multirow{4}{*}{ PS J0147+4630 } 	 & A	 & $0.41 \pm 0.03$	 & $0.55 \pm 0.04$	 & $0.37 \pm 0.17$ \\
 					 & B	 & $0.39 \pm 0.03$	 & $0.52 \pm 0.04$	 & $0.37 \pm 0.17$ \\
 					 & C	 & $0.43 \pm 0.03$	 & $0.45 \pm 0.03$	 & $0.27 \pm 0.12$ \\
 					 & D	 & $0.84 \pm 0.06$	 & $0.99 \pm 0.08$	 & $0.41 \pm 0.18$ \\
\hline
\multirow{4}{*}{ SDSS J0248+1913 } 	 & A	 & $0.63 \pm 0.05$	 & $0.87 \pm 0.06$	 & $0.002 \pm 0.001$ \\
 					 & B	 & $0.26 \pm 0.03$	 & $0.50 \pm 0.02$	 & $0.003 \pm 0.002$ \\
 					 & C	 & $0.20 \pm 0.03$	 & $0.31 \pm 0.02$	 & $0.006 \pm 0.003$ \\
 					 & D	 & $0.87 \pm 0.07$	 & $1.04 \pm 0.09$	 & $0.011 \pm 0.003$ \\
\hline
\multirow{4}{*}{ ATLAS J0259-1635 } 	 & A	 & $0.41 \pm 0.03$	 & $0.42 \pm 0.03$	 & $0.06 \pm 0.03$ \\
 					 & B	 & $0.66 \pm 0.05$	 & $0.67 \pm 0.05$	 & $0.25 \pm 0.15$ \\
 					 & C	 & $0.36 \pm 0.03$	 & $0.36 \pm 0.02$	 & $0.05 \pm 0.03$ \\
 					 & D	 & $0.64 \pm 0.04$	 & $0.65 \pm 0.04$	 & $0.21 \pm 0.11$ \\
\hline
\multirow{4}{*}{ DES J0405-3308 } 	 & A	 & $0.48 \pm 0.04$	 & $0.46 \pm 0.03$	 & $0.06 \pm 0.02$ \\
 					 & B	 & $0.53 \pm 0.04$	 & $0.53 \pm 0.04$	 & $0.20 \pm 0.08$ \\
 					 & C	 & $0.52 \pm 0.04$	 & $0.51 \pm 0.04$	 & $0.19 \pm 0.08$ \\
 					 & D	 & $0.49 \pm 0.03$	 & $0.47 \pm 0.03$	 & $0.06 \pm 0.02$ \\
\hline
\multirow{4}{*}{ DES J0408-5354 } 	 & A	 & $0.34 \pm 0.02$	 & $0.31 \pm 0.02$	 & $0.17 \pm 0.06$ \\
 					 & B	 & $0.45 \pm 0.03$	 & $0.35 \pm 0.02$	 & $0.24 \pm 0.09$ \\
 					 & C	 & $0.57 \pm 0.04$	 & $0.57 \pm 0.04$	 & $0.17 \pm 0.07$ \\
 					 & D	 & $0.75 \pm 0.05$	 & $0.68 \pm 0.05$	 & $0.32 \pm 0.13$ \\
\hline
\multirow{4}{*}{ DES J0420-4037 } 	 & A	 & $0.56 \pm 0.04$	 & $0.56 \pm 0.04$	 & $0.12 \pm 0.05$ \\
 					 & B	 & $0.50 \pm 0.04$	 & $0.44 \pm 0.03$	 & $0.07 \pm 0.03$ \\
 					 & C	 & $0.45 \pm 0.03$	 & $0.41 \pm 0.03$	 & $0.05 \pm 0.02$ \\
 					 & D	 & $0.56 \pm 0.04$	 & $0.51 \pm 0.04$	 & $0.16 \pm 0.06$ \\
\hline
\multirow{4}{*}{ PS J0630-1201 } 	 & A	 & $0.52 \pm 0.04$	 & $0.45 \pm 0.03$	 & $0.07 \pm 0.05$ \\
 					 & B	 & $0.49 \pm 0.03$	 & $0.52 \pm 0.04$	 & $0.07 \pm 0.05$ \\
 					 & C	 & $0.45 \pm 0.03$	 & $0.55 \pm 0.04$	 & $0.07 \pm 0.04$ \\
 					 & D	 & $1.39 \pm 0.10$	 & $1.23 \pm 0.08$	 & $0.03 \pm 0.02$ \\
\hline
\multirow{4}{*}{ SDSS J1251+2935 } 	 & A	 & $0.57 \pm 0.04$	 & $0.57 \pm 0.04$	 & $0.21 \pm 0.08$ \\
 					 & B	 & $0.50 \pm 0.04$	 & $0.41 \pm 0.03$	 & $0.17 \pm 0.07$ \\
 					 & C	 & $0.63 \pm 0.04$	 & $0.57 \pm 0.04$	 & $0.21 \pm 0.09$ \\
 					 & D	 & $0.33 \pm 0.02$	 & $0.25 \pm 0.02$	 & $0.12 \pm 0.05$ \\
\hline
\multirow{4}{*}{ SDSS J1330+1810 } 	 & A	 & $0.48 \pm 0.08$	 & $0.43 \pm 0.04$	 & $0.04 \pm 0.02$ \\
 					 & B	 & $0.59 \pm 0.09$	 & $0.52 \pm 0.05$	 & $0.16 \pm 0.08$ \\
 					 & C	 & $0.36 \pm 0.06	$ & $0.43 \pm 0.04$	 & $0.02 \pm 0.01$ \\
 					 & D	 & $0.74 \pm 0.12	$ & $0.71 \pm 0.06	$ & $0.23 \pm 0.12$ \\
\hline
\multirow{4}{*}{ SDSS J1433+6007 } 	 & A	 & $0.35 \pm 0.02$	 & $0.20 \pm 0.01$	 & $0.08 \pm 0.03$ \\
 					 & B	 & $0.38 \pm 0.03$	 & $0.35 \pm 0.02$	 & $0.10 \pm 0.04$ \\
 					 & C	 & $0.78 \pm 0.06$	 & $0.72 \pm 0.05$	 & $0.16 \pm 0.06$ \\
 					 & D	 & $1.20 \pm 0.09$	 & $1.16 \pm 0.08$	 & $0.22 \pm 0.09$ \\
\hline
\multirow{4}{*}{ PS J1606-2333 } 	 & A	 & $0.46 \pm 0.03$	 & $0.25 \pm 0.02$	 & $0.50 \pm 0.20$ \\
 					 & B	 & $0.49 \pm 0.03$	 & $0.22 \pm 0.02$	 & $0.53 \pm 0.22$ \\
 					 & C	 & $0.77 \pm 0.06$	 & $0.75 \pm 0.05$	 & $0.61 \pm 0.25$ \\
 					 & D	 & $0.57 \pm 0.04$	 & $0.66 \pm 0.05$	 & $0.77 \pm 0.31$ \\
\hline
\multirow{4}{*}{ DES J2038-4008 } 	 & A	 & $0.21 \pm 0.02$	 & $0.43 \pm 0.03$	 & $0.73 \pm 0.27$ \\
 					 & B	 & $0.22 \pm 0.02$	 & $0.49 \pm 0.03$	 & $0.73 \pm 0.27$ \\
 					 & C	 & $0.45 \pm 0.03$	 & $0.89 \pm 0.06$	 & $0.72 \pm 0.28$ \\
 					 & D	 & $0.59 \pm 0.04$	 & $1.11 \pm 0.08$	 & $0.74 \pm 0.26$ \\
\hline
\multirow{4}{*}{ WISE J2344-3056 } 	 & A	 & $0.79 \pm 0.06$	 & $0.69 \pm 0.06$	 & $0.54 \pm 0.22$ \\
 					 & B	 & $0.37 \pm 0.03$	 & $0.38 \pm 0.03$	 & $0.62 \pm 0.25$ \\
 					 & C	 & $0.37 \pm 0.03$	 & $0.38 \pm 0.03$	 & $0.66 \pm 0.26$ \\
 					 & D	 & $0.82 \pm 0.07$	 & $0.72 \pm 0.06$	 & $0.60 \pm 0.24$ \\
\hline
	\end{tabular}
	\end{adjustbox}
\end{table}

\section{Time delays} \label{app:time_delay}
The time delay between two images ${\rm I}$ and ${\rm J}$ is given by
\begin{equation}
\Delta t_{\rm IJ} = \frac{D_{\Delta t}}{c} \left[ \frac{1}{2} (\bm{\theta}_{\rm I} - \bm{\beta})^2 - \frac{1}{2} (\bm{\theta}_{\rm J} - \bm{\beta})^2 - \psi(\bm{\theta_{\rm I}}) + \psi (\bm{\theta}_{\rm J}) \right],
\end{equation}
where $\bm{\theta}$ is the image position, $\bm{\beta}$ is the source position, $\psi$ is the lensing potential, $c$ is the speed of light, and $D_{\Delta t}$ is the time-delay distance given by
\begin{equation}
D_{\Delta t} = \left( 1 + z_{\rm d}\right) \frac{D_{\rm d} D_{\rm s}}{D_{\rm ds}}.	
\end{equation}
Here $z_{\rm d}$ is the deflector redshift, $D_{\rm d}$, $D_{\rm s}$, and $D_{\rm ds}$ are the angular diameter distances between the observer and the deflector, between the observer and the source, and between the deflector and the source, respectively. The predicted time delays between the images for the quads are given in Table \ref{tab:time_delays}.

\begin{table}
	\centering
	\caption{Predicted time-delays between the quasar images. The reported uncertainties are systematic and statistical uncertainties added in quadrature. We adopt fiducial redshifts $z_{\rm d}=0.5$ and $z_{\rm s}=2$ where the redshifts are not measured yet.}
	\label{tab:time_delays}
	\begin{adjustbox}{width=\columnwidth}
	\begin{tabular}{%{lccccc} % 13 columns, alignment for each
		 @{\extracolsep{\fill}}
		  l
		  S[table-format=1.3(0)]
		  S[table-format=1.3(0)]
		  r@{\,\( \pm \)\,}
		  l
		  r@{\,\( \pm \)\,}
		  l
		  r@{\,\( \pm \)\,}
		  l
		  @{}	
	}
		\hline
		System name &  $z_{\rm d}$ & $z_{\rm s}$ & \multicolumn{2}{c}{$\Delta t_{\rm AB}$} & \multicolumn{2}{c}{$\Delta t_{\rm AC}$} & \multicolumn{2}{c}{$\Delta t_{\rm AD}$} \\	
		& & & \multicolumn{2}{c}{(days)} & \multicolumn{2}{c}{(days)} & \multicolumn{2}{c}{(days)} \\
		\hline
PS J0147+4630 	 & 0.572	 & 2.341	 & -2.1 & 0.3	 & -7 & 1 & -193 & 18 \\
SDSS J0248+1913 	 & 0.5	 & 2.0	 & 2.7 & 0.2	 & 20 & 2 & -5.9 & 0.4 \\
ATLAS J0259-1635 	 & 0.5	 & 2.16	 & -3.6 & 0.3	 & 7 & 1 & -2.7 & 0.2 \\
DES J0405-3308 	 & 0.5	 & 1.713	 & -1.7 & 0.2	 & -0.9 & 0.2 & -0.3 & 0.2 \\
DES J0408-5354 	 & 0.597	 & 2.375	 & -100 & 9	 & -105 & 9 & -140 & 13 \\
DES J0420-4037 	 & 0.5	 & 2.0	 & 1.8 & 0.2	 & 7 & 1 & 1.4 & 0.1 \\
PS J0630-1201 	 & 0.5	 & 3.34	 & -0.12 & 0.02	 & -0.09 & 0.02 & -108 & 10  \\
SDSS J1251+2935 	 & 0.41	 & 0.802	 & 0.6 & 0.1	 & -0.43 & 0.04 & 36 & 3 \\
SDSS J1330+1810 	 & 0.373	 & 1.393	 & -0.20 & 0.02	 & 6 & 1 & -11 & 1 \\
SDSS J1433+6007 	 & 0.407	 & 2.737	 & -24 & 2	 & -36 & 3 & -100 & 9 \\
PS J1606-2333 	 & 0.5	 & 2.0	 & -3.8 & 0.4	 & -11 & 1 & -7 & 1 \\
DES J2038-4008 	 & 0.23	 & 0.777	 & -6 & 1	 & -11 & 1 & -27 & 2 \\
WISE J2344-3056 	 & 0.5	 & 2.0	 & 3.3 & 0.4	 & 3.4 & 0.4 & -0.6 & 0.2 \\
		\hline
	\end{tabular}
	\end{adjustbox}
\end{table}
  
\section{Lens models} \label{app:lens_models}
\begin{figure*}
	\includegraphics[width=1.05\columnwidth]{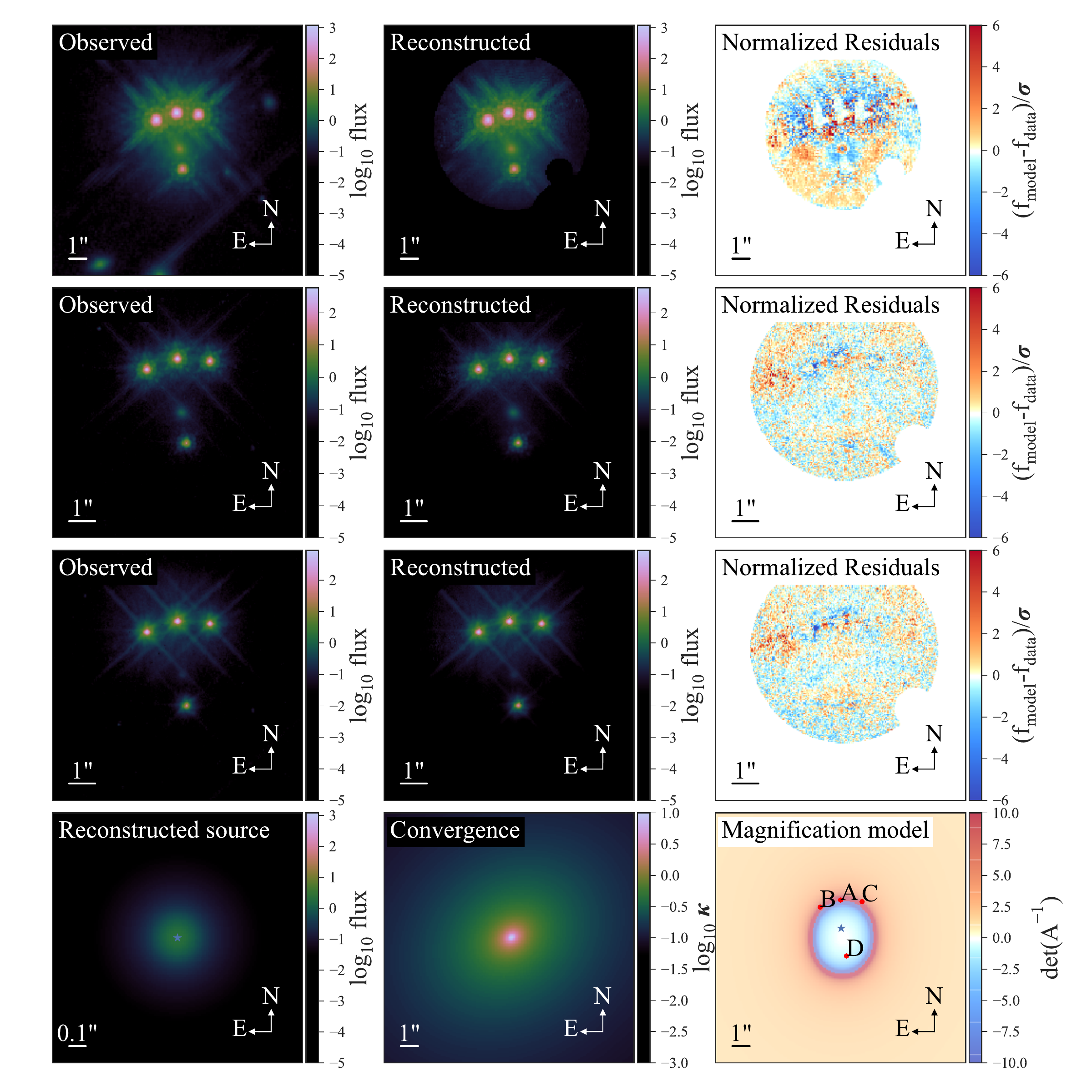}%{figures/PSJ0147+4630_model.pdf} 
	\includegraphics[width=1.05\columnwidth]{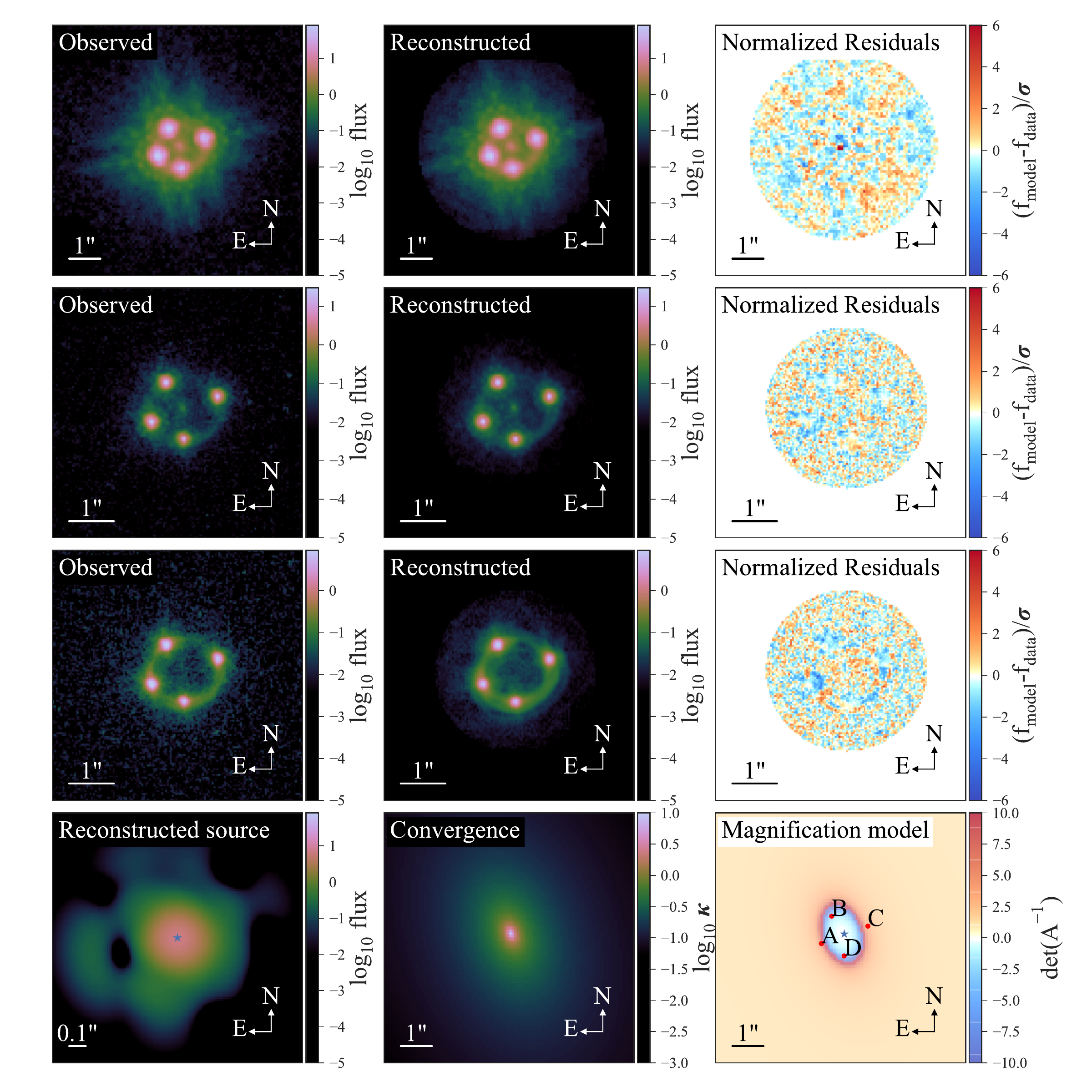} \\ %{figures/WISEJ0259-1635_model.pdf} \\
	\includegraphics[width=1.05\columnwidth]{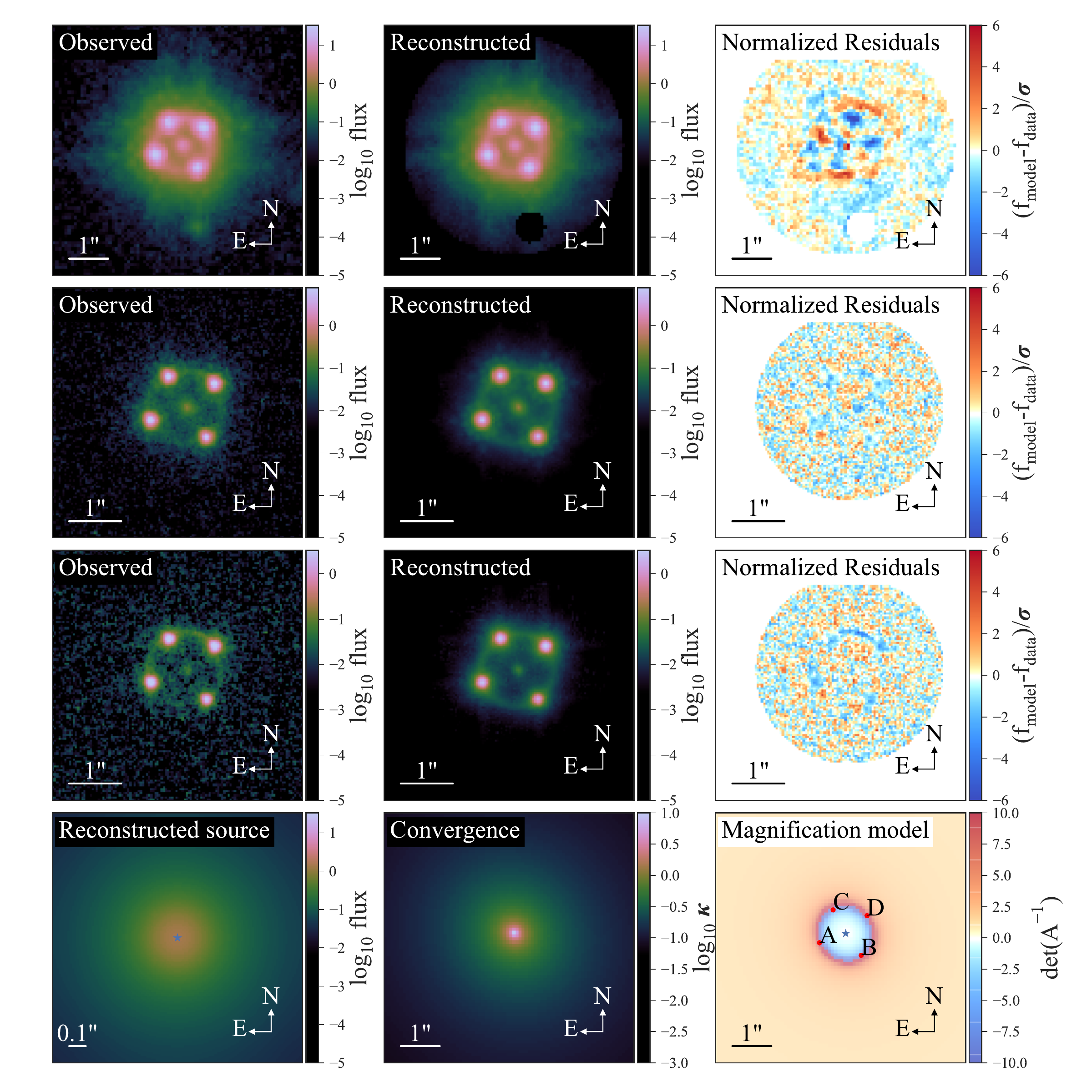}%{figures/DESJ0405-3308_model.pdf} 
	\includegraphics[width=1.05\columnwidth]{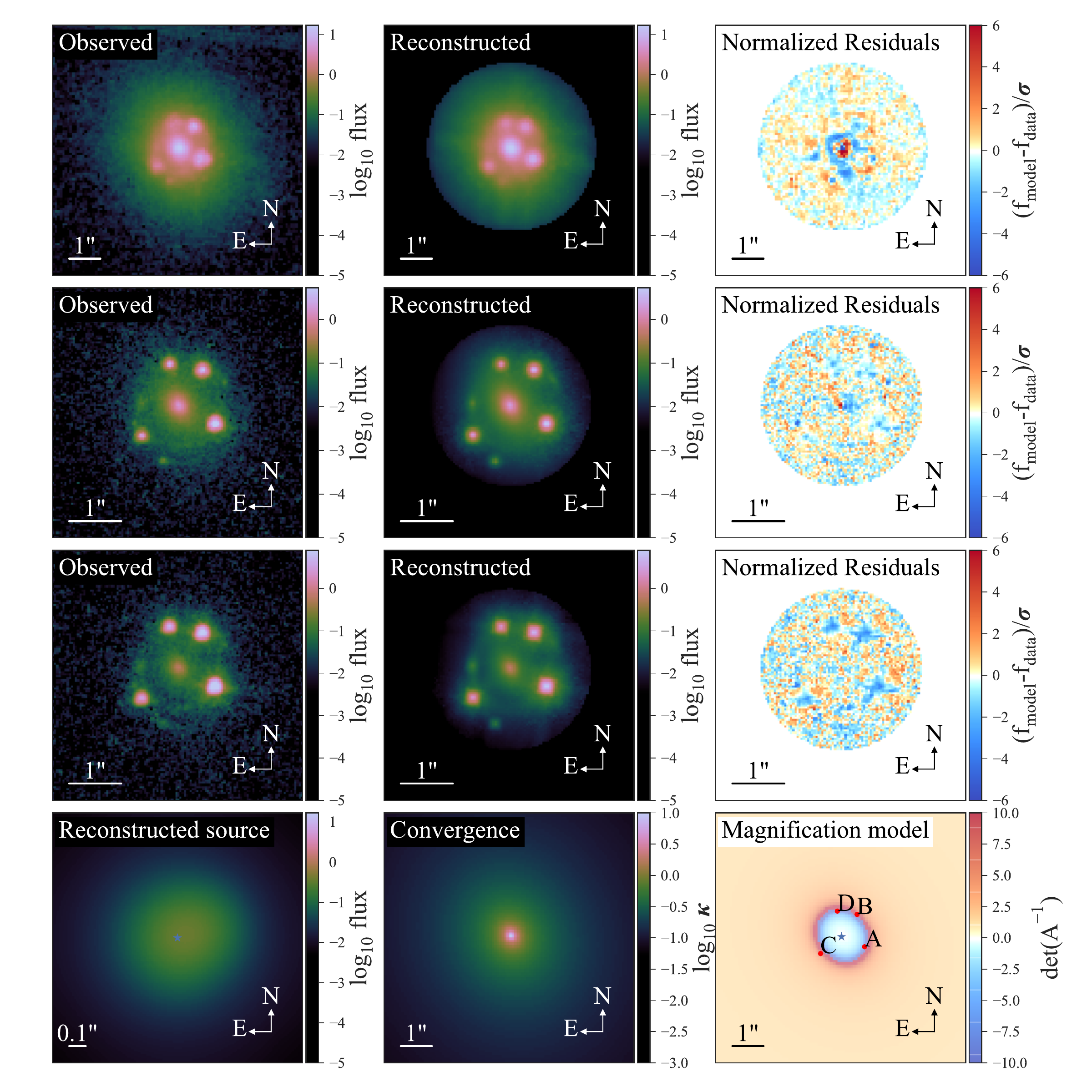}%{figures/DESJ0420-4037_model.pdf}
	\caption{
	Best fit models for PS J0147+4630 (top left), ATLAS J0259-1635 (top right), DES J0405-3308 (bottom left), and DES J0420-4037 (bottom right). The first three rows for each lens system show the observed image, reconstructed lens image, and the normalized residuals in three \textit{HST} bands: F160W, F814W, and F475X, respectively. The fourth row shows the reconstructed source in the F160W band, the convergence, and the magnification model.
	\label{fig:lens_model_breakdown_2}
	}
	\end{figure*}

\begin{figure*}
	\includegraphics[width=1.05\columnwidth]{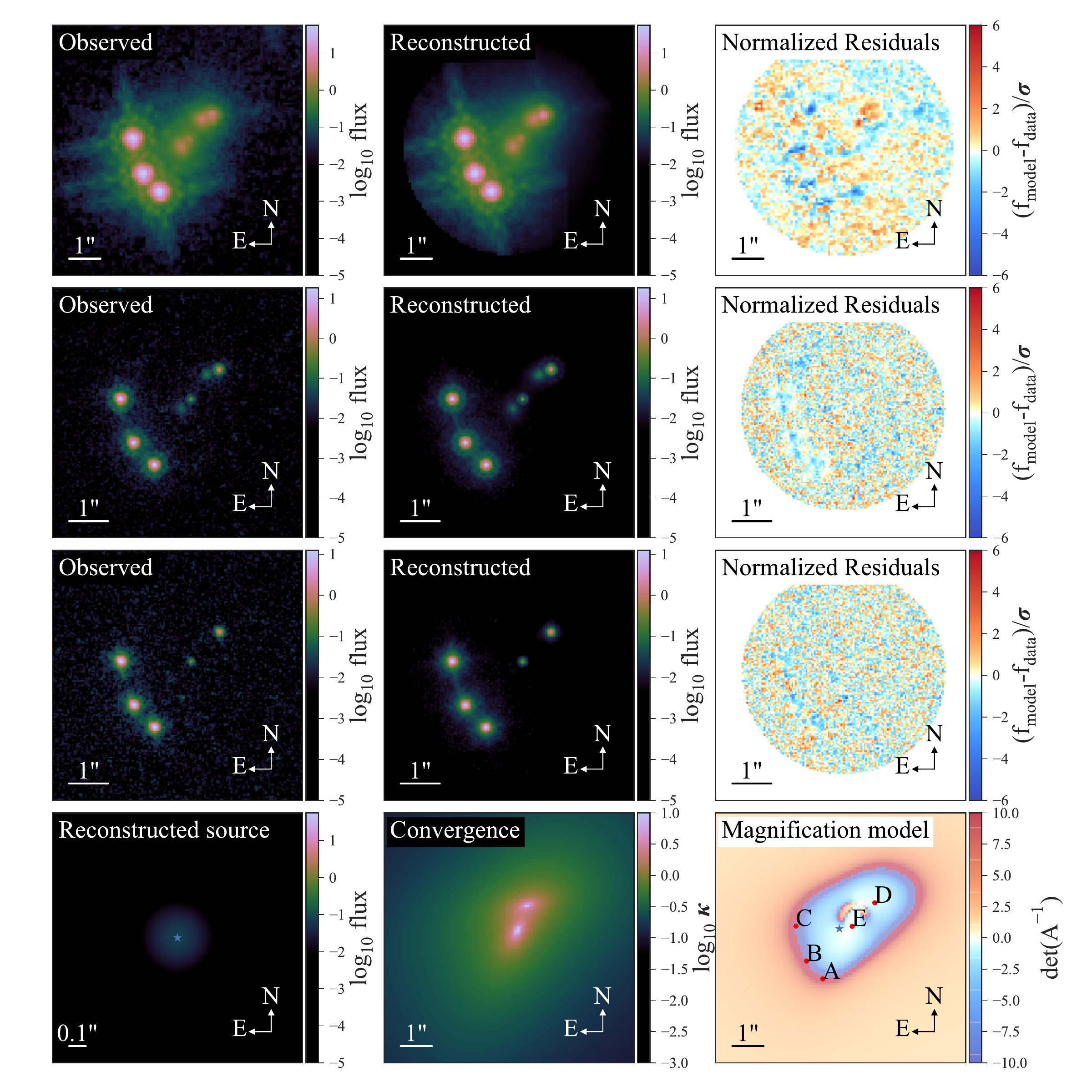}%{figures/PSJ0630-1201_model.pdf} 
	\includegraphics[width=1.05\columnwidth]{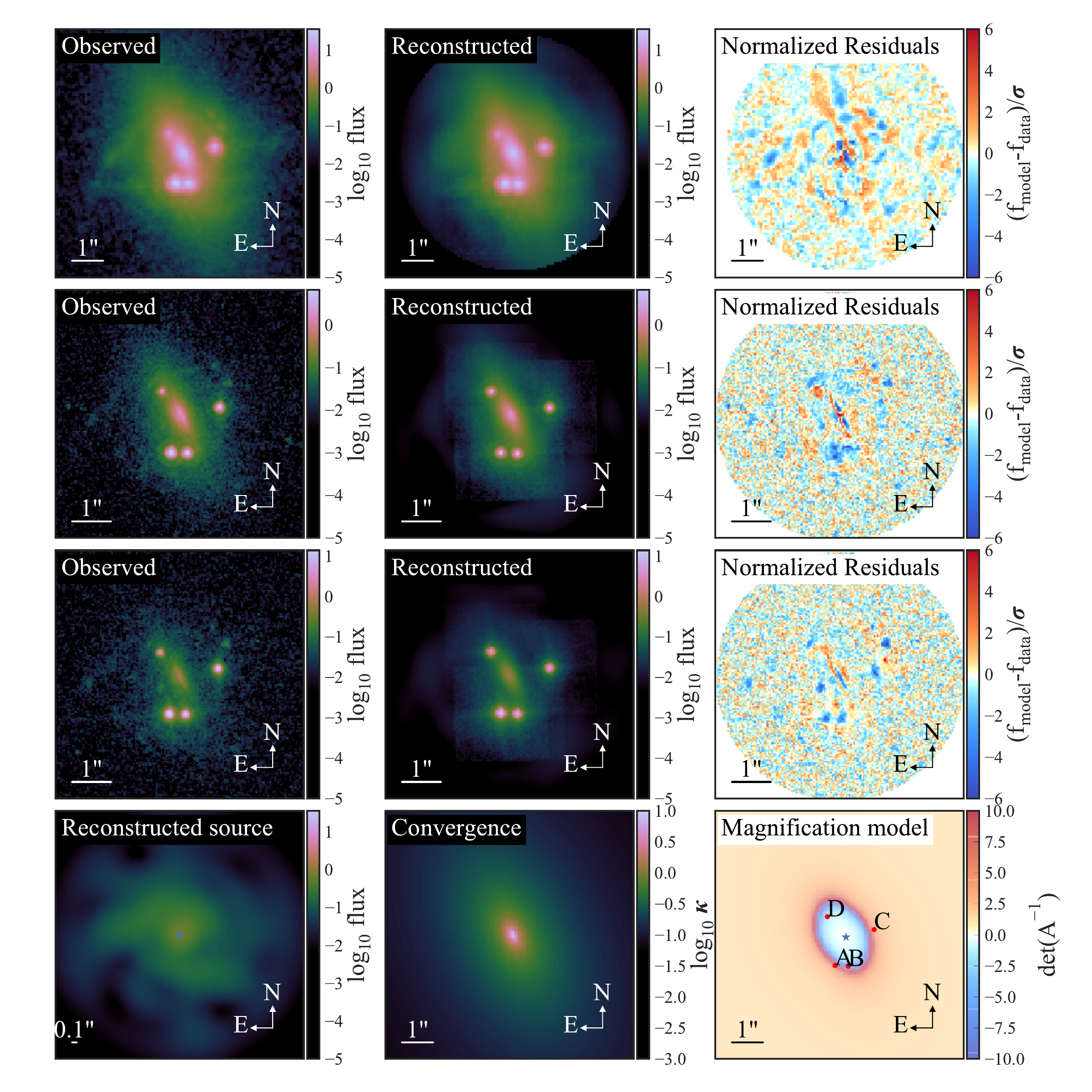} \\ %{figures/PSJ1606-2333_model.pdf} \\
	\includegraphics[width=1.05\columnwidth]{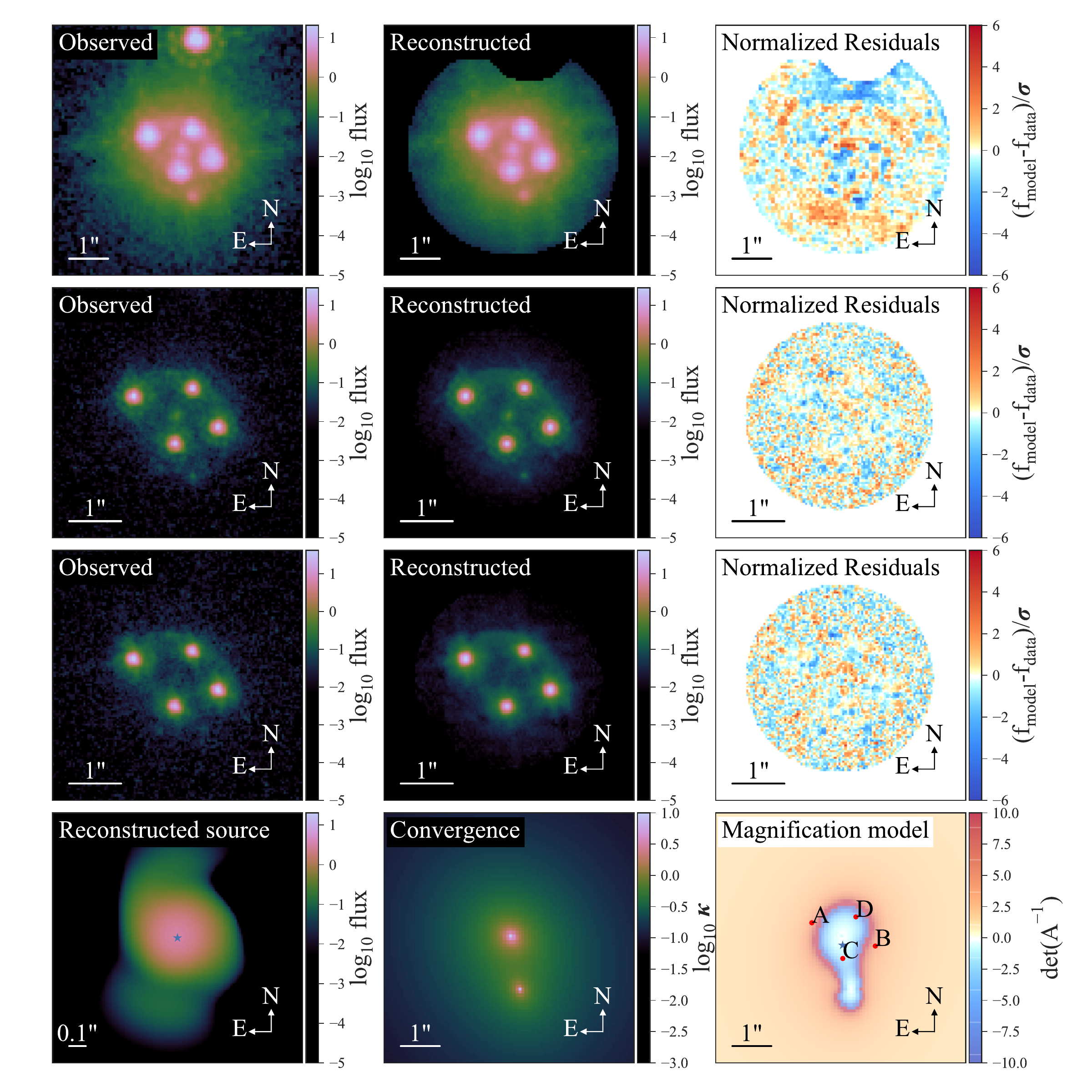}%{figures/DESJ2038-4008_model.pdf} 
	\includegraphics[width=1.05\columnwidth]{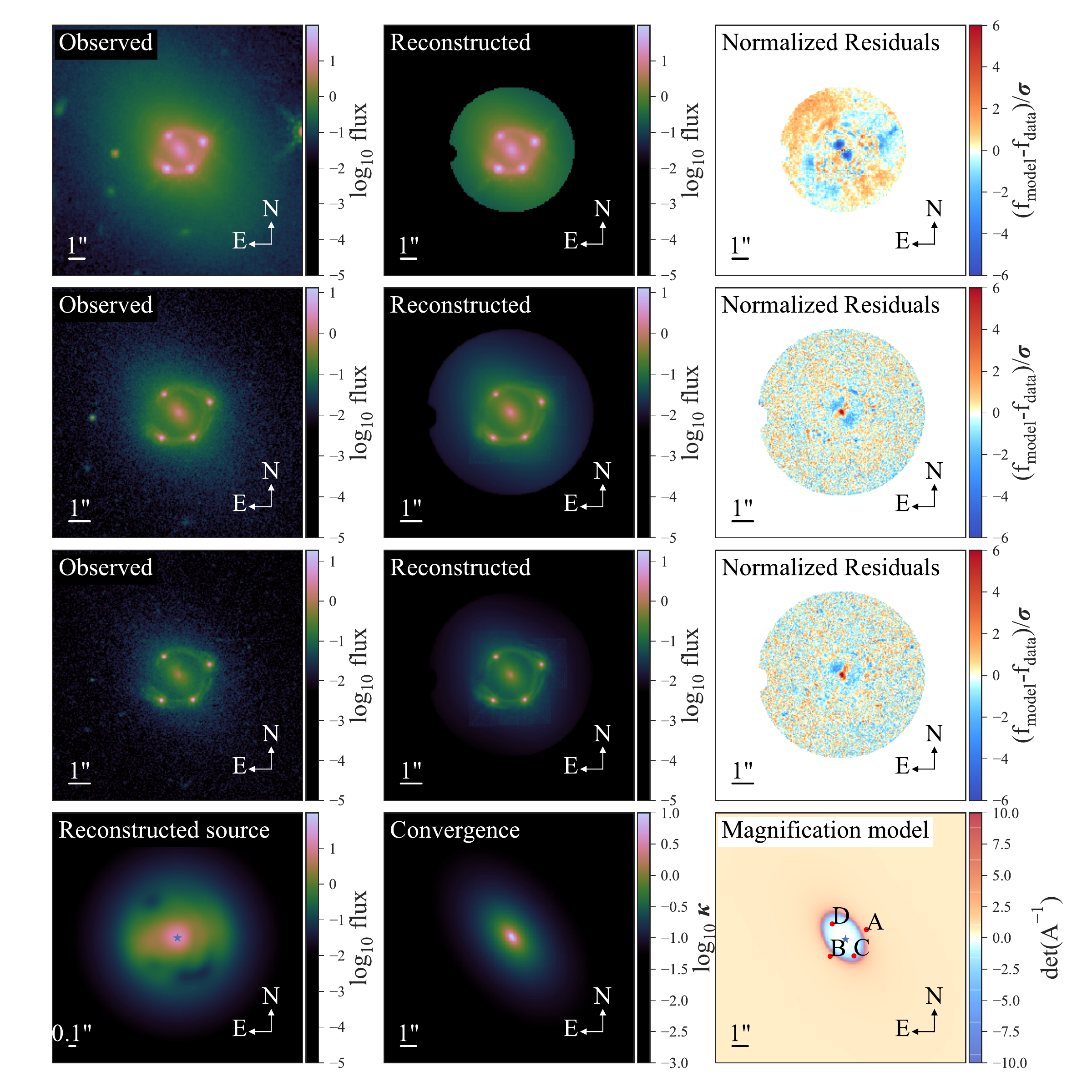}%{figures/ATLASJ2344-3056_model.pdf}
	\caption{
	Best fit models for PS J0630-1201 (top left), SDSS J1330+1810 (top right), PS J1606-2333 (bottom left), and DES J2038-4008 (bottom right). The first three rows for each lens system show the observed image, reconstructed lens image, and the normalized residuals in three \textit{HST} bands: F160W, F814W, and F475X, respectively. The fourth row shows the reconstructed source in the F160W band, the convergence, and the magnification model.
	\label{fig:lens_model_breakdown_3}
	}
	\end{figure*}

\begin{figure}
	\includegraphics[width=1.05\columnwidth]{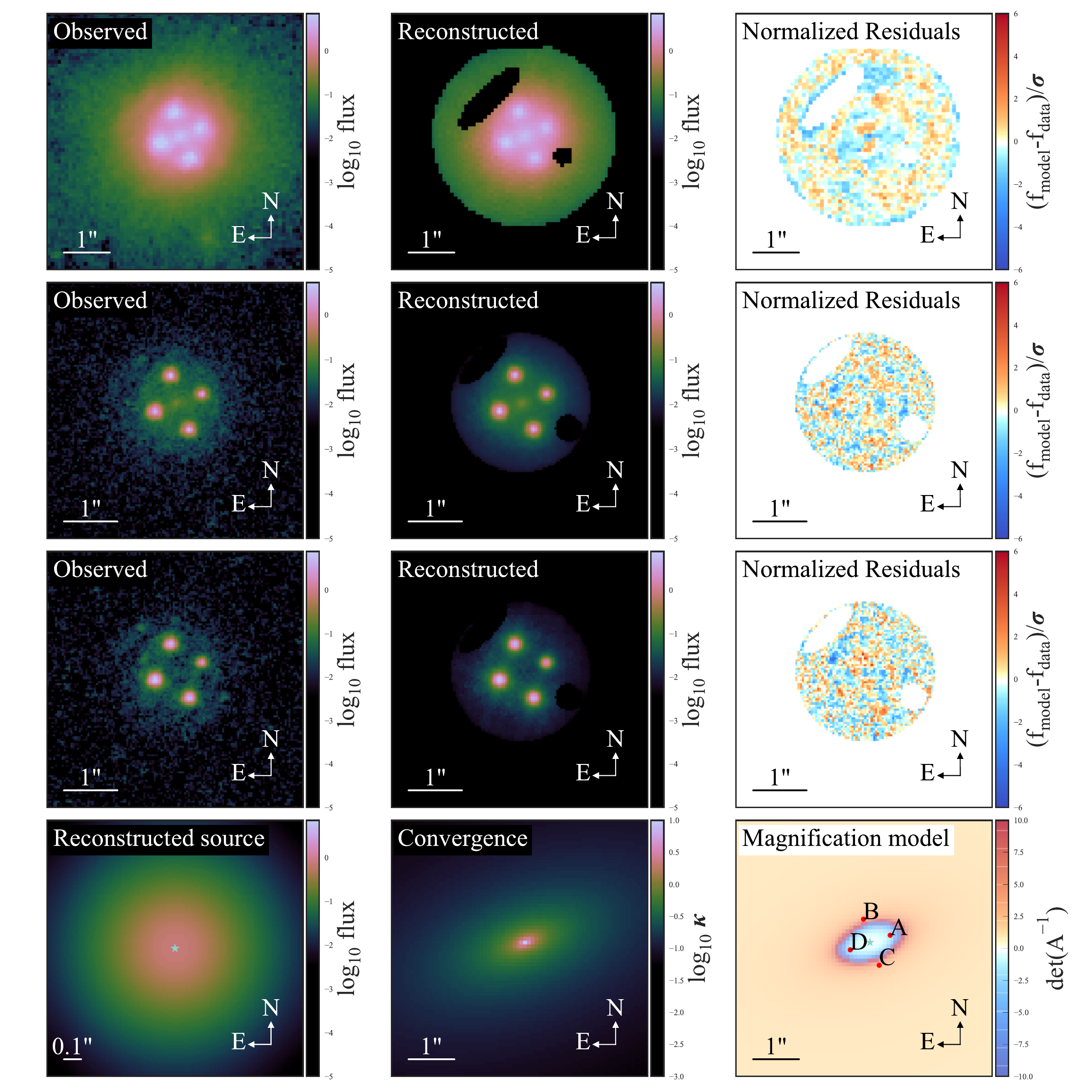}%{figures/PSJ0147+4630_model.pdf} 
	\caption{
	Best fit models for WISE J2344-3056. The first three rows show the observed image, reconstructed lens image, and the normalized residuals in three \textit{HST} bands: F160W, F814W, and F475X, respectively. The fourth row shows the reconstructed source in the F160W band, the convergence, and the magnification model.
	\label{fig:lens_model_breakdown_4}
	}
	\end{figure}

In this section, we provide rest of the lens models in Figure \ref{fig:lens_model_breakdown_2}, \ref{fig:lens_model_breakdown_3} and  \ref{fig:lens_model_breakdown_4} that were not included in Figure \ref{fig:lens_model_breakdown}.

% List of institutions
%\parbox{\columnwidth}
%{
\ \\
$^{1}$Department of Physics and Astronomy, University of California, Los Angeles, CA 90095-1547, USA (Email: ajshajib@astro.ucla.edu) \\
$^{2}$Packard Fellow \\ 
$^{3}$Institute of Astronomy, Madingley Road, Cambridge CB3 0HA, UK \\
$^{4}$European Southern Observatory, Karl-Schwarzschild-Strasse 2, 85748 Garching bei Muenchen, Germany \\
$^{5}$Departamento de Ciencias Fisicas, Universidad Andres Bello Fernandez Concha 700, Las Condes, Santiago, Chile\\
$^{6}$Millennium Institute of Astrophysics, Chile\\
$^{7}$Fermi National Accelerator Laboratory, P. O. Box 500, Batavia, IL 60510, USA \\
$^{8}$Institute of Physics, Laboratoire d'Astrophysique, Ecole Polytechnique F\'ed\'erale de Lausanne (EPFL), Observatoire de Sauverny, CH-1290 Versoix, Switzerland \\
$^{9}$Institute of Cosmology and Gravitation, University of Portsmouth, Portsmouth, PO1 3FX, UK \\
$^{10}$Physics Dept. University of California, Davis, 1 Shields Ave., Davis, CA 95161 \\
$^{11}$Kavli Institute for Cosmological Physics, The University of Chicago, Chicago, IL 60637 \\
$^{12}$Department of Liberal Arts, Tokyo University of Technology, Ota-ku, Tokyo 144-8650, Japan \\
$^{13}$Kavli Institute for Particle Astrophysics and Cosmology, P.O. Box 20450, MS29, Stanford, CA 94309, USA \\
$^{14}$Kavli IPMU (WPI), UTIAS, The University of Tokyo, Kashiwa, Chiba 277-8583, Japan \\
$^{15}$Staples High School, Westport, CT 06880, USA \\
$^{16}$Instituto de F\'{\i}sica y Astronom\'{\i}a, Universidad de Valpara\'{\i}so, Avda. Gran Breta\~na 1111, Playa Ancha, Valpara\'{\i}so 2360102, Chile \\
$^{17}$Research Center for the Early Universe, The University of Tokyo, 7-3-1 Hongo, Bunkyo-ku, Tokyo, 113-0033, Japan\\
$^{18}$Department of Physics, The University of Tokyo, 7-3-1 Hongo, Bunkyo-ku, Tokyo 113-0033, Japan\\
$^{19}$Kavli Institute for the Physics and Mathematics of the Universe (WPI), The University of Tokyo, 5-1-5 Kashiwanoha, Kashiwa, Chiba 277-8583, Japan\\
$^{20}$Subaru Fellow \\
$^{21}$Subaru telescope, National Astronomical Observatory of Japan, 650 North Aohoku Place, Hilo, HI 96720, USA \\
$^{22}$MIT Kavli Institute for Astrophysics and Space Research, Cambridge, MA 02139, USA \\
$^{23}$Dept. of Physics, Durham University, South Road, Durham DH1 3LE, England \\
$^{24}$Max-Planck-Institut f{\"u}r Astrophysik, Karl-Schwarzschild-Str.~1, 85748 Garching, Germany \\
$^{25}$Physik-Department, Technische Universit\"at M\"unchen, James-Franck-Stra\ss{}e~1, 85748 Garching, Germany \\
$^{26}$Institute of Astronomy and Astrophysics, Academia Sinica, P.O.~Box 23-141, Taipei 10617, Taiwan \\
$^{27}$Cerro Tololo Inter-American Observatory, National Optical Astronomy Observatory, Casilla 603, La Serena, Chile \\
%$^{27}$ Institute of Cosmology \& Gravitation, University of Portsmouth, Portsmouth, PO1 3FX, UK \\
$^{28}$CNRS, UMR 7095, Institut d'Astrophysique de Paris, F-75014, Paris, France \\
$^{29}$Sorbonne Universit\'es, UPMC Univ Paris 06, UMR 7095, Institut d'Astrophysique de Paris, F-75014, Paris, France \\
$^{30}$Department of Physics \& Astronomy, University College London, Gower Street, London, WC1E 6BT, UK \\
$^{31}$Laborat\'orio Interinstitucional de e-Astronomia - LIneA, Rua Gal. Jos\'e Cristino 77, Rio de Janeiro, RJ - 20921-400, Brazil \\
$^{32}$Observat\'orio Nacional, Rua Gal. Jos\'e Cristino 77, Rio de Janeiro, RJ - 20921-400, Brazil \\
$^{33}$Department of Astronomy, University of Illinois at Urbana-Champaign, 1002 W. Green Street, Urbana, IL 61801, USA \\
$^{34}$National Center for Supercomputing Applications, 1205 West Clark St., Urbana, IL 61801, USA \\
$^{35}$Institut de F\'{\i}sica d'Altes Energies (IFAE), The Barcelona Institute of Science and Technology, Campus UAB, 08193 Bellaterra (Barcelona) Spain \\
$^{36}$Kavli Institute for Particle Astrophysics \& Cosmology, P. O. Box 2450, Stanford University, Stanford, CA 94305, USA \\
%$^{37-31}$Laborat\'orio Interinstitucional de e-Astronomia - LIneA, Rua Gal. Jos\'e Cristino 77, Rio de Janeiro, RJ - 20921-400, Brazil \\
%$^{38-32}$Observat\'orio Nacional, Rua Gal. Jos\'e Cristino 77, Rio de Janeiro, RJ - 20921-400, Brazil \\
$^{37}$Centro de Investigaciones Energ\'eticas, Medioambientales y Tecnol\'ogicas (CIEMAT), Madrid, Spain \\
$^{38}$Department of Physics, IIT Hyderabad, Kandi, Telangana 502285, India \\
%$^{41-30}$Department of Physics \& Astronomy, University College London, Gower Street, London, WC1E 6BT, UK \\
%$^{42}$Fermi National Accelerator Laboratory, P. O. Box 500, Batavia, IL 60510, USA \\
$^{39}$Institut d'Estudis Espacials de Catalunya (IEEC), 08193 Barcelona, Spain \\
$^{40}$Institute of Space Sciences (ICE, CSIC),  Campus UAB, Carrer de Can Magrans, s/n,  08193 Barcelona, Spain \\
$^{41}$Instituto de Fisica Teorica UAM/CSIC, Universidad Autonoma de Madrid, 28049 Madrid, Spain \\
$^{42}$Department of Astronomy, University of Michigan, Ann Arbor, MI 48109, USA \\
$^{43}$Department of Physics, University of Michigan, Ann Arbor, MI 48109, USA \\
%$^{48-36}$Kavli Institute for Particle Astrophysics \& Cosmology, P. O. Box 2450, Stanford University, Stanford, CA 94305, USA \\
$^{44}$SLAC National Accelerator Laboratory, Menlo Park, CA 94025, USA \\
%$^{50-33}$Department of Astronomy, University of Illinois at Urbana-Champaign, 1002 W. Green Street, Urbana, IL 61801, USA \\
%$^{51-34}$National Center for Supercomputing Applications, 1205 West Clark St., Urbana, IL 61801, USA \\
%$^{52}$Fermi National Accelerator Laboratory, P. O. Box 500, Batavia, IL 60510, USA \\
%$^{53}$Department of Physics \& Astronomy, University College London, Gower Street, London, WC1E 6BT, UK \\
$^{45}$Department of Physics, ETH Zurich, Wolfgang-Pauli-Strasse 16, CH-8093 Zurich, Switzerland \\
$^{46}$Santa Cruz Institute for Particle Physics, Santa Cruz, CA 95064, USA \\
$^{47}$Max Planck Institute for Extraterrestrial Physics, Giessenbachstrasse, 85748 Garching, Germany \\
$^{48}$Universit\"ats-Sternwarte, Fakult\"at f\"ur Physik, Ludwig-Maximilians Universit\"at M\"unchen, Scheinerstr. 1, 81679 M\"unchen, Germany \\
$^{49}$Harvard-Smithsonian Center for Astrophysics, Cambridge, MA 02138, USA \\
$^{50}$Australian Astronomical Observatory, North Ryde, NSW 2113, Australia \\
%$^{60}$Fermi National Accelerator Laboratory, P. O. Box 500, Batavia, IL 60510, USA \\
%$^{61}$Department of Physics \& Astronomy, University College London, Gower Street, London, WC1E 6BT, UK \\
$^{51}$Departamento de F\'isica Matem\'atica, Instituto de F\'isica, Universidade de S\~ao Paulo, CP 66318, S\~ao Paulo, SP, 05314-970, Brazil \\
%$^{63}$Laborat\'orio Interinstitucional de e-Astronomia - LIneA, Rua Gal. Jos\'e Cristino 77, Rio de Janeiro, RJ - 20921-400, Brazil \\
%$^{64}$Laborat\'orio Interinstitucional de e-Astronomia - LIneA, Rua Gal. Jos\'e Cristino 77, Rio de Janeiro, RJ - 20921-400, Brazil \\
%$^{65}$Observat\'orio Nacional, Rua Gal. Jos\'e Cristino 77, Rio de Janeiro, RJ - 20921-400, Brazil \\
$^{52}$Department of Physics and Astronomy, University of Pennsylvania, Philadelphia, PA 19104, USA \\
$^{53}$George P. and Cynthia Woods Mitchell Institute for Fundamental Physics and Astronomy, and Department of Physics and Astronomy, Texas A\&M University, College Station, TX 77843,  USA \\
$^{54}$Department of Astrophysical Sciences, Princeton University, Peyton Hall, Princeton, NJ 08544, USA \\
%$^{69}$Department of Astronomy, University of Illinois at Urbana-Champaign, 1002 W. Green Street, Urbana, IL 61801, USA \\
%$^{70}$National Center for Supercomputing Applications, 1205 West Clark St., Urbana, IL 61801, USA \\
$^{55}$Instituci\'o Catalana de Recerca i Estudis Avan\c{c}ats, E-08010 Barcelona, Spain \\
%$^{72}$Institut de F\'{\i}sica d'Altes Energies (IFAE), The Barcelona Institute of Science and Technology, Campus UAB, 08193 Bellaterra (Barcelona) Spain \\
$^{56}$Jet Propulsion Laboratory, California Institute of Technology, 4800 Oak Grove Dr., Pasadena, CA 91109, USA \\
%$^{74}$Centro de Investigaciones Energ\'eticas, Medioambientales y Tecnol\'ogicas (CIEMAT), Madrid, Spain \\
%$^{75}$Fermi National Accelerator Laboratory, P. O. Box 500, Batavia, IL 60510, USA \\
%$^{76}$Centro de Investigaciones Energ\'eticas, Medioambientales y Tecnol\'ogicas (CIEMAT), Madrid, Spain \\
$^{57}$School of Physics and Astronomy, University of Southampton,  Southampton, SO17 1BJ, UK \\
$^{58}$Brandeis University, Physics Department, 415 South Street, Waltham MA 02453 \\
$^{59}$Instituto de F\'isica Gleb Wataghin, Universidade Estadual de Campinas, 13083-859, Campinas, SP, Brazil \\
%$^{80}$Laborat\'orio Interinstitucional de e-Astronomia - LIneA, Rua Gal. Jos\'e Cristino 77, Rio de Janeiro, RJ - 20921-400, Brazil \\
$^{60}$Computer Science and Mathematics Division, Oak Ridge National Laboratory, Oak Ridge, TN 37831 \\
%$^{82}$National Center for Supercomputing Applications, 1205 West Clark St., Urbana, IL 61801, USA \\
%$^{83}$Department of Physics, University of Michigan, Ann Arbor, MI 48109, USA \\
%$^{84}$Cerro Tololo Inter-American Observatory, National Optical Astronomy Observatory, Casilla 603, La Serena, Chile \\

%}

%%%%%%%%%%%%%%%%%%%%%%%%%%%%%%%%%%%%%%%%%%%%%%%%%%

% Don't change these lines
\bsp	% typesetting comment
\label{lastpage}
\end{document}